\documentclass[12pt]{iopart}
\usepackage{iopams}
\usepackage{setstack}
\usepackage{graphicx}

\begin{document}
\title[Dark energy interacting with dark matter] {Reconstruction
of the interaction term between dark matter and dark energy using SNe Ia, BAO,
CMB, H(z) and X-ray gas mass fraction}
\author{Freddy Cueva Solano and Ulises Nucamendi}

\address{Instituto de F\'{\i}sica y Matem\'aticas\\
Universidad Michoacana de San Nicol\'as de Hidalgo \\
Edificio C-3, Ciudad Universitaria, CP. 58040\\
Morelia, Michoac\'an, M\'exico\\ }


\eads{\mailto{freddy@ifm.umich.mx}, \mailto{ulises@ifm.umich.mx}}

\begin{abstract}
Recently, in \cite{Cueva-Nucamendi2012} we developed a parametric
reconstruction method to a homogeneous, isotropic and spatially flat
Friedmann-Robertson-Walker (FRW) cosmological model filled of a
fluid of dark energy (DE) with constant equation of state (EOS)
parameter interacting with dark matter (DM).
The reconstruction method is based on expansions of the
general interaction term and the relevant cosmological variables in
terms of Chebyshev polynomials which form a complete set orthonormal
functions. This interaction term describes an exchange of energy
flow between the DE and DM within dark sector. In this article, we
reconstruct the interaction function expanding it in terms of only
the first four Chebyshev polynomials
and obtain the best estimation for the coefficients of the expansion
assuming three models: (a) a DE equation of the state parameter $w
=-1$ (an interacting cosmological $\Lambda$), (b) a DE equation of
the state parameter $w =$ constant with a dark matter density parameter
fixed, (c) a DE equation of the state parameter $w =$ constant with
a free constant dark matter density parameter to be estimated, and using the
recent five test data:
Union2 SNe Ia data set from ``The Supernova Cosmology Project'' (SCP)
composed by 557 type Ia supernovae, the Baryonic Acoustic Oscillations (BAO),
the CMB anisotropies from 7-year WMAP experiment, the Hubble expansion rate data
and the X-ray gas mass fraction.
In all the cases, the preliminary reconstruction
shows that in the best scenario there exist the possibility of a
crossing of the noninteracting line $Q=0$ in the recent past within
the $1\sigma$ and $2\sigma$ errors from positive values at early
times to negative values at late times. This means that, in this
reconstruction, there is an energy transfer from DE to DM at early
times and an energy transfer from DM to DE at late times. We
conclude that this fact is an indication of the possible existence
of a crossing behavior in a general interaction coupling between
dark components. Finally, we conclude that in this scenario, the observations put
strong constraints on the strength of the interaction so that its magnitude can not
solve the coincidence problem or at least alleviate significantly.
\end{abstract}

 \submitto{Journal of Cosmology
and Astroparticle physics} \maketitle

\section{Introduction}

Recent observations of the apparent magnitude of Supernova Ia (SNIa) suggest
that the universe is in a stage of recent acceleration
\cite{Riess1998}-\cite{AmanullahUnion22010}. This has now been confirmed by
another sets of independent observational data as such the measurements of
the galaxy power spectrum and the Baryon Acoustic Oscillations (BAO)
detected in the large-scale correlation function of luminous red galaxies in the
experiment Sloan Digital Sky Survey (SDSS) \cite{Eisenstein}-\cite{ReidSDSS},
the power spectrum of the cosmic microwave background (CMB) anisotropies
measured in the experiment Wilkinson Microwave Anisotropy Probe (WMAP)
\cite{WMAP}-\cite{Komatsu2011}. The most simplest explanation to the recent
acceleration is the assumption of the existence of a cosmological constant
$\Lambda$ assumed to stand for the vacuum density energy ($\rho_{\Lambda} \approx \Lambda/8\pi G$).
However, its inferred value from observations
$\rho_{\Lambda}^{\rm{obs}} \approx (10^{-12} \rm{Gev})^{4}$ is too tiny when is compared
with the theoretical value $\rho_{\Lambda}^{\rm{Pl}} = (10^{18} \rm{Gev})^{4}$
estimated from the application of the quantum field theory to
the Planck scale. This is known as the cosmological constant problem
\cite{Weinberg1989}-\cite{Peebles2003}. In addition to this problem, the cosmological
constant model has a new one known in the literature as the cosmic coincidence problem
\cite{Steinhardt1997}, \cite{Sahni-2002}-\cite{Copeland-Sami-2006} which can be
established as: why are both dark densities (dark matter and dark energy) of the
same order of magnitude at the present whereas that they were so different in most of
the past evolution of the universe?. In order to avoid the problem of fine tuning of
the cosmological constant, sometimes it is assumed a null contribution of the vacuum
energy density to the cosmological constant which is fixed to zero and therefore it does not
contribute to the gravitational sector. In this scenario, we need a different explanation for
the present acceleration of the universe. In this direction, it has been proposed a new form
of matter-energy named dark energy (DE) which can be a perfect fluid \cite{AmanullahUnion22010},
\cite{ReidSDSS}, \cite{Komatsu2011}, \cite{Vikhlinin2009}-\cite{Rozo2010}
or a slowly evolving scalar field named quintessence \cite{Peebles1988}-\cite{Copeland-Sami-2006},
both with a equation of state (EOS) parameter $w = P_{DE}/ \rho_{DE}$ varying with the redshift.
In most of the cosmological models considered
in the literature, it is generally assumed that dark matter (DM) and DE interact only gravitationally.
However, it has been argued that, in the absence of an underlying symmetry suppressing a possible
coupling between dark matter and dark energy, a phenomenological interaction between them is not
only possible but necessary \cite{campo-herrera-pavon2008}-\cite{campo-herrera-pavon2009}.
In addition to, it has been speculated that an interaction between dark components is able
to alleviate or inclusive to solve the problem of the Cosmic Coincidence
\cite{campo-herrera-pavon2008}-\cite{olivares-atrio2006}. As a consequence,
a large body of work dealing with such a possibility has been explored in the
literature \cite{Amendola2000}-\cite{campo-herrera2005}.

However, the existence or not of some class of interaction between
dark components is to be discerned observationally. To this respect,
constraints on the strength of such interaction have been put using
different observations \cite{pasqui}-\cite{He-Zhang}.


Recently, it has been suggested that an interacting term $Q(z)$
dependent of the redshift crosses the noninteracting line $Q(z)=0$
\cite{Cai-Su}-\cite{He-Zhang}. In \cite{Cai-Su}, this conclusion
have been obtained using observational data samples in the range $z
\in [0, 1.8]$ in order to fit a scenario in which the whole redshift
range is divided into a determined numbers of bins and the
interaction function is set to be a constant in each bin. They found
an oscillatory behavior of the interaction function $Q(z)$ changing
its sign several times during the evolution of the universe. On the
other hand, in \cite{He-Zhang} is reported a crossing of the
noninteracting line $Q(z) = 0$ under the assumption that the
interacting term $Q(z)$ is a linearly dependent interacting function
of the scale factor with two free parameters to be estimated. They
found a crossing from negative values at the past (energy transfers
from dark matter to dark energy) to positive values at the present
(energy transfers from dark energy to dark matter) at $z \simeq
0.2-0.3$.

In order to shed light on the question of if an interaction term can solved the
\textit{The Cosmic Coincidence problem} or if in this scenario such crossing really
exists, in a previous article \cite{Cueva-Nucamendi2012} we propose the reconstruction
of a quite general nongravitational interaction $Q$ between dark components which
we introduce phenomenologically in the equations of motion. This reconstruction of $Q$
as a function of the redshift was done in terms of an expansion of Chebyshev
Polynomials which constitute a complete orthonormal basis on the finite
interval [-1,1] and have the nice property to be the minimax approximating
polynomial (this technique has been applied to the reconstruction of the DE
potential in \cite{Simon2005}-\cite{Martinez2008}). To this respect, we did the
reconstruction of the interaction function using the data from the Union2 Ia Supernova (SNIa)
data test (557 data) \cite{AmanullahUnion22010}. In this article, we use a set of data
samples covering the wide range of redshift $z \in [0, 1089]$ in order to reconstruct
the interaction function $Q$ from late to early times.


A summary of the paper is as follows. In the second section, we introduce the general formalism
and the equations of motion representing a DE fluid interacting with a DM fluid
in Einstein gravity. In the third section, we write the cosmological equations of motion for the
interacting dark fluids in a Friedmann-Robertson-Walker (FRW) spacetime. We continue in the section
fourth with the introduction and development of the method for reconstructing the interaction
function in terms of an expansion in the base of Chebyshev polynomials. The fifth section is dedicated
to describe the parametric bayesian statistical method used in the reconstruction of the interaction
function together with the chi-square test and the priors used on the free parameters of each model.
Additionally, we present the observations used in this article in order to fit the free parameters of the
models considered. These observations consist in the data from the Union2 Ia Supernova (SNIa)
data test (557 data) \cite{AmanullahUnion22010}, the BAO from SDSS DR7 \cite{ReidSDSS},
the CMB from 7-year WMAP experiment \cite{Komatsu2011}, the Hubble expansion rate (15 data)
\cite{Riess-Hubble}-\cite{Stern2010-1}, \cite{Gaztanaga} and the X-ray gas mass fraction
(42 data) \cite{Allen2008}.

Then, the sixth section is dedicated to the reconstruction of the interaction function and,
to put constraints on the parameters of each model and on the coefficients of the Chebyshev expansion
Then we give a discussion of the results of the reconstruction and the best estimated values
of the parameters of every model fitting the observations. Finally, in the last section we present
a brief resume of our principal results and our conclusions.

\section{General equations of motion for dark energy interacting with dark matter.}
\label{General equations of motion for dark energy interacting with dark
matter}

We assume an universe formed by four components: the baryonic matter
fluid $(b)$, the radiation fluid $(r)$, the dark matter fluid $(DM)$
and the dark energy fluid $(DE)$. Moreover all these constituents
are interacting gravitationally and additionally only the dark
components interact nongravitationally through an energy exchange
between them mediated by the interaction term defined below.

The gravitational equations of motion are the Einstein field
equations
\\
\begin{equation}
\label{EinsteinEquations} G_{\mu\nu} = 8\pi G \left[ T^{b}_{\mu\nu}
+ T^{r}_{\mu\nu} + T^{DM}_{\mu\nu} + T^{DE}_{\mu\nu} \right] ,
\end{equation}
whereas that the equations of motion for each fluid are
\begin{equation}
\label{FluidsEquationsb} \nabla^{\nu} T^{b}_{\mu\nu} = 0 ,
\end{equation}
\begin{equation}
\label{FluidsEquationsr} \nabla^{\nu} T^{r}_{\mu\nu} = 0 ,
\end{equation}
\begin{equation}
\label{FluidsEquationsDM} \nabla^{\nu} T^{DM}_{\mu\nu} = - F_{\mu} ,
\end{equation}
\begin{equation}
\label{FluidsEquationsDE} \nabla^{\nu} T^{DE}_{\mu\nu} = F_{\mu} ,
\end{equation}
where the respective energy-momentum tensor for the fluid $i$ is
defined as $(i = b, r, DM, DE)$,
\begin{equation}\label{Energy_momentum_tensor}
T^{i}_{\mu\nu}=\rho_{i} \,u_\mu u_\nu + (g_{\mu\nu}+u_\mu
u_\nu)P_{i},
\end{equation}
here $u_\mu$ is the velocity of the fluids (assumed to be the same
for each one) where as $\rho_{i}$ and $P_{i}$ are respectively the
density and pressure of the fluid $i$ measured by an observer with
velocity $u^\mu$. $F_{\mu}$ is the cuadrivector of interaction
between dark components and its form is not known a priori because
in general we do not have fundamental theory, in case of existing,
to predict its structure.
We project the equations
(\ref{FluidsEquationsb})-(\ref{FluidsEquationsDE}) in a part
parallel to the velocity $u^\mu$,
\begin{equation}
\label{FluidsEquationsbP} u^\mu \nabla^{\nu} T^{b}_{\mu\nu} = 0 ,
\end{equation}
\begin{equation}
\label{FluidsEquationsrP} u^\mu \nabla^{\nu} T^{r}_{\mu\nu} = 0 ,
\end{equation}
\begin{equation}
\label{FluidsEquationsDMP} u^\mu \nabla^{\nu} T^{DM}_{\mu\nu} = -
u^\mu F_{\mu} ,
\end{equation}
\begin{equation}
\label{FluidsEquationsDEP} u^\mu \nabla^{\nu} T^{DE}_{\mu\nu} =
u^\mu F_{\mu} ,
\end{equation}
and in other part orthogonal to the velocity using the projector
$h_{\beta\mu} = g_{\beta\mu} + u_{\beta}u_{\mu}$ acting on the
hypersurface orthogonal to the velocity $u^\mu$,
\begin{equation}
\label{FluidsEquationsbO} h^{\mu\beta} \nabla^{\nu} T^{b}_{\mu\nu} =
0 ,
\end{equation}
\begin{equation}
\label{FluidsEquationsrO} h^{\mu\beta} \nabla^{\nu} T^{r}_{\mu\nu} =
0 ,
\end{equation}
\begin{equation}
\label{FluidsEquationsDMO} h^{\mu\beta} \nabla^{\nu} T^{DM}_{\mu\nu}
= - h^{\mu\beta} F_{\mu} ,
\end{equation}
\begin{equation}
\label{FluidsEquationsDEO} h^{\mu\beta} \nabla^{\nu} T^{DE}_{\mu\nu}
= h^{\mu\beta} F_{\mu} ,
\end{equation}
using (\ref{Energy_momentum_tensor}) in
(\ref{FluidsEquationsbP})-(\ref{FluidsEquationsDEP}) we obtain the
mass energy conservation equations for each fluid,
\begin{equation}
\label{MECB} u^{\mu} \nabla_{\mu} \rho_{b} + \left( \rho_{b} + P_{b}
\right) \nabla_{\mu} u^{\mu} = 0,
\end{equation}
\begin{equation}
\label{MECr} u^{\mu} \nabla_{\mu} \rho_{r} + \left( \rho_{r} + P_{r}
\right) \nabla_{\mu} u^{\mu} = 0,
\end{equation}
\begin{equation}
\label{MECDM} u^{\mu} \nabla_{\mu} \rho_{DM} + \left( \rho_{DM} +
P_{DM} \right) \nabla_{\mu} u^{\mu} = u^{\mu}F_{\mu},
\end{equation}
\begin{equation}
\label{MECDE} u^{\mu} \nabla_{\mu} \rho_{DE} + \left( \rho_{DE} +
P_{DE} \right) \nabla_{\mu} u^{\mu} = - u^{\mu}F_{\mu}.
\end{equation}
On the other hand, introducing (\ref{Energy_momentum_tensor}) in
(\ref{FluidsEquationsbO})-(\ref{FluidsEquationsDEO}) it permits to
have the Euler equations for every fluid,
\begin{equation}
\label{EEB} h^{\mu\beta} \nabla_{\mu} P_{b} + \left( \rho_{b} +
P_{b} \right) u^{\mu} \nabla_{\mu} u^{\beta} = 0,
\end{equation}
\begin{equation}
\label{EEr} h^{\mu\beta} \nabla_{\mu} P_{r} + \left( \rho_{r} +
P_{r} \right) u^{\mu} \nabla_{\mu} u^{\beta} = 0,
\end{equation}

\begin{equation}
\label{EEDM} h^{\mu\beta} \nabla_{\mu} P_{DM} + \left( \rho_{DM} +
P_{DM} \right) u^{\mu} \nabla_{\mu} u^{\beta} = -
h^{\mu\beta}F_{\mu}  ,
\end{equation}
\begin{equation}
\label{EEDE} h^{\mu\beta} \nabla_{\mu} P_{DE} + \left( \rho_{DE} +
P_{DE} \right) u^{\mu} \nabla_{\mu} u^{\beta} = h^{\mu\beta} F_{\mu}.
\end{equation}
Finally we closed the system of equations assuming the following
state equations for the respectively baryonic, dark matter,
radiation components,
\begin{eqnarray}\label{State EquationB}
P_{b} &=& 0, \,
\\
\label{State EquationDM} P_{DM} &=& 0,  \,
\\
\label{State EquationR} P_{r} &=& \frac{1}{3} \, \rho_{r}, \,
\end{eqnarray}
while for the dark energy we assume a state equation with constant
parameter $w$,
\begin{eqnarray}\label{State EquationDE}
P_{DE} &=& w \rho_{DE}. \,
\end{eqnarray}

\section{Cosmological Equations of motion for dark energy interacting with dark matter.}
\label{Equations of motion for dark energy interacting with dark
matter}
We assumed that the background metric is described by the flat
Friedmann-Robertson-Walker (FRW) metric written in comoving
coordinates as supported by the anisotropies of the cosmic microwave
background (CMB) radiation measured by the WMAP experiment
\cite{WMAP}
\begin{equation}
\label{metricFRW} ds^{2}=-dt^{2} + a^{2}(t) \left( dr^{2} +
r^{2}d\Omega^2 \right ),
\end{equation}
where $a(t)$ is the scale factor and $t$ is the cosmic time.
In these coordinates we choose for the normalized velocity,
\begin{equation}
\label{velocity}
u^{\mu} = (1,0,0,0),
\end{equation}
and therefore we have,
\begin{eqnarray}
\label{expansion}
\nabla_{\mu} u^{\mu} &=& 3\, \frac{\dot{a}}{a} \equiv 3H,
\\
\label{derivative velocity}
u^{\mu} \nabla_{\mu} u^{\beta} &=& 0,
\end{eqnarray}
where $H$ is the Hubble parameter and the point means derivative respect to the cosmic time.
In congruence with the symmetries of spatial isotropy and homogeneity of the FRW spacetime,
the densities and pressures of the fluids are depending only of the cosmic time, $\rho_{i}(t)$, $P_{i}(t)$,
and at the same time, the parallel and orthogonal components of the cuadrivector of interaction
with respect to the velocity are respectively,
\begin{eqnarray}
\label{nonullparallel}
u^{\mu} F_{\mu} &=& Q(a),
\\
\label{nullorthogonal}
 h^{\mu\beta} F_{\mu} &=& 0,
\end{eqnarray}
where $Q(a)$ is known as the interaction function depending on the scale factor.
The introduction of the state equations (\ref{State EquationB})-(\ref{State EquationDE}),
the metric (\ref{metricFRW}) and the expressions (\ref{velocity})-(\ref{nullorthogonal})
in the equations of mass energy conservation for the fluids (\ref{MECB})-(\ref{MECDE}) produces,
\begin{equation}
\label{EoFB}
{\dot \rho}_{b} + 3 H \rho_{b} = 0,
\end{equation}
\begin{equation}
\label{EoFr}
{\dot \rho}_{r} + 4 H \rho_{r} = 0,
\end{equation}
\begin{equation}
\label{EoFDM}
{\dot \rho}_{DM} + 3 H \rho_{DM} = Q,
\end{equation}
\begin{equation}
\label{EoFDE}
{\dot \rho}_{DE} + 3 \left(1 + w \right) H \rho_{DE} = - Q.
\end{equation}
On the other hand, the Euler equations (\ref{EEB})-(\ref{EEDE}) are
satisfied identically and do not produce any new equation. From the
Einstein equation (\ref{EinsteinEquations}) we complete the
equations of motion with the first Friedmann equation,
\begin{equation}
\label{eq:hubble1} H^{2}\left(a \right) = \frac{8\pi G}{3}
\left(\rho_{b} + \rho_{r} + \rho_{DM} + \rho_{DE} \right).
\end{equation}
Its convenient to define the following dimensionless density
parameters $\Omega^{\star}_{i}$, for $i=b, r, DM, DE$, as the energy
densities normalized by the critical density at the actual epoch,
\begin{equation}
\label{densityparameters} \Omega^{\star}_{i} \equiv
\frac{\rho_{i}}{\rho_{crit}^{0}} ,\\
\end{equation}
and the corresponding dimensionless density parameters at the
present,
\begin{equation}
\label{densityparameterstoday} \Omega^{0}_{i} \equiv
\frac{\rho^{0}_{i}}{\rho_{crit}^{0}},\\
\end{equation}
where $\rho_{crit}^{0} \equiv 3H^2_0/8\pi G$ is the critical density
today and $H_0$ is the Hubble constant. Solving (\ref{EoFB}) and
(\ref{EoFr}) in terms of the redshift $z$, defined as $a = 1/(1+z)$,
we obtain the known solutions for the baryonic matter and radiation
density parameters respectively:
\begin{equation}
\label{Omegabs} \Omega^{\star}_{b}(z) =
\Omega_{b}^{0}{(1+z)}^{3},\\
\end{equation}
\begin{equation}
\label{Omegars} \Omega^{\star}_{r}(z) =
\Omega_{r}^{0}{(1+z)}^{4}.\\
\end{equation}
The energy conservation equations (\ref{EoFDM}) and (\ref{EoFDE})
for both dark components are rewritten in terms of the redshift as:
\begin{eqnarray}
\label{EDFDM} \frac{\mathrm{d}{\rho}_{DM}}{\mathrm{d}z} -
\frac{3}{1+z} \,\rho_{DM}= -
\frac{Q(z)}{(1 + z) \cdot H(z)},\\
\label{EDFDE} \frac{\mathrm{d}{\rho}_{DE}}{\mathrm{d}z} -
\frac{3(1+w)}{1 + z}\, \rho_{DE} = \frac{Q(z)}{(1+z) \cdot H(z)}.
\end{eqnarray}
Phenomenologically, we choose to describe the interaction between
the two dark fluids as an exchange of energy at a rate proportional
to the Hubble parameter: \\
\begin{eqnarray}
\label{Interaction} Q(z) &\equiv& \rho_{crit}^{0}\cdot(1+z)^{3}
\cdot H(z)\cdot {\rm I}_{\rm Q}(z).
\end{eqnarray}
The term $\rho_{crit}^{0}\cdot(1+z)^{3}$ has been introduced by
convenience in order to mimic a rate proportional to the behavior of
a matter density without interaction. Let be note that the
dimensionless interaction function ${\rm I}_{\rm Q}(z)$ depends
of the redshift and it will be the function to be reconstructed.
With the help of (\ref{Interaction}), we rewrite the equations for
the dark fluids (\ref{EDFDM})-(\ref{EDFDE}) as,
\begin{eqnarray}
\label{EDFDMOmega} \frac{\mathrm{d}
\Omega^{\star}_{DM}}{\mathrm{d}z} -
\frac{3}{1+z} \,\Omega^{\star}_{DM}= - (1 + z)^2 \cdot {\rm I}_{\rm Q}(z),\\
\label{EDFDEOmega} \frac{\mathrm{d}
\Omega^{\star}_{DE}}{\mathrm{d}z} - \frac{3(1+w)}{1 + z}\,
\Omega^{\star}_{DE}= (1 + z)^2 \cdot {\rm I}_{\rm Q}(z).
\end{eqnarray}

\section{General Reconstruction of the interaction using Chebyshev
polynomials.} \label{General Reconstruction}

We do the parametrization of the dimensionless coupling ${\rm
I}_{\rm Q}(z)$ in terms of the Chebyshev polynomials, which form a
complete set of orthonormal functions on the interval $[-1, 1]$.
They also have the property to be the minimax approximating
polynomial, which means that has the smallest maximum deviation from
the true function at any given order \cite{Simon2005}-\cite{Martinez2008}. Without loss of generality,
we can then expand the coupling ${\rm I}_{\rm Q}(z)$ in the redshift
representation as:
\begin{equation}
\label{eq:Coupling} {\rm I}_{\rm Q}(z) \equiv
\sum_{n=0}^{N}\lambda_{n} \cdot T_{n}(z),
\end{equation}
where $T_{n}(z)$ denotes the Chebyshev polynomials of order $n$ with
$n \in [0,N]$ and $N$ a positive integer. The coefficients of the
polynomial expansion $\lambda_{n}$ are real free dimensionless
parameters. Then the interaction function can be rewritten as
\begin{equation}
\label{Coupling} Q(z) = \rho_{crit}^{0}\cdot(1+z)^{3} \cdot
H(z)\cdot \sum_{n=0}^{N}\lambda_{n}\cdot T_{n}(z).
\end{equation}
We introduce (\ref{eq:Coupling}) in
(\ref{EDFDMOmega})-(\ref{EDFDEOmega}) and integrate both equations
obtaining the solutions,
\begin{eqnarray}
\label{eq:Omega12} \Omega^{\star}_{DM}(z) &=& (1+z)^{3}\left[
{\Omega_{DM}^0} - \frac{z_{max}}{2} \sum_{n=0}^{N}
\lambda_{n} \, \cdot K_{n}(x, 0) \right],\\
\label{eq:Omega13} \Omega^{\star}_{DE}(z) &=& (1+z)^{3(1+w)}\left[
{\Omega_{DE}^0} + \frac{z_{max}}{2} \sum_{n=0}^{N} \lambda_{n} \,
\cdot K_{n}(x, w) \right]  \,,
\end{eqnarray}
where we have defined the integrals
\begin{eqnarray}
\label{IntegralK} K_{n}(x, w) \equiv \int_{-1}^{x}
\frac{T_{n}(\tilde{x})}{(a+b\tilde{x})^{(1+3w)}} \, d\tilde{x} \,\,,
\end{eqnarray}
and the quantities,
\begin{eqnarray}
x \equiv \frac{2 \,z}{z_{max}} - 1 , \\
a \equiv 1\,+\, \frac{z_{max}}{2} , \\
b \equiv \frac{z_{max}}{2},
\end{eqnarray}
here $z_{max}$ is the maximum redshift at which observations are
available so that $x \in [-1, 1]$ and $ \vert T_{n}(x)
\vert\leq 1$, \,for all $n \in [0,N]$.\\
Finally, using the solutions (\ref{Omegabs})-(\ref{Omegars}) and
(\ref{eq:Omega12})-(\ref{eq:Omega13}) we rewrite the Friedmann
equation (\ref{eq:hubble1}) as \vspace{0.5cm}
\begin{equation}
\label{eq:hubble2} H^{2}\left(z \right) = H^2_{0}
\left[\Omega_{b}^{0}{(1+z)}^{3} + \Omega_{r}^{0}{(1+z)}^{4} +
\Omega^{\star}_{DM}(z) + \Omega^{\star}_{DE}(z) \right].
\end{equation}
\vspace{0.5cm}
The Hubble parameter depends of the
parameters ($H_0$, $\Omega_{b}^0$, $\Omega_{r}^0$, $\Omega_{DM}^0$,
$\Omega_{DE}^0$, $w$) and the dimensionless coefficients
$\lambda_{n}$. However one of the parameters depends of the
others due to the Friedmann equation evaluated at the present,
\begin{equation}
\label{HubblePresent} \Omega^{0}_{DE} = 1 - \Omega_{b}^{0} -
\Omega_{r}^{0} - \Omega^{0}_{DM}.
\end{equation}
At the end, for the reconstruction, we have the five parameters
($H_0$, $\Omega_{b}^0$, $\Omega_{r}^0$, $\Omega_{DM}^0$, $w$) and
the dimensionless coefficients $\lambda_{n}$.

To do a general reconstruction in
(\ref{eq:Omega12})-(\ref{eq:Omega13}) we must take $N\rightarrow
\infty$ and to obtain the solutions in a closed form. The details of
the calculation of the integrals $K_{n}(x, w)$ in the right hand
side of (\ref{eq:Omega12})-(\ref{eq:Omega13}) are shown in detail in
the Appendix A which shows the closed forms
(A.9)-(A.10) for the integrals with odd and even integer $n$ subindex,
and valid for $w\neq n/3$, where $n\geq 0$ (see ref. \cite{Cueva-Nucamendi2012}).
Finally, we point out the formulas we use for the reconstruction of
other important cosmological properties of the universe:
\begin{itemize}
    \item The age of the universe at redshift $z$ in \cite{Saini2000}:
\begin{eqnarray}
t_{0}(z)=\int_{z}^{\infty}\frac{d\tilde z}{(1+\tilde z).H(\tilde z)}.
\end{eqnarray}
    \item The deceleration parameter:
\begin{eqnarray}
q(z) =-1 + \frac{(1 + z)}{H(z)} \cdot \frac{d H(z)}{dz}.
\end{eqnarray}
\end{itemize}

\section{Current observational data and cosmological constraint.} \label{SectionAllTest}
To simplify our analysis, we reconstruct the coupling function ${\rm I}_{\rm Q}(z)$ to
different orders ($N=1,2,3,4$), up to order $N=4$. The details of this reconstruction are described
in the Appendix B. (see ref. \cite{Cueva-Nucamendi2012}).
In order to do it, we test our cosmological models using the observational data currently available.
In this section, we then describe how we use these data.


\subsection{Type Ia supernovae.} \label{SNIa}
We test and constrain the coupling function ${\rm
I}_{\rm Q}(z)$ using the ``Union2'' SNe Ia data set from ``The
Supernova Cosmology Project'' (SCP) composed by 557 type Ia
supernovae \cite{AmanullahUnion22010}. \noindent As it is usual, we
use the definition of luminosity distance $d_L$ (see ref. \cite{Riess1998}) in a flat cosmology,
\begin{eqnarray}\label{luminosity_distance1}
d_L(z, \mathbf{X}) &=& c(1+z) \int_0^z \frac{dz'}{H(z', \mathbf{X})},
\end{eqnarray}
\noindent where $H(z, \mathbf{X})$ is the Hubble parameter, i.e.,
the expression (\ref{eq:hubble2}), ''$c$'' is the speed of light given
in units of km/sec and $\mathbf{X}$ represents the parameters of the model,
\begin{eqnarray}
\label{parametersm} \mathbf{X} & \equiv & (H_0, \Omega_{b}^0, \Omega_{r}^0,
\Omega_{DM}^0, w, \lambda_{1},..., \lambda_{N}).
\end{eqnarray}
\noindent The \emph{theoretical distance moduli} for the $k$-th
supernova with redshift $z_k$ is defined as
\begin{eqnarray}\label{distanceModuli}
\mu^{{\rm th}}(z_k, \mathbf{X}) & \equiv & m(z)-M = 5\log_{10}
\left[\frac{d_L(z_k, \mathbf{X})}{{\rm Mpc}} \right] +25,
\end{eqnarray}
\noindent where $m$ and $M$ are the apparent and absolute magnitudes
of the SNe Ia respectively, and the superscript ``th'' stands for
\textit{``theoretical''}. We construct the statistical $\chi^{2}_{\bf SN}$
function as
\begin{eqnarray}\label{ChiSquareDefinition}
\chi^{2}_{\bf SN} (\mathbf{X}) & \equiv & \sum_{k = 1}^{557} \frac{\left[\mu^{{\rm
th}} (z_k, \mathbf{X}) - \mu_k \right]^2}{\sigma_k^2},
\end{eqnarray}
\noindent where $\mu_k$ is the \emph{observational} distance moduli
for the $k$-th supernova, $\sigma_k^2$ is the variance of the
measurement.
With this function $ \chi^{2}_{\bf SN}$, we construct the probability density
function (\textbf{pdf}) as
\begin{equation}\label{expChi1}
{{\rm {\bf pdf}}}_{\bf SN}(\mathbf{X}) = {\rm{A}}_{1} \cdot{{\rm e}}^{-\chi^{2}_{\bf SN}(\mathbf{X}) \,/2},
\end{equation}
\noindent where ${\rm A}_{1}$ is a integration constant.


\subsection{Baryon acoustic oscillations} \label{BAO}
The baryon acoustic oscillations (BAO) are detected in the clustering of the combined 2dFGRS and SDSS main galaxy samples,
and measure the distance-redshift relation at $z=0.2$. Additionally, BAO in the clustering of the SDSS luminous red galaxies
measure the distance-redshift relation at $z=0.35$. The observed scale of the BAO calculated from these samples, as well as
from the combined sample, are jointly analyzed using estimates of the correlated errors to constrain the form of the distance
measure $D_{v}(z)$ \cite{Eisenstein}-\cite{Percival}
\begin{equation} \label{Dv}
D_{v}(z, \mathbf{X})\equiv \left(\left(\int_{0}^{z}\frac{dz'}{H(z', \mathbf{X})}\right)^{2}\frac{z}{H(z, \mathbf{X})} \right)^{1/3}.
\end{equation}
where $\mathbf{X}$ represents the parameters of the model, see equation (\ref{parametersm}).
The peak position of the BAO depends on the ratio of $D_{v}(z)$ to the sound horizon size at the drag epoch (where baryons were released from photons)
$z_{d}$, which can be obtained by using a fitting formula \cite{Eisenstein}:
\begin{equation} \label{zd}
z_{d}=\frac{1291(\Omega_{M})^{-0.419}}{1+0.659(\Omega_{M})^{0.828}} \left[1+ b_{1}(\Omega_{b}^{0})^{b_{2}} \right],
\end{equation}
where $\Omega_{M}=\Omega^{0}_{DM} + \Omega_{b}^{0} $ and
\begin{eqnarray}
\label{b1} b_{1}=0.313(\Omega_{M})^{-0.419} \left[1+ 0.607(\Omega_{M})^{0.674} \right], \\
\label{b2} b_{2}=0.238(\Omega_{M})^{0.223}.
\end{eqnarray}
We use the distance radio $d_{z}$ at $z=0.2$ and $z=0.35$ \cite{Eisenstein}-\cite{Percival}
\begin{equation} \label{dvalues}
d_{0.2}(\mathbf{X})=\frac{r_{s}(z_{d})}{D_{V}(0.2, \mathbf{X})},\hspace{0.3cm} d_{0.35}(\mathbf{X})=\frac{r_{s}(z_{d})}{D_{V}(0.35, \mathbf{X})},
\end{equation}
where $r_{s}(z_{d}, \mathbf{X})$ is the comoving sound horizon size at the baryon drag epoch.
\begin{equation}\label{rz}
r_{s}(z, \mathbf{X})=\frac{1}{\sqrt3} {\int}^{1/(1+z)}_{0} \frac{dz'}{H(z', \mathbf{X})\sqrt{1 + \frac{3 \Omega_{b}^{0}}{4(1+z') \Omega_{r}^{0}}}}.
\end{equation}
Using the data of BAO of the table \ref{tableBAO} and the following inverse covariance matrix of BAO in \cite{Percival}
\begin{table}
\centering
\begin{tabular}{|c|c|}
\multicolumn{2}{c}{\textbf{The observational $d_{z}$ data}}\\
\hline
 $z$ & $d_{z}$  \\
 \hline
 $0.2$ & $0.1905\pm0.0061$ \\
 \hline
 $0.35$ & $0.1097\pm0.0036$ \\
\hline
\end{tabular}
\caption{Summary of the BAO data set.}
 \label{tableBAO}
\end{table}
\begin{equation} \label{MatrixBAO}
C^{-1}_{\bf BAO}=\left( \begin{array}{cc}
+30124  & -17227 \\
-17227  & +86977 \\
\end{array} \right)
\end{equation}
thus, the $\chi^{2}$ function of the BAO data is constructed as:
\begin{equation}\label{X2BAO}
\chi_{\bf BAO}^{2}(\mathbf{X})=\left(d^{th}_{i}(\mathbf{X})-d^{obs}_{i}\right)^{t}\left(C^{-1}_{\bf BAO}\right)_{ij}\left(d^{th}(\mathbf{X})_{j}-d^{obs}_{j}\right),
\end{equation}
where $(d^{th}-d^{obs})$ is a column vector formed from the values of theory minus the corresponding observational data, with
\begin{equation}
d^{th}_{i}(\mathbf{X})-d^{obs}_{i}=\left(\begin{array}{c}
 d_{0.20}(\mathbf{X})-0.1905 \\
 d_{0.35}(\mathbf{X})-0.1097 \\
\end{array}\right)
\end{equation}
and "t`` denotes its transpose.
With this $ \chi^{2}_{\bf BAO}$ function, we construct the probability density
function (\textbf{pdf}) as
\begin{equation}\label{expChi2}
{{\rm {\bf pdf}}}_{\bf BAO}(\mathbf{X}) = {\rm{A}}_{2} \cdot{{\rm e}}^{-\chi^{2}_{\bf BAO}(\mathbf{X}) \,/2}
\end{equation}
\noindent where ${\rm A}_{2}$ is a integration constant.


\subsection{Cosmic microwave background (CMB)} \label{CMB}
The CMB shift parameter \rm{\textit{R}} is provided by \cite{Bond-Tegmark1997}
\begin{equation} \label{Shiftparameter}
\rm{R}(z_{*}, \mathbf{X})\equiv \sqrt{\Omega_{M}H^{2}_{0}}(1+z_{*})(D_{A}(z_{*}, \mathbf{X})/c),
\end{equation}
where $\mathbf{X}$ represents the parameters of the model (see equation (\ref{parametersm}) and the second distance ratio $D_{A}$ is given by:
\begin{eqnarray} \label{DA}
 D_{A}(z_{*}, \mathbf{X})=\frac{1}{(1+z_{*})}{\int}^{z_{*}}_{0}\frac{dz'}{H(z', \mathbf{X})}.
\end{eqnarray}
The redshift $z_{*}$ (the decoupling epoch of photons) is obtained using the fitting function \cite{Hu-Sugiyama1996}
\begin{equation} \label{Redshift_decoupling}
z_{*}=1048 \left[1+0.00124(\Omega^{0}_{b})^{-0.738}\right]\left[1+g_{1}(\Omega_{M})^{g_{2}} \right],
\end{equation}
where $\Omega_{M}=\Omega^{0}_{DM} + \Omega_{b}^{0} $ and the functions $g_{1}$ and $g_{2}$ are
\begin{eqnarray}\label{g1g2}
g_{1}=\frac{0.0783(\Omega_{b0})^{-0.238}}{1+39.5(\Omega_{b0})^{0.763}},\\
g_{2}=\frac{0.560}{1+21.1(\Omega_{b0})^{1.81}}. \,
\end{eqnarray}
In addition, the acoustic scale $l_{A}$ describes the ratio $D_{A}(z_{*})/r_{s}(z_{*})$, and is defined as
\begin{equation}\label{Acoustic_scale}
l_{A}(\mathbf{X})\equiv(1+z_{*})\frac{\pi D_{A}(z_{*}, \mathbf{X})}{r_{s}(z_{*}, \mathbf{X})},
\end{equation}
where a factor of $(1+z_{*})$ arises because $D_{A}(z_{*}, \mathbf{X})$ is the proper angular diameter distance, whereas
$r_{s}(z_{*}, \mathbf{X})$ is the comoving sound horizon at $z_{*}$. The fitting formula of $r_{s}(z, \mathbf{X})$ is given by the equation (\ref{rz}).
The seven year WMAP observations \cite{Komatsu2011} give the maximum likelihood values according to table \ref{tableCMB}.
\begin{table}
\centering
\begin{tabular}{|c|c|}
 \hline
 $l_{A}(z_{*})$ & $302.40$  \\
 \hline
 $R(z_{*})$ & $1.7239$ \\
 \hline
 $z_{*}$ & $1090.89$ \\
 \hline
\end{tabular}
\caption{The observational CMB data.}
\label{tableCMB}
\end{table}
Following \cite{Komatsu2011}, the $\chi^{2}$ function for the CMB data is
\begin{equation}\label{X2CMB}
\chi_{\bf CMB}^{2}(\mathbf{X})=\left(x^{th}_{i}(\mathbf{X})-x^{obs}_{i}\right)^{t}\left(C^{-1}_{\bf CMB}\right)_{ij}\left(x^{th}_{j}(\mathbf{X})-x^{obs}_{j}\right),
\end{equation}
where $(x^{th}-x^{obs})$ is a column vector formed from the values of theory minus the corresponding observational data, with
\begin{equation}
x^{th}_{i}(\mathbf{X})-x^{obs}_{i}=\left(
\begin{array}
 l_{A}(z_{*})-302.40 \\
 R(z_{*})-1.7239 \\
 z_{*}-1090.89\\
\end{array}\right)
\end{equation}
"t`` denotes its transpose and $(C^{-1}_{\bf CMB})_{ij}$ is the inverse covariance matrix.
In \cite{Komatsu2011}, the inverse covariance matrix is also given as follows
\begin{equation}\label{MatrixCMB}
C^{-1}_{\bf CMB}=\left(
\begin{array}{ccc}
+2.3050 & +29.6980 & -1.3330 \\
+29.698 & +6825.27 & -113.18 \\
-1.3330 & -113.180 & +3.4140 \\
\end{array}
\right).
\end{equation}
With this function $ \chi^{2}_{\bf CMB}$, we construct the probability density
function (\textbf{pdf}) as
\begin{equation}\label{expChi3}
{{\rm {\bf pdf}}}_{\bf CMB}(\mathbf{X}) = {\rm{A}}_{3} \cdot{{\rm e}}^{-\chi^{2}_{\bf CMB}(\mathbf{X}) \,/2}
\end{equation}
\noindent where ${\rm A}_{3}$ is a integration constant.


\subsection{Hubble expansion rate (H)} \label{OHD}
The Hubble parameter can be written as the following form:
\begin{equation}\label{Hubble}
H(z)=-\frac{1}{1+z}\frac{dz}{dt}.
\end{equation}
So,through measuring $dz/dt$, we can obtain $H(z)$. In \cite{Jimenez2002}, \cite{Jimenez2003} and \cite{Simon2005},
the authors indicated that it is possible to use absolute ages of passively evolving galaxies to compute values of
$dz/dt$. The galactic spectral data used by \cite{Simon2005} come from the Gemini Deep Survey \cite{Abraham2004} and the archival data
\cite{Dunlop1996}-\cite{Nolan2003}. Detailed calculations of $dz/dt$ can be found in \cite{Simon2005}. Simon obtained $H(z)$ in the range of [0,1.8].
The twelve observational Hubble data from \cite{Riess-Hubble},\cite{Stern2010-1} are list in the table \ref{tableOHD}. In addition, in
\cite{Gaztanaga}, the authors took the BAO scale as a standard ruler in the radial direction, obtain three more additional data:
$H(z=0.24)=79.69 \pm 2.32$,\, $H(z=0.34)=83.80 \pm 2.96$ and $H(z=0.43)=86.45 \pm 3.27$ (in units of $Km \,s^{-1} Mpc^{-1}$).


\begin{table}
  \centering
\begin{tabular}{| c | c | c | c | c | c | c | c | c | c | c | c | c |}
\hline
$z$ & $0.0$ & $0.1$ & $0.17$ & $0.27$ & $0.4$ & $0.48$ & $0.88$ & $0.90$ & $1.30$ & $1.43$ & $1.53$ & $1.73$ \\
\hline
$H(z)(Km \, s^{-1}Mpc^{-1})$ & $74.2$ & $69$ & $83$ & $77$ & $95$ & $97$ & $90$ & $117$ & $168$ & $177$ & $140$ & $202$ \\
\hline
$1\sigma$ & $\pm3.6$ & $\pm12$ & $\pm8$ & $\pm14$ & $\pm17$ & $\pm60$ & $\pm40$ & $\pm23$ & $\pm17$ & $\pm18$ & $\pm14$ & $\pm40$ \\
\hline
\end{tabular}
  \caption{The observational $H(z)$ data.}
  \label{tableOHD}
\end{table}
   We can use these data to constraint different kinds of dark energy models and determine the best fit values of the model parameters by
minimizing \cite{Lazkoz2007}
\begin{eqnarray}\label{X2OHD}
\chi^2_{H}(\mathbf{X}) & \equiv & \sum_{i = 1}^{15} \frac{\left[H^{{\rm th}}(\mathbf{X},z_{i},)-H^{obs}(z_{i}) \right]^2}{\sigma^2(z_{i})}.
\end{eqnarray}
where $\mathbf{X}$ represents the parameters of the model (see equation (\ref{parametersm}), $H^{{\rm th}}$ is the theoretical value  for the Hubble parameter, $H^{obs}$ is the observed value, $\sigma(z_{i})$ is the standard
deviation of the measurement, and the summation is over the 15 observational Hubble data points.
This test has already been used to constrain several cosmological models \cite{Yi2007}-\cite{Xu2010}.
With this function $ \chi^{2}_{\bf H}$ we construct the probability density
function (\textbf{pdf}) as
\begin{equation}\label{expChi4}
{{\rm {\bf pdf}}}_{\bf H}(\mathbf{X}) = {\rm{A}}_{4} \cdot{{\rm e}}^{-\chi^{2}_{\bf H}(\mathbf{X}) \,/2}
\end{equation}
\noindent where ${\rm A}_{4}$ is a integration constant.


\subsection{X-ray gas mass fraction (X-ray)} \label{Xray}
According to the X-ray cluster gas mass fraction observations, the baryon mass fraction in clusters of galaxies can be used to constrain cosmological
parameters. The X-ray cluster gas mass fraction, $f_{gas}$, is defined as the radio of the X-ray gas mass to the total mass of a cluster, which is a
constant and independent of the redshift. In the framework of the $\Lambda CDM$ reference cosmology, the X-ray cluster gas mass fraction is presented as
\cite{Allen2008}
\begin{eqnarray}\label{fgas}
f_{gas}(z)=\frac{K A \gamma b(z)}{1+s(z)}\left( \frac{\Omega_{b}^{0}}{\Omega_{b}^{0}+\Omega^{0}_{DM}} \right) \left(\frac{D^{\Lambda CDM}_{A}(z,\mathbf{X} )}{D_{A}(z,\mathbf{X})} \right)^{1.5}
\end{eqnarray}
where $\mathbf{X}$ represents the parameters of the model.
The factor $K$ is used to describe the combined effects of the residual uncertainties, such as the instrumental calibration and certain X-ray
modeling issues, and a Gaussian prior for the calibration factor is considered by $K=1.0\pm0.1$ \cite{Allen2008}. The parameter $\gamma$ denotes
permissible departures from the assumption of hydrostatic equilibrium, due to non-thermal pressure support; the bias factor $b(z)= b_{0}(1+{\alpha}_{b}Z)$
accounts for uncertainties in the cluster depletion factor; $s(z)= s_{0}(1+{\alpha}_{s}Z)$ accounts for uncertainties of the baryon mass fraction in stars
and a Gaussian prior for $s_{0}$ is employed, with $s_{0}=0.16\pm0.05$ \cite{Allen2008} and $A$ is the angular correction factor,
which is caused by the change in angle for the current test model $\Theta_{2500}$ in comparison with that of the reference cosmology
$\Theta^{\Lambda CDM}_{2500}$:
\begin{eqnarray}\label{Afactor}
A=\left(\frac{\Theta^{\Lambda CDM}_{2500}}{\Theta_{2500}} \right){\eta}\approx \left(\frac{H(z, \mathbf{X})D_{A}(z, \mathbf{X})}{\left[H(z, \mathbf{X})D_{A}(z, \mathbf{X})\right]^{\Lambda CDM}}\right)^{\eta}, \\
\end{eqnarray}
here, the index $\eta$ is the slope of the $f_{gas}(r/r_{2500})$ data within the radius $r_{2500}$, with the best-fit average value
$\eta=0.214\pm0.022$ \cite{Allen2008}. And the proper angular diameter distance $D_{A}(z)$ is given by the equation (\ref{DA}).
Following the method in \cite{Allen2008} and \cite{Nesseris2007} and adopting the updated 42 observational $f_{gas}$ data in \cite{Allen2008}, the best fit values
of the model parameters for the X-ray cluster gas mass fraction analysis are determined by minimizing,
\begin{eqnarray}\label{Xgas}
{{\chi}^{2}}_{X_{ray}}(\mathbf{X}) & \equiv & \sum_{i = 1}^{42} \frac{\left[\left(f_{gas} \right)^{{\rm
th}}(z_{i}, \mathbf{X})- \left(f_{gas}(z_{i}) \right)^{obs} \right]^2}{\sigma^2(z_{i})}.
\end{eqnarray}
With this function $ \chi^{2}_{\bf X_{ray}}$, we construct the probability density
function (\textbf{pdf}) as
\begin{equation}\label{expChi5}
{{\rm {\bf pdf}}}_{\bf X_{ray}}(\mathbf{X}) = {\rm{A}}_{5} \cdot{{\rm e}}^{-\chi^{2}_{\bf X_{ray}}(\mathbf{X}) \,/2}
\end{equation}
\noindent where ${\rm A}_{5}$ is a integration constant.
We have considered the following values to do the fit
\begin{eqnarray*}\label{values1}
K=+1.0, \, \, \gamma=+0.9980,\, \,s_{0}=+0.16, \\
\label{values2}
\alpha_{s}=+0.150,\, \,b_{0}=+0.787, \, \, \alpha_{b}=-0.080. \,\end{eqnarray*}
Therefore, using the equations (\ref{expChi1}), (\ref{expChi2}), (\ref{expChi3}), (\ref{expChi4}) and (\ref{expChi5}) we build the total probability density function as
\begin{equation}\label{TotalPdf}
{{\rm{\bf Pdf}}}={{\rm{\bf pdf}}}_{\bf SN}\cdot{{\rm{\bf pdf}}}_{\bf BAO}\cdot{{\rm{\bf pdf}}}_{\bf CMB}\cdot{{\rm{\bf pdf}}}_{\bf H}\cdot{{\rm{\bf pdf}}}_{\bf X_{ray}},
\end{equation}
We can then rewrite the above equation as:
\begin{equation}\label{RewriteTotalPdf}
{{\rm{\bf Pdf}}}(\mathbf{X})={\rm{A}}\cdot{{\rm e}}^{-\chi^{2}(\mathbf{X}) \,/2}, \,
\end{equation}from here we define the total function $\chi^{2}_{T}$ as,
\begin{eqnarray}\label{TotalX2}
{{\chi}^{2}}\equiv{{\chi}^{2}}_{\bf SN}+{{\chi}^{2}}_{\bf BAO}+{{\chi}^{2}}_{\bf CMB}+{{\chi}^{2}}_{\bf H}+{{\chi}^{2}}_{X_{\bf ray}}.
\end{eqnarray}
Thus the constraints on cosmological models from a combination of the above discussed observational datasets can be obtained by minimizing the equation (\ref{TotalX2}).

\subsection{Priors on the total probability density function (\bf{Pdf}).} \label{PriorsSS}
In the models I, II and III shown in the Table \ref{tablemodels}, we
marginalize the parameters $\mathbf{Y} = (H_0, \Omega_{DM}^0,
\Omega_{b}^0, \Omega_{r}^0$) in the \textbf{pdf} (\ref{TotalPdf})
choosing priors on them. In order to it, we must compute the
following integration,
\begin{eqnarray}\label{MarginalizationIntegral2}
 {{\rm {\bf Pdf}}}(\mathbf{V}) = \int^{\infty}_{0}
 \int^{\infty}_{0} \int^{\infty}_{0} \int^{\infty}_{0}
 {{\rm {\bf Pdf}}}(\mathbf{X}) \,{{\rm {\bf Pdf}}}
 (\mathbf{Y}) \, dH_0 \,
 d\Omega_{DM}^0 \, d\Omega_{b}^0 \, d\Omega_{r}^0,
\end{eqnarray}
\noindent where $\mathbf{V} = (w, \lambda_{1},..., \lambda_{N})$
represents the nonmarginalized parameters, ${{\rm {\bf
Pdf}}}(\mathbf{X}) $ is given by (\ref{TotalPdf}) and ${{\rm {\bf
Pdf}}}(\mathbf{Y})$ is the \textit{prior} probability distribution
function for the parameters ($H_0, \Omega_{DM}^0, \Omega_{b}^0,
\Omega_{r}^0$) which are chosen as Dirac delta priors around the
specific values $\mathbf{\tilde{Y}} = (\tilde{H}_0,
\tilde{\Omega}_{DM}^0, \tilde{\Omega}_{b}^0, \tilde{\Omega}_{r}^0$)
measured by some other independent observations,
\begin{equation}\label{Priors}
{{\rm {\bf Pdf}}}(\mathbf{Y}) = \delta(H_0 - \tilde{H}_0)
 \cdot \delta(\Omega_{DM}^0 -  \tilde{\Omega}_{DM}^0)
 \cdot \delta(\Omega_{b}^0 -  \tilde{\Omega}_{b}^0)
 \cdot \delta(\Omega_{r}^0 -  \tilde{\Omega}_{r}^0).
\end{equation}
\noindent Introducing (\ref{Priors}) in
((\ref{MarginalizationIntegral2}) it produces,
\begin{equation}\label{expChinew}
{{\rm {\bf Pdf}}}(\mathbf{V}) = \rm{A} \cdot{{\rm
e}}^{-\tilde{\chi}^2(\mathbf{V}) \,/2},
\end{equation}
\noindent where we have defined a new function $\tilde{\chi}^2$
depending only on the parameters $\mathbf{V} = (w, \lambda_{1},...,
\lambda_{N})$ as,
\begin{eqnarray}\label{ChiSquareew}
\tilde{\chi}^{2}(\mathbf{V}) & \equiv & \sum_{k = 1}^n
\frac{\left[\mu^{{\rm th}} (z_k, \mathbf{V}, \mathbf{\tilde{Y}}) -
\mu_k \right]^2}{\sigma_k^2}.
\end{eqnarray}
The specific values chosen for the Dirac delta priors are,
\begin{itemize}
\item
$\tilde{H}_0 = 70.4 \; ({{\rm km}}/{{\rm s}}){{\rm Mpc}}^{-1}$ as
suggested by the observations in \cite{Komatsu2011}.
\item $\tilde{\Omega}_{DM}^0 = 0.227$,
\item $\tilde{\Omega}_{b}^0 = 0.0456$,
\item $\tilde{\Omega}_{r}^0 = 8.42095\times 10^{-5}$.
\end{itemize}

Once constructed the function $\tilde{\chi}^2$ (\ref{TotalX2}),
we numerically minimize it to compute the ``\textit{best
estimates}'' for the free parameters of the model: $\mathbf{V} = (w,
\lambda_{1},..., \lambda_{N})$. The minimum value of the
$\tilde{\chi}^2$ function gives the best estimated values of
$\mathbf{V}$ and measures the goodness-of-fit of the model to data.
For the Model IV, we leave too the parameter $\Omega_{DM}^0$ free to
vary and estimated it from the minimization of the $\tilde{\chi}^2$
function. In this case, the parameters to be marginalized are
$\mathbf{Y} = (H_0, \Omega_{b}^0, \Omega_{r}^0$). Then, the
marginalization will be as,
\begin{eqnarray}\label{MarginalizationIntegral3}
 {{\rm {\bf Pdf}}}(\mathbf{V}) =
 \int^{\infty}_{0} \int^{\infty}_{0} \int^{\infty}_{0}
 {{\rm {\bf Pdf}}}(\mathbf{X}) \,{{\rm {\bf Pdf}}}
 (\mathbf{Y}) \, dH_0 \, d\Omega_{b}^0 \, d\Omega_{r}^0
\end{eqnarray}
\noindent where now $\mathbf{V} = (w, \Omega_{DM}^0,
\lambda_{1},..., \lambda_{N})$ represents the nonmarginalized
parameters to be estimated, ${{\rm {\bf Pdf}}}(\mathbf{X}) $ is
given by (\ref{TotalPdf}) and ${{\rm {\bf Pdf}}}(\mathbf{Y})$ is the
\textit{prior} probability distribution function for the parameters
($H_0, \Omega_{b}^0, \Omega_{r}^0$) which are chosen as Dirac delta
priors around the specific values $\mathbf{\tilde{Y}} =
(\tilde{H}_{0}, \tilde{\Omega}_{b}^0, \tilde{\Omega}_{r}^0$) given
above.
In the models II, III and IV the interaction function ${\rm I}_{\rm
Q}(z)$ will be reconstructed up to order $N=4$ in the expansion in
terms of Chebyshev polynomials.

\begin{table}
  \centering
\begin{tabular}{| c | c | c | c | }
 \multicolumn{4}{c}{\textbf{Models}} \\
\hline
Models & $\Omega_{DM}^0$ & EOS parameter $w$ & Interaction function \\
\hline \hline
Model I & 0.227 (fixed) &  constant &  ${\rm I}_{\rm Q}(z) \equiv 0$ \\
Model II & 0.227 (fixed) & -1 &  ${\rm I}_{\rm Q}(z) \neq 0$ \\
Model III & 0.227 (fixed) & constant & ${\rm I}_{\rm Q}(z) \neq 0$ \\
Model IV & free parameter & constant & ${\rm I}_{\rm Q}(z) \neq 0$ \\
\hline \hline
\end{tabular}
\caption{Summary of the models studied in this work. In the models
II,  III and IV the interaction function ${\rm I}_{\rm Q}(z)$ will
be reconstructed. Additionally, in the model IV the parameter
$\Omega_{DM}^0$ is estimated.} \label{tablemodels}
\end{table}

\section{Results of the reconstruction of the interaction function.} \label{Results}
Now, we present the results of the fit of the models listed in the
Table \ref{tablemodels} with the ``Union2'' SNe Ia data set
\cite{AmanullahUnion22010}, the baryon acoustic oscillation (BAO),
the cosmic microwave background (CMB) data from 7-year WMAP,
the observational Hubble data and the cluster X-ray gas mass fraction,
using the priors described in the Section \ref{PriorsSS}.
For the noninteracting model I, the only free
parameter to be estimated is $\theta = \{w\}$, whilst for the
interacting models II, III and IV the free parameters are $\theta =
\{\lambda_{0},..., \lambda_{N}\}$, $\theta = \{w, \lambda_{0},...,
\lambda_{N}\}$ and $\theta = \{w, \Omega_{DM}^0, \lambda_{0},...,
\lambda_{N}\}$ respectively, where $N$ is taking the values $N = 1,
2, 3, 4$. In every case, we obtain the best fitted parameters and
the corresponding $\tilde{\chi}^2_{min}$.

\begin{table}
  \centering
\begin{tabular}{| c | c | c |}
\multicolumn{3}{c}{\textbf{Model I: $w=$ constant.}} \\
\multicolumn{3}{c}{Best estimate for the EOS parameter $w$.} \\
\hline
    Errors & $\pm 1\sigma$ & $\pm 2\sigma$\\
\hline
$\omega$&${-1.0130}^{+0.0168}_{-0.0163}$&${-1.0130}^{+0.0329}_{-0.0334}$\, \\
\hline
\end{tabular}
  \caption{The best estimates of the dark energy EOS parameter $w$ for the Model I. It was computed through a Bayesian statistical
   analysis using SNeIa + BAO + CMB + H + X-ray data sets giving $\tilde{\chi}^{2}_{min} = 719.2342$}.
  \label{BestestimationI}
\end{table}

\begin{table}
\centering
\begin{tabular}{| c | c | c | }
\multicolumn{3}{c}{\textbf{Age of the universe}}\\
\hline
Model I & $\pm 1\sigma$ (Gyr) & $\pm 2\sigma$ (Gyr)\\
\hline
Age & ${15.8679}^{+0.0189}_{-0.0201}$ & ${15.8679}^{+0.0379}_{-0.0400}$\\
\hline
\end{tabular}
\caption{Best estimates for the age of the universe and their errors at $1\sigma$ and $2\sigma$ for the model I using SNeIa + BAO + CMB + H + X-ray data sets.}
 \label{ageuniverse2}
\end{table}

\begin{table}
  \centering
\begin{tabular}{| c | c | c | c | c | c |}
\multicolumn{6}{c}{\textbf{Model II: $w=-1$.}} \\
\multicolumn{6}{c}{Best estimates for the parameters $\lambda_n$.} \\
\hline $\lambda_n$  & $N=0$ & $N=1$ & $N=2$ & $N=3$ & $N=4$ \\
\hline
$\lambda_{0}$&$-1.00\times 10^{-5}$&$-1.0296\times 10^{-5}$&$-1.0108\times 10^{-5}$&
$-1.0105\times 10^{-5}$&$-1.0120\times 10^{-5}$ \\
$\lambda_{1}$&$0.0$&$2.0027\times 10^{-4}$&$2.0024\times 10^{-4}$&$2.0014\times 10^{-4}$&$2.0006\times 10^{-4}$ \\
$\lambda_{2}$&$0.0$&$0.0$&$1.9990\times 10^{-6}$&$1.9935\times 10^{-6}$&$1.9918\times 10^{-6}$ \\
$\lambda_{3}$&$0.0$&$0.0$&$0.0$&$1.00\times 10^{-11}$&$1.018\times 10^{-14}$ \\
$\lambda_{4}$&$0.0$&$0.0$&$0.0$&$0.0$&$1.00\times 10^{-16}$ \\
\hline $\tilde{\chi}^{2}_{min}$&$719.8193$&$719.8154$&$719.8134$&$719.8131$&$719.8127$ \\
\hline
\end{tabular}
 \caption{Summary of the best estimates of the dimensionless coefficients $\lambda_n$ of the polynomial expansion of ${\rm I}_{\rm Q}(z)$
  for the Model II, corresponding to a interacting dark energy EOS parameter $w=-1$.
  They were computed through a Bayesian statistical analysis using the SNeIa+BAO+CMB+H+X-ray data sets. The number $N$ in the top of every column
  indicates the maximum number of Chebyshev polynomials used in the expansion of ${\rm I}_{\rm Q}(z)$ starting from $N=1$ to $N=4$. From the
  Figure \ref{TotalExchangeModels} to Figure \ref{TotalRelationModels} show the reconstruction of several cosmological variables using these best fitted values.}
 \label{BestestimationII}
 \end{table}

\begin{table}
  \centering
\begin{tabular}{| c | c | c |}
\multicolumn{3}{c}{\textbf{Model II: $w=-1$.}} \\
\hline Errors & $\pm 1\sigma$ & $\pm 2\sigma$\\
\hline
$\lambda_{0}$&${-1.0108\times 10^{-5}}^{+0.0105\times 10^{-5}}_{-0.0104\times 10^{-5}}$&${-1.0108\times 10^{-5}}^{+0.0282\times 10^{-5}}_{-0.0244\times 10^{-5}}$\, \\
\hline
$\lambda_{1}$&${+2.0024\times 10^{-4}}^{+0.0253\times 10^{-4}}_{-0.0223\times 10^{-4}}$&${+2.0024\times 10^{-4}}^{+0.0408\times 10^{-4}}_{-0.0335\times 10^{-4}}$\, \\
\hline
$\lambda_{2}$&${+1.9990\times 10^{-6}}^{+0.0011\times 10^{-6}}_{-0.0017\times 10^{-6}}$&${+1.9990\times 10^{-6}}^{+0.0028\times 10^{-6}}_{-0.0047\times 10^{-6}}$\, \\
\hline
\end{tabular}
  \caption{Summary of the $1\sigma$ and $2\sigma$ errors of the best estimates for $N=2$.}
  \label{BestestimationIIerrors}
\end{table}

\begin{table}
\centering
\begin{tabular}{| c | c | c | }
\multicolumn{3}{c}{\textbf{Age of the universe}}\\
\hline
Model II & $\pm 1\sigma$ (Gyr) & $\pm 2\sigma$ (Gyr)\\
\hline
Age & $15.85028^{+0.000022}_{-0.000025}$ & $15.85028^{+0.000032}_{-0.000039}$\\
\hline
\end{tabular}
\caption{Best estimates for the age of the universe and their errors at $1\sigma$ and $2\sigma$ for the model II using SNeIa + BAO + CMB + H + X-ray data sets.}
 \label{ageuniverse2}
\end{table}

\begin{center}
\begin{figure}
 \includegraphics[width=8cm,height=60mm,scale=0.90]{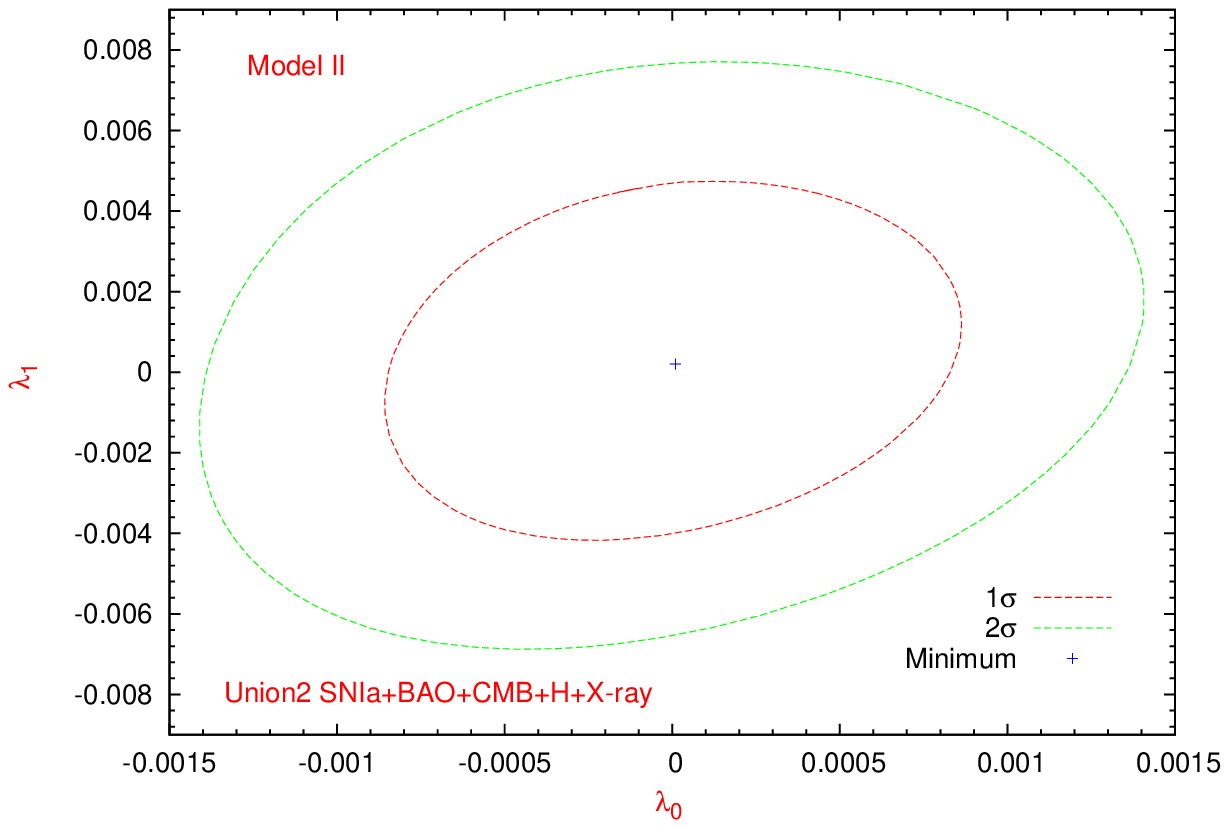}
 \includegraphics[width=8cm,height=60mm,scale=0.90]{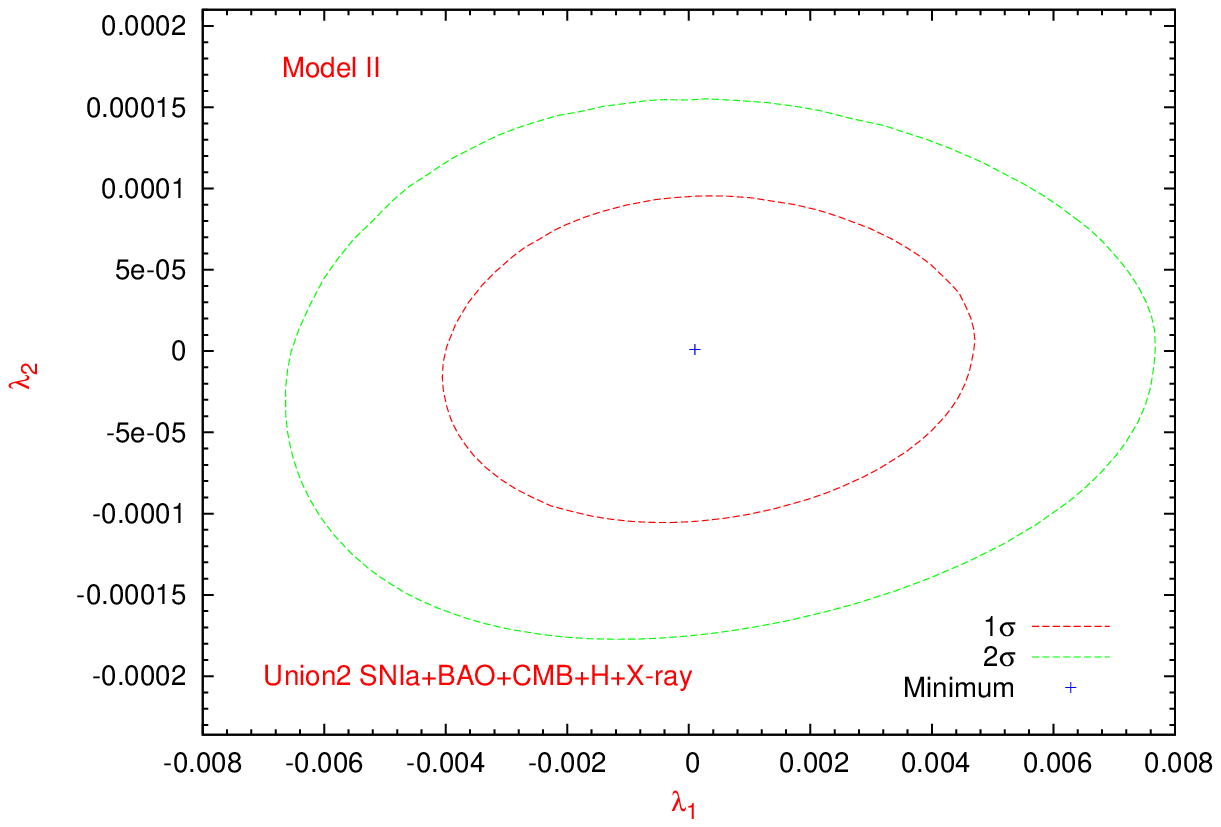}
  \caption{Contours correspond to $1\sigma$, $2\sigma$ confidence levels constrained from joint SNeIa + BAO + CMB + H + X-ray data analysis
    for the marginalized probability densities of the model II, using an expansion in terms of the first $N=2$ Chebyshev polynomials in the
    ($\lambda_0$ - $\lambda_1$) and ($\lambda_1$ - $\lambda_2$) planes. It is clear that before marginalization, we have three free parameters
    ($\lambda_0$, $\lambda_1$, $\lambda_2$). In every figure, we marginalized on one of the parameters.}
 \label{TotalModel2}
\end{figure}
\end{center}

\begin{table}
  \centering
\begin{tabular}{| c | c | c | c | c | c | }
\multicolumn{6}{c}{\textbf{Model III: $w=$} constant.} \\
\multicolumn{6}{c}{Best estimates for the parameters $\lambda_n$ and $w$.} \\
\hline $\lambda_n$  & $N=0$ & $n=1$ & $n=2$ & $n=3$ & $n=4$ \\
\hline
$\lambda_{0}$&$-1.00\times 10^{-5}$&$-1.0040\times 10^{-5}$& $-1.0015\times 10^{-5}$&
$-1.0001\times 10^{-5}$&$-1.00\times 10^{-5}$ \\
$\lambda_{1}$&$0.0$&$2.0021\times 10^{-4}$&$2.0040\times 10^{-4}$&$2.0120\times 10^{-4}$&$2.0390\times 10^{-4}$ \\
$\lambda_{2}$&$0.0$&$0.0$&$9.9980\times 10^{-6}$&$9.9970\times 10^{-6}$&$9.9962\times 10^{-6} $ \\
$\lambda_{3}$&$0.0$&$0.0$&$0.0$&$1.0010\times 10^{-9}$&$1.0020\times 10^{-10}$ \\
$\lambda_{4}$&$0.0$&$0.0$&$0.0$&$0.0$&$1.00\times 10^{-12}$ \\
$\omega$&$-1.0089$&$-1.0088$&$-1.0086$&$-1.0089$&$-1.0092$ \\
\hline $\tilde{\chi}^{2}_{min}$&$719.2984$&$719.2861$&$719.1833$&$719.1779$&$719.1731$ \\
\hline
\end{tabular}
 \caption{Summary of the best estimates of the dimensionless coefficients $\lambda_n$ of the polynomial expansion of ${\rm I}_{\rm Q}(z)$
  for the Model III, corresponding to a interacting dark energy EOS parameter $w=$ constant and $\Omega_{DM}^0 = 0.227$.
  They were computed through a Bayesian statistical analysis using the SNeIa+BAO+CMB+H+X-ray data sets. The number $N$ in the top of every column
  indicates the maximum number of Chebyshev polynomials used in the expansion of ${\rm I}_{\rm Q}(z)$ starting from $N=1$ to $N=4$. From the
  Figure \ref{TotalExchangeModels} to Figure \ref{TotalRelationModels} show the reconstruction of several cosmological variables using these best fitted values.}
 \label{BestestimationIII}
\end{table}

\begin{table}
  \centering
\begin{tabular}{| c | c | c | }
\multicolumn{3}{c}{\textbf{Model III: $w=$constant.}} \\
\hline Errors & $\pm 1\sigma$ & $\pm 2\sigma$\\
\hline \hline
$\lambda_{0}$&${-1.0015\times 10^{-5}}^{+0.0860\times 10^{-5}}_{-0.0473\times 10^{-5}}$&${-1.0015\times 10^{-5}}^{+0.1000\times 10^{-5}}_{-0.1339\times 10^{-5}}$\, \\
$\lambda_{1}$&${+2.0040\times 10^{-4}}^{+0.5885\times 10^{-4}}_{-0.1020\times 10^{-4}}$&${+2.0040\times 10^{-4}}^{+0.8034\times 10^{-4}}_{-0.1813\times 10^{-4}}$\, \\
$\lambda_{2}$&${+9.9980\times 10^{-6}}^{+2.0691\times 10^{-6}}_{-0.1588\times 10^{-6}}$&${+9.9980\times 10^{-6}}^{+2.7073\times 10^{-6}}_{-0.3744\times 10^{-6}}$\, \\
$\omega$ & ${-1.0086}^{+0.0041}_{-0.0292}$&${-1.0086}^{+0.0046}_{-0.0312}$\, \\
\hline
\end{tabular}
  \caption{Summary of the $1\sigma$ and $2\sigma$ errors of the best estimates for $N=2$.}
  \label{BestestimationIIIerrors}
\end{table}

\begin{table}
\centering
\begin{tabular}{| c | c | c |}
\multicolumn{3}{c}{\textbf{Age of the universe}}\\
\hline
Model III & $\pm 1\sigma$ (Gyr) & $\pm 2\sigma$ (Gyr)\\
\hline
Age & $15.86064^{+0.0337}_{-0.0054}$ & $15.86064^{+0.0360}_{-0.0062}$\\
\hline
\end{tabular}
\caption{Best estimates for the age of the universe and their errors at $1\sigma$ and $2\sigma$ for the model III using SNeIa + BAO + CMB + H + X-ray data sets.}
 \label{ageuniverse3}
\end{table}

\begin{center}
\begin{figure}
   \includegraphics[width=8cm, height=60mm, scale=0.90]{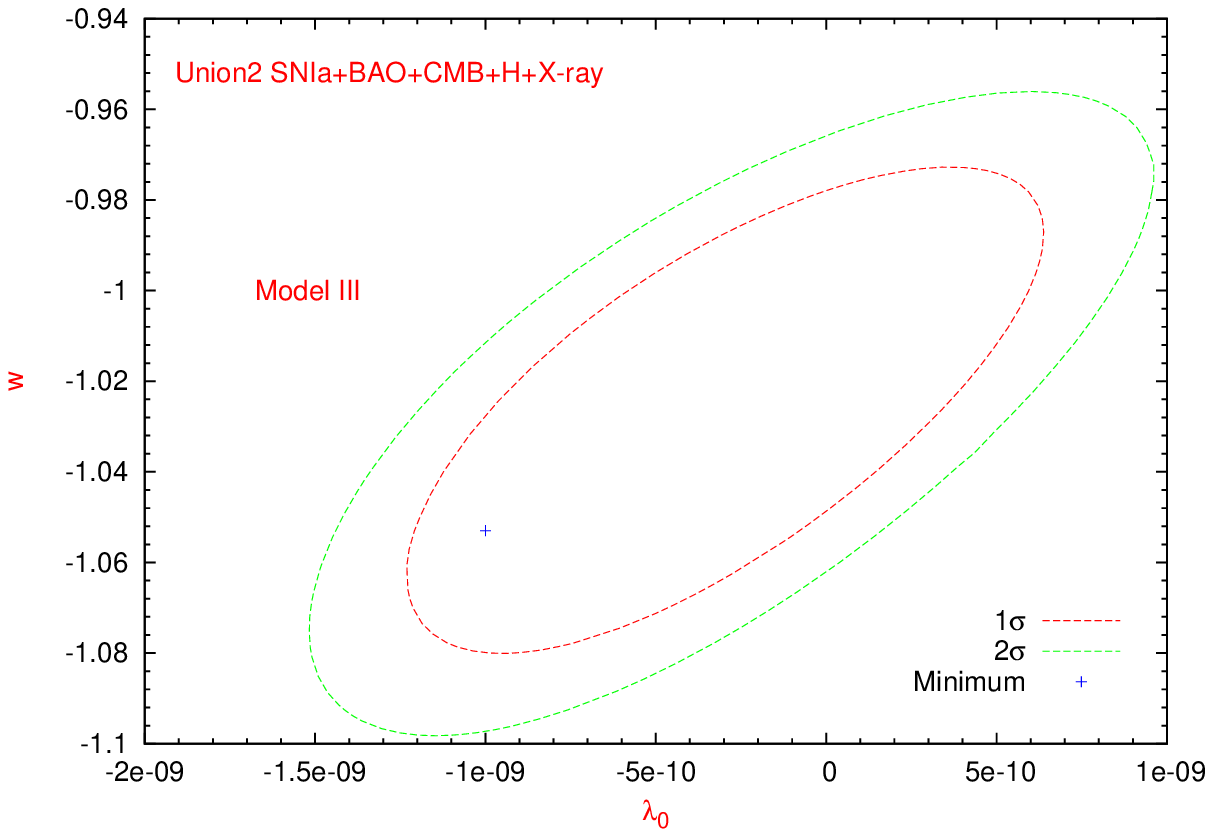}
   \includegraphics[width=8cm, height=60mm, scale=0.90]{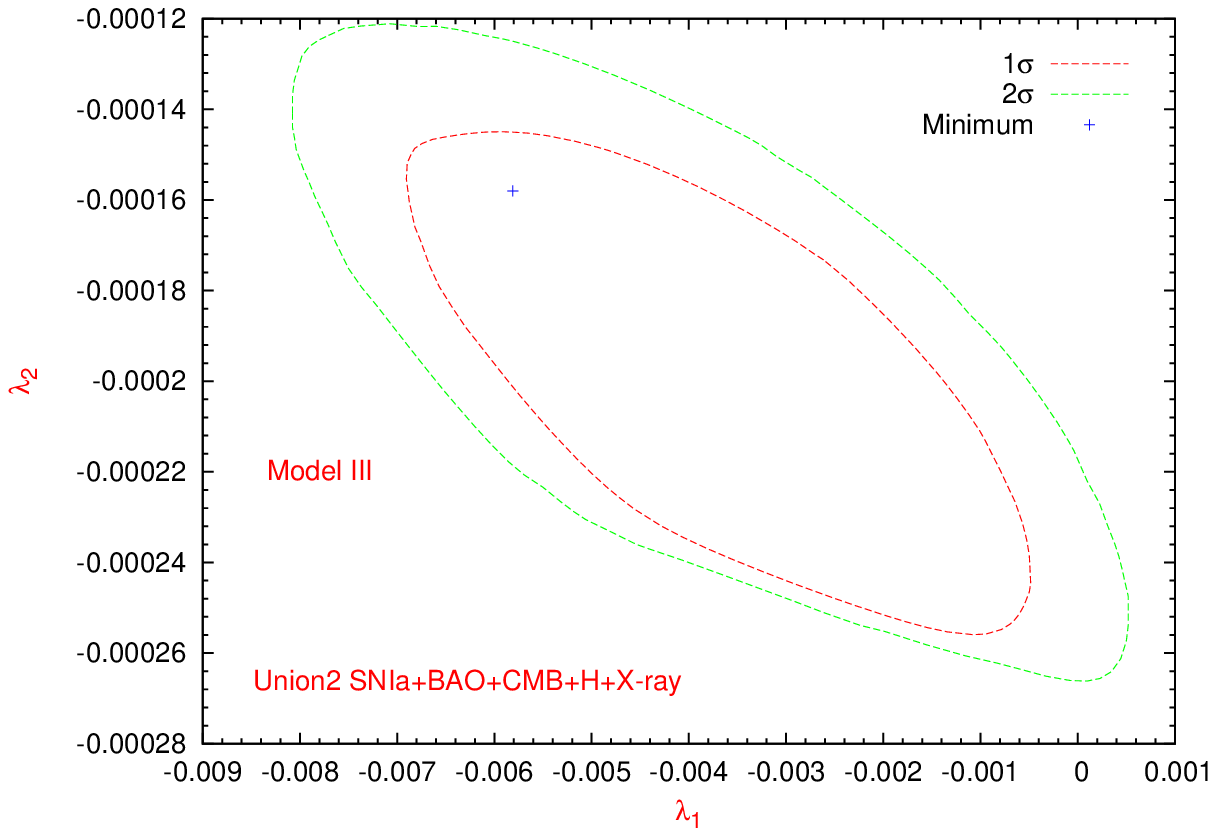}
  \caption{Contours correspond to $1\sigma$, $2\sigma$ confidence levels constrained from joint SNeIa + BAO + CMB + H + X-ray data analysis
    for the marginalized probability densities of the model III, using an expansion in terms of the first $N=2$ Chebyshev polynomials in the
    ($\lambda_0$-$\omega$) and ($\lambda_1$-$\lambda_2$) planes. It is clear that before marginalization, we have four free parameters
    ($\lambda_0$, $\lambda_1$, $\lambda_2$, $\omega$). In every figure, we marginalized on the last two remaining parameters. Note that the preferred
    region for the EOS parameter $w$ is the phantom region and LCDM model is in the $1\sigma$ error region.}
  \label{TotalModel3}
\end{figure}
\end{center}

\begin{table}
  \centering
\begin{tabular}{| c | c | c | c | c | c |}
\multicolumn{6}{c}{\textbf{Model IV: $w=$constant and $\Omega_{DM}^0=$constant }} \\
\multicolumn{6}{c}{Best estimates for the parameters $\lambda_n$, $w$ and $\Omega_{DM}^0$.} \\
\hline $\lambda_n$  & $n=0$ & $n=1$ & $n=2$ & $n=3$ & $n=4$ \\
\hline
$\lambda_{0}$&$-1.00\times 10^{-5}$&$-1.00\times 10^{-5}$& $-1.0010\times 10^{-5}$&
$-1.0006\times 10^{-5}$&$-1.0002\times 10^{-5}$ \\
$\lambda_{1}$&$0.0$&$2.0314\times 10^{-4}$&$2.0331\times 10^{-4}$&$2.0385\times 10^{-4}$&$2.0220\times 10^{-4}$ \\
$\lambda_{2}$&$0.0$&$0.0$&$9.9990\times 10^{-6}$&$9.9925\times 10^{-6}$&$9.9915\times 10^{-6} $ \\
$\lambda_{3}$&$0.0$&$0.0$&$0.0$&$1.0068\times 10^{-8}$&$1.0125\times 10^{-10}$ \\
$\lambda_{4}$&$0.0$&$0.0$&$0.0$&$0.0$&$1.00\times 10^{-12}$ \\
$w$&$-1.0089$&$-1.0089$&$-1.0088$&$-1.0088$&$-1.0089$ \\
$\Omega_{DM}^0$&$+0.2278$&$+0.2278$&$+0.2278$&$+0.2279$&$+0.2279$ \\
\hline $\tilde{\chi}^{2}_{min}$&$718.6851$&$718.6602$&$718.6236$&$718.5704$&$718.5617$ \\
\hline
\end{tabular}
 \caption{Summary of the best estimates of the dimensionless coefficients $\lambda_n$ of the polynomial expansion of ${\rm I}_{\rm Q}(z)$
  for the Model IV, corresponding to a interacting dark energy EOS parameter $w=$ constant and $\Omega_{DM}^0$ as a free parameter to be estimated.
  They were computed through a Bayesian statistical analysis using the SNeIa+BAO+CMB+H+X-ray data sets. The number $N$ in the top of every column
  indicates the maximum number of Chebyshev polynomials used in the expansion of ${\rm I}_{\rm Q}(z)$ starting from $N=1$ to $N=4$. From the
  Figure \ref{TotalExchangeModels} to Figure \ref{TotalRelationModels} show the reconstruction of several cosmological variables using these best fitted values.}
 \label{BestestimationIV}
 \end{table}

\begin{table}
  \centering
\begin{tabular}{| c | c | c |}
\multicolumn{3}{c}{\textbf{Model IV: $w=$constant and $\Omega_{DM}^0=$constant.}} \\
\hline Errors & $\pm 1\sigma$ & $\pm 2\sigma$\\
\hline
$\lambda_{0}$&${-1.0010\times 10^{-5}}^{+0.0616\times 10^{-5}}_{-0.01015\times 10^{-5}}$&${-1.0010\times 10^{-5}}^{+0.1122\times 10^{-5}}_{-0.0434\times 10^{-5}}$\, \\
$\lambda_{1}$&${+2.0331\times 10^{-4}}^{+0.0756\times 10^{-4}}_{-0.0173\times 10^{-4}}$&${+2.0331\times 10^{-4}}^{+0.1887\times 10^{-4}}_{-0.0681\times 10^{-4}}$\, \\
$\lambda_{2}$&${+9.9990\times 10^{-6}}^{+0.0891\times 10^{-6}}_{-0.0489\times 10^{-6}}$&${+9.9990\times 10^{-6}}^{+0.1550\times 10^{-6}}_{-0.1073\times 10^{-6}}$\, \\
$\omega$&${-1.0088}^{+3.9588\times 10^{-8}}_{-0.0658}$&${-1.0088}^{+1.8109\times 10^{-7}}_{-0.0848}$\, \\
$\Omega^{0}_{DM}$&${+0.2278}^{+8.0936\times 10^{-5}}_{-0.0056}$&${+0.2278}^{+2.58411\times 10^{-4}}_{-0.0079}$\, \\
\hline
\end{tabular}
  \caption{Summary of the $1\sigma$ and $2\sigma$ errors of the best estimates for $N=2$.}
  \label{BestestimationIVerrors}
\end{table}

\begin{table}
\centering
\begin{tabular}{| c | c | c |}
\multicolumn{3}{c}{\textbf{Age of the universe}}\\
\hline
Model IV & $\pm 1\sigma$ (Gyr) & $\pm 2\sigma$ (Gyr)\\
\hline
Age & $15.8515^{+0.1409}_{-0.0010}$ & $15.8515^{+0.1907}_{-0.0032}$\\
\hline
\end{tabular}
\caption{Best estimates for the age of the universe and their errors at $1\sigma$ and $2\sigma$ for the model IV using SNeIa + BAO + CMB + H + X-ray data sets.}
 \label{ageuniverse4}
\end{table}

\begin{center}
\begin{figure}
  \includegraphics[width=8cm, height=60mm, scale=0.90]{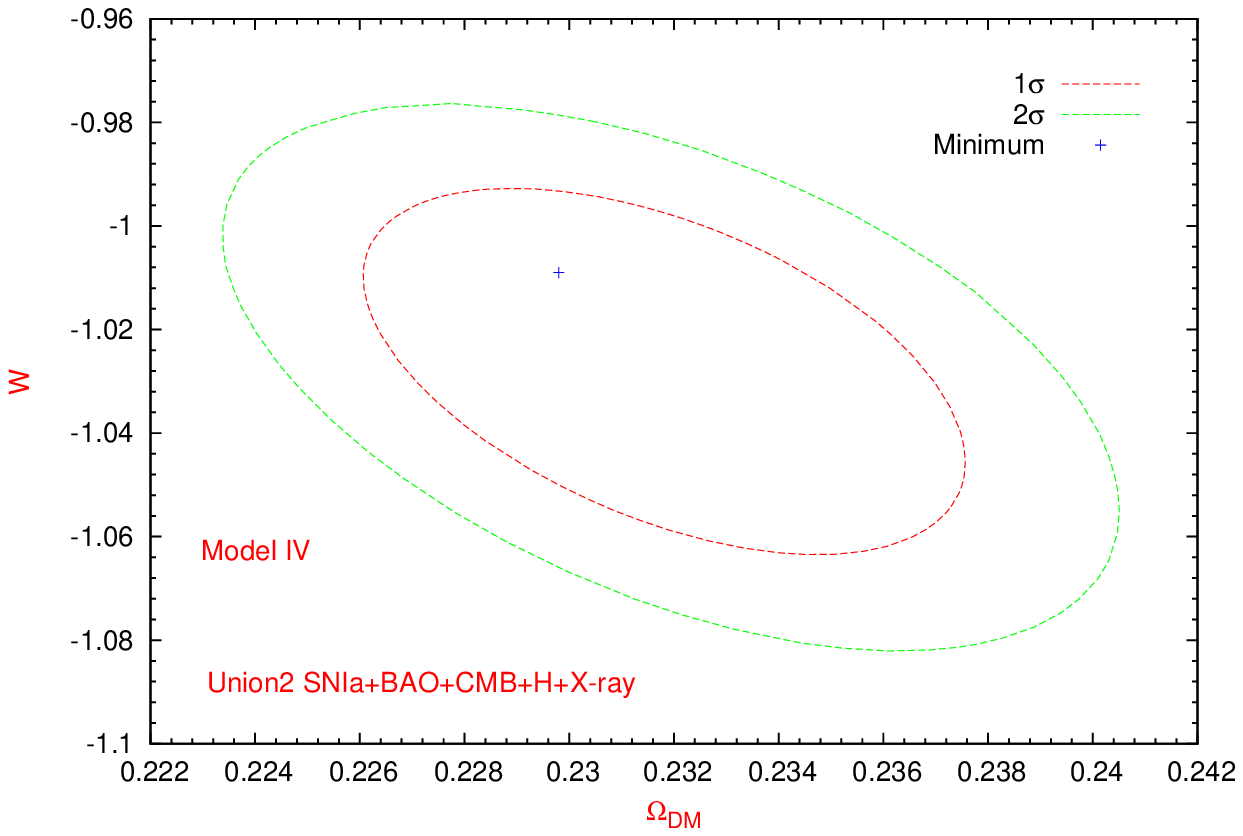}
  \includegraphics[width=8cm, height=60mm, scale=0.90]{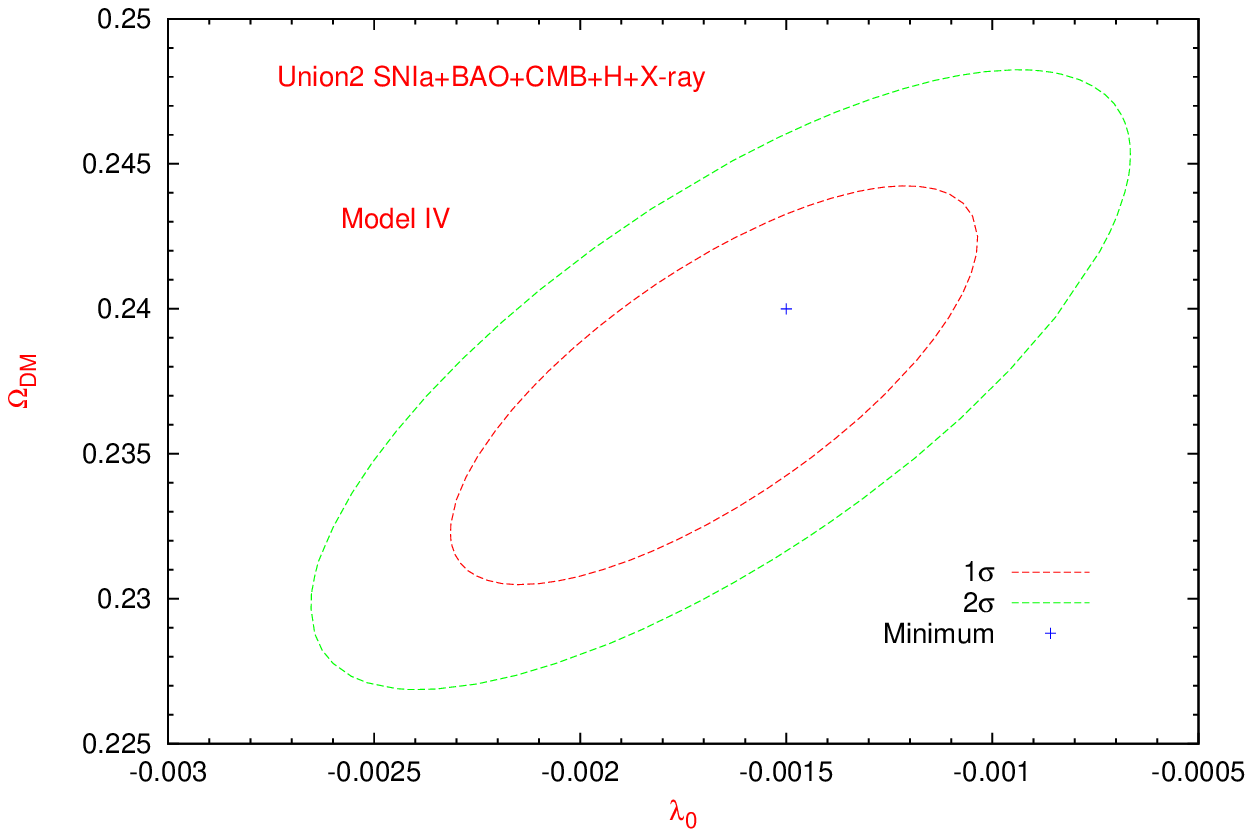}
  \includegraphics[width=8cm, height=60mm, scale=0.90]{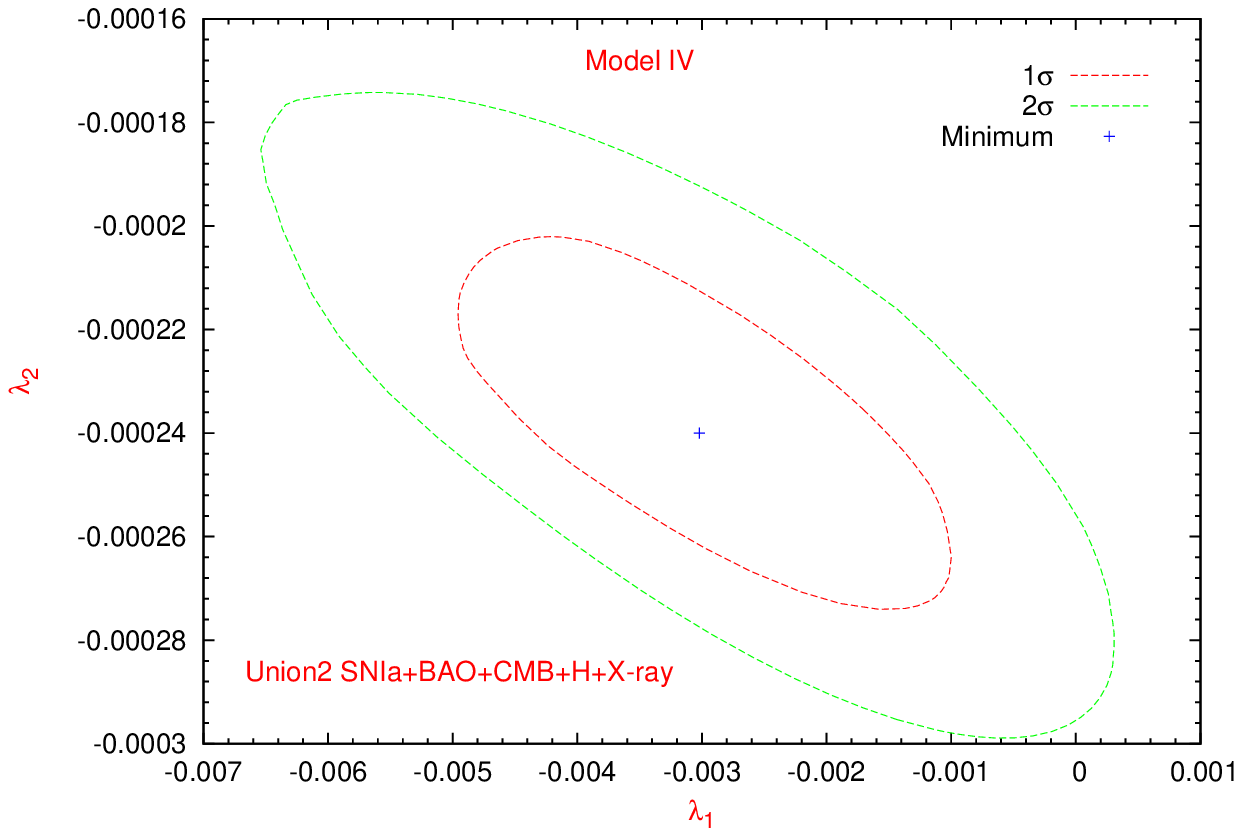}
  \caption{Contours correspond to $1\sigma$, $2\sigma$ confidence levels constrained from joint SNeIa + BAO + CMB + H + X-ray data analysis
    for the marginalized probability densities of the model IV, using an expansion in terms of the first $N=2$ Chebyshev polynomials in the
    ($\Omega_{DM}$ - $\omega$), ($\lambda_0$ - $\Omega_{DM}$) and ($\lambda_1$ - $\lambda_2$) planes. It is clear that before marginalization, we have
    five free parameters ($\lambda_0$, $\lambda_1$, $\lambda_2$, $\omega$, $\Omega_{DM}$). In every figure, we marginalized on the last three remaining parameters. Note that the preferred region for the EOS parameter $w$ is the phantom region. Our result is consistent with the LCDM model in the $1\sigma$ error region.}
  \label{TotalModel4}
\end{figure}
\end{center}

\begin{center}
\begin{figure}
  \includegraphics[width=8cm, height=60mm, scale=0.90]{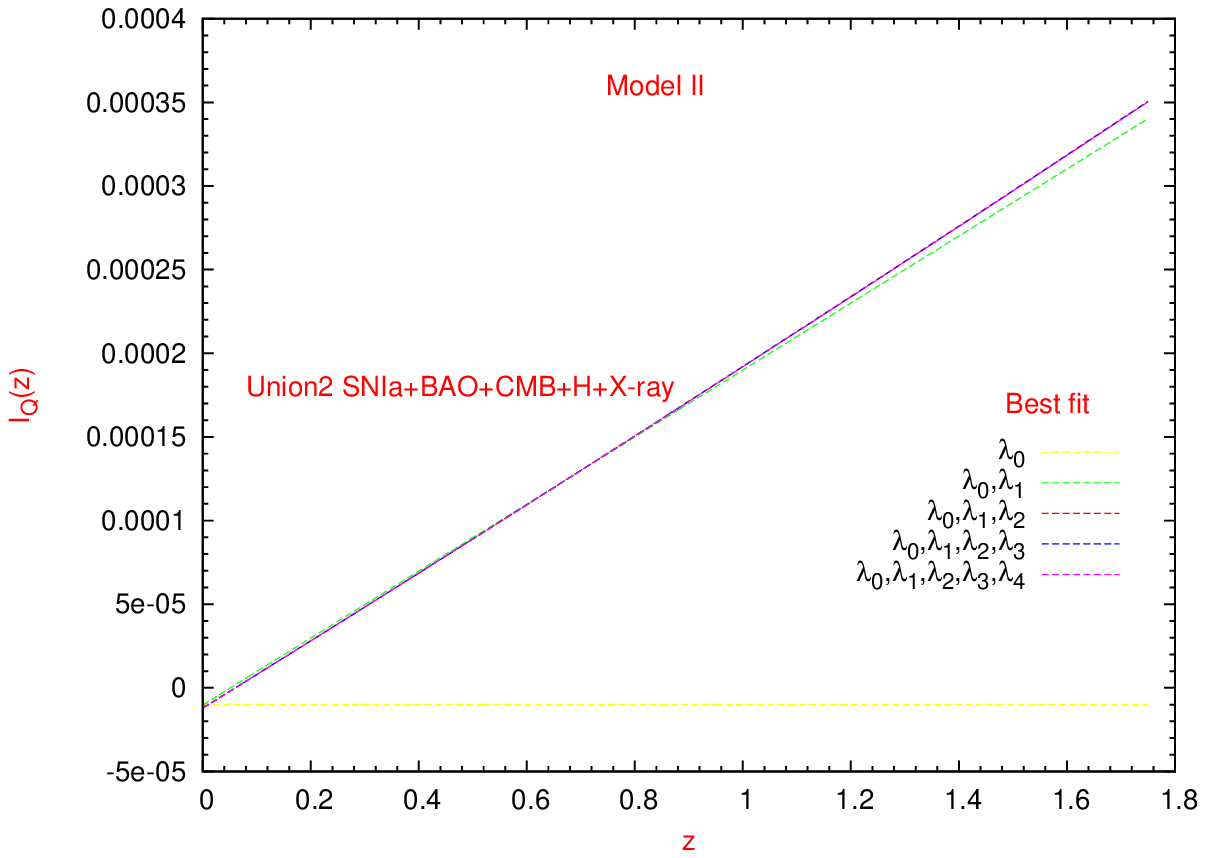}
  \includegraphics[width=8cm, height=60mm, scale=0.90]{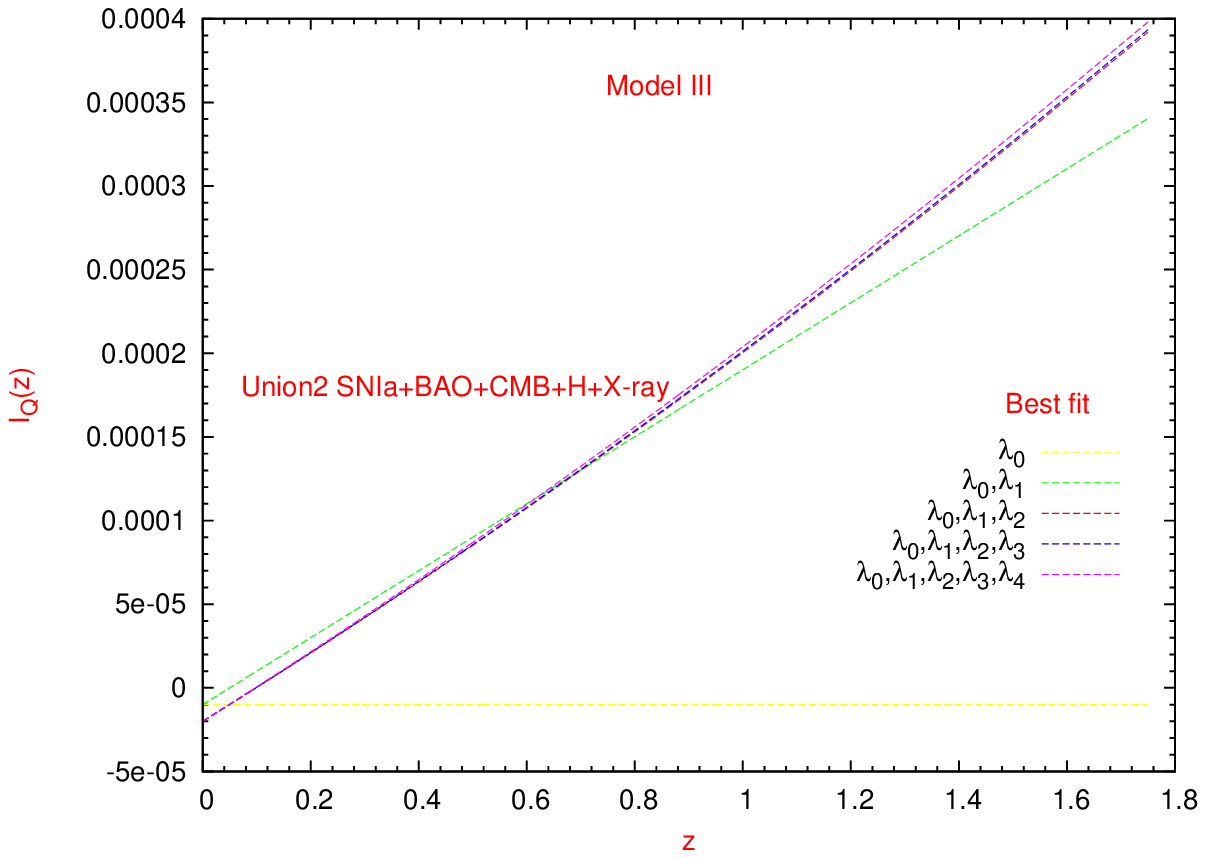}
  \includegraphics[width=8cm, height=60mm, scale=0.90]{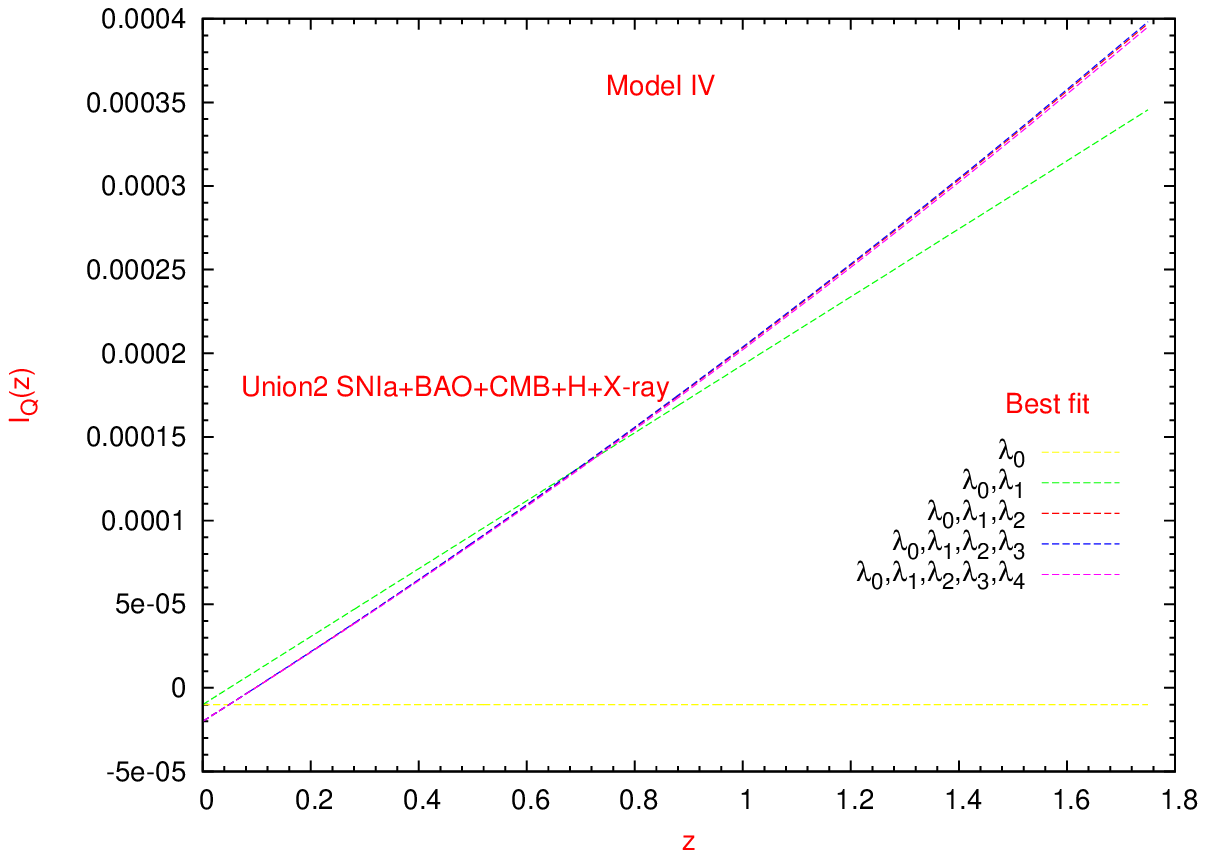}
 \caption{Reconstruction for the dimensionless interaction function ${\rm I}_{\rm Q}(z)$ as a function of the redshift for the models II (left above panel),
III (right above panel) and IV (left below panel) corresponding to a dark energy equation of state parameter (II) $w=-1$, (III) $w=$ constant
(both with $\Omega_{DM}^0=0.227$), and (IV) $w=$ constant, $\Omega_{DM}^0=$ constant, respectively. The curves with different colors show the best estimates using
the expansion of ${\rm I}_{\rm Q}(z)$ in terms of the parameters $\lambda_n$ corresponding to the Chebyshev polynomial expansion ranging from $N=1$ to $4$.
Note the fast convergence of the curves when the number of polynomials $N$ involved in the expansion increases. The reconstruction is derived from the best estimation
obtained from the combination of SNeIa + BAO + CMB + H + X-ray data sets. Note that the best estimated values of the strength of the interaction cross marginally the
noninteracting line ${\rm I}_{\rm Q}(z)=0$ only at the present changing sign from positive values at the past (energy transfers from dark energy to dark matter)
to negative values almost at the present (energy transfers from dark matter to dark energy).}
  \label{TotalExchangeModels}
\end{figure}
\end{center}

The Figure \ref{TotalExchangeModels} shows the reconstruction of the dimensionless interaction function ${\rm I}_{\rm Q}(z)$ as a
function of the redshift for the models II (corresponding to a dark energy EOS parameter $w=-1$), III (corresponding to a dark energy
EOS parameter $w=$ constant and $\Omega_{DM}^0$ fixed) and IV (corresponding to a dark energy EOS parameter $w=$ constant and
$\Omega_{DM}^0$ as a free parameter to be estimated) respectively.

\begin{center}
\begin{figure}
   \includegraphics[width=8cm, height=60mm, scale=0.90]{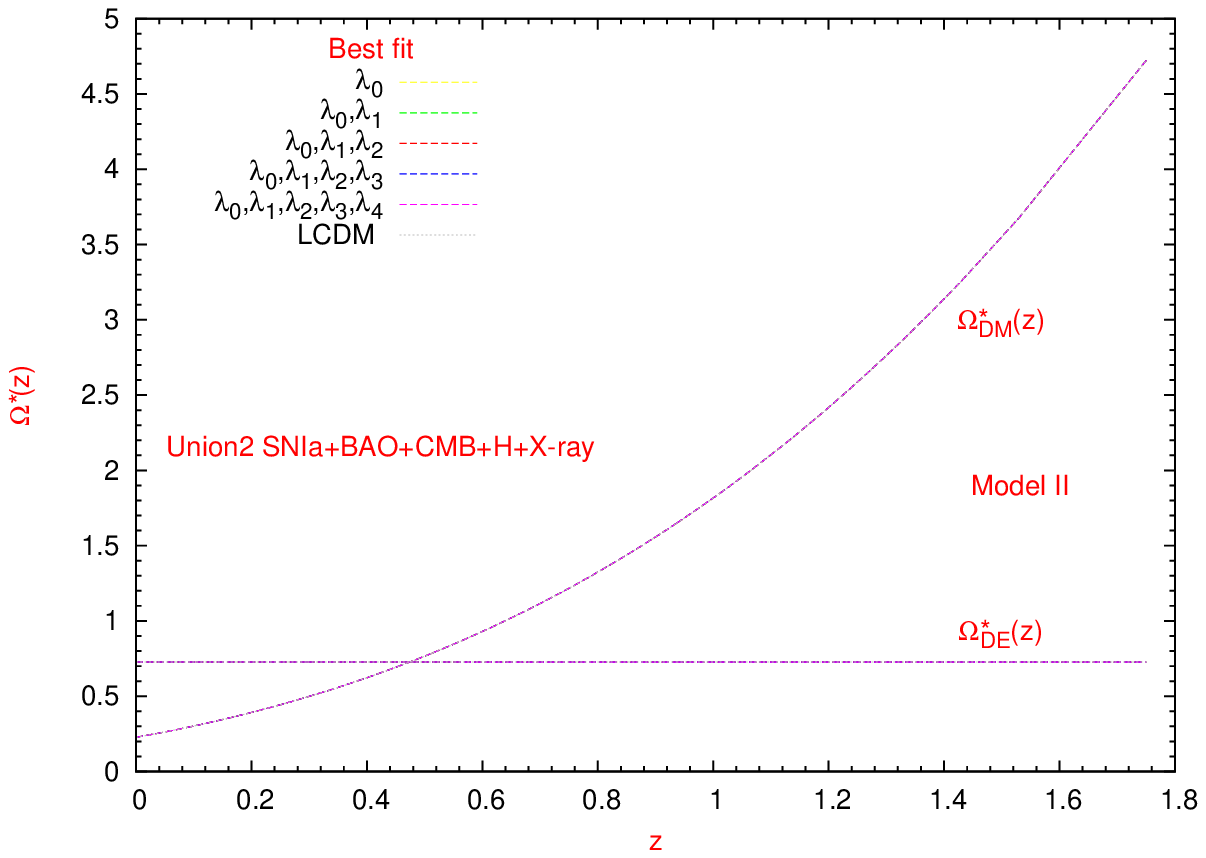}
   \includegraphics[width=8cm, height=60mm, scale=0.90]{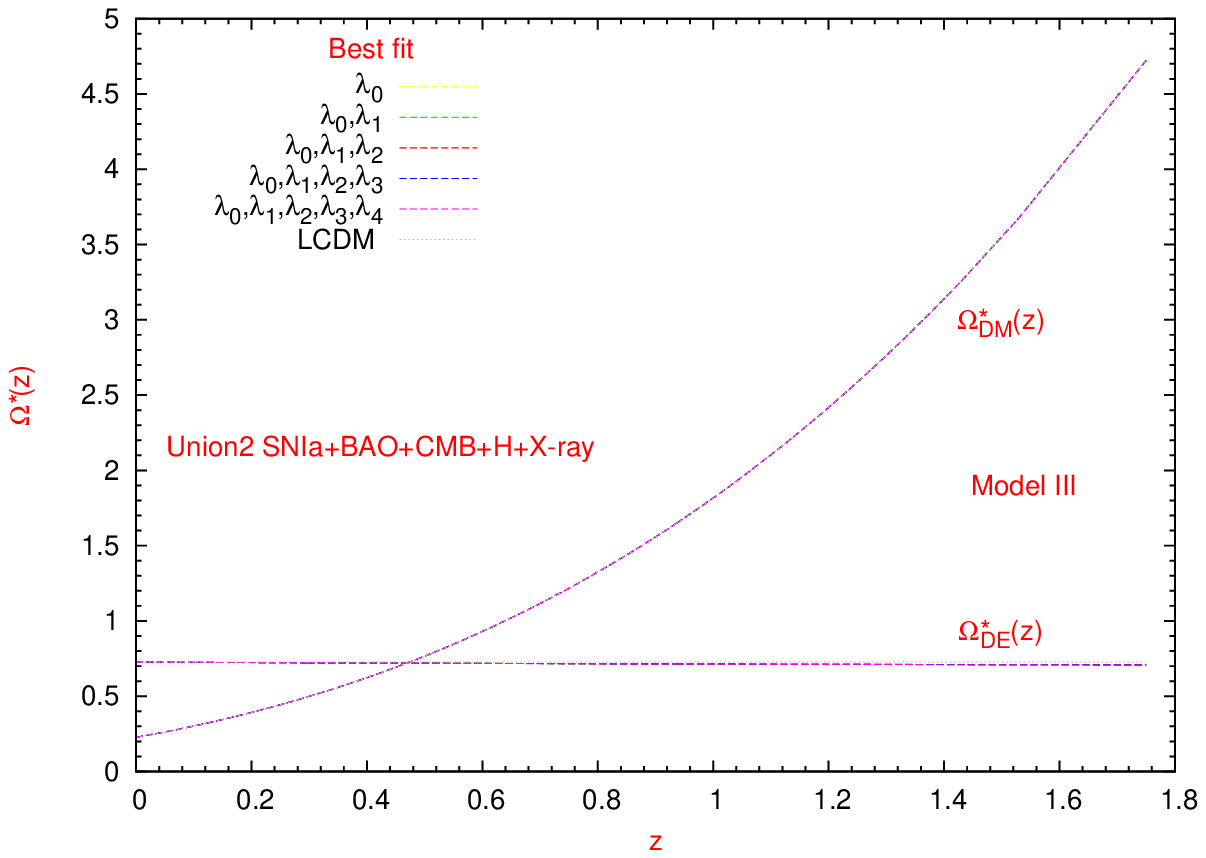}
   \includegraphics[width=8cm, height=60mm, scale=0.90]{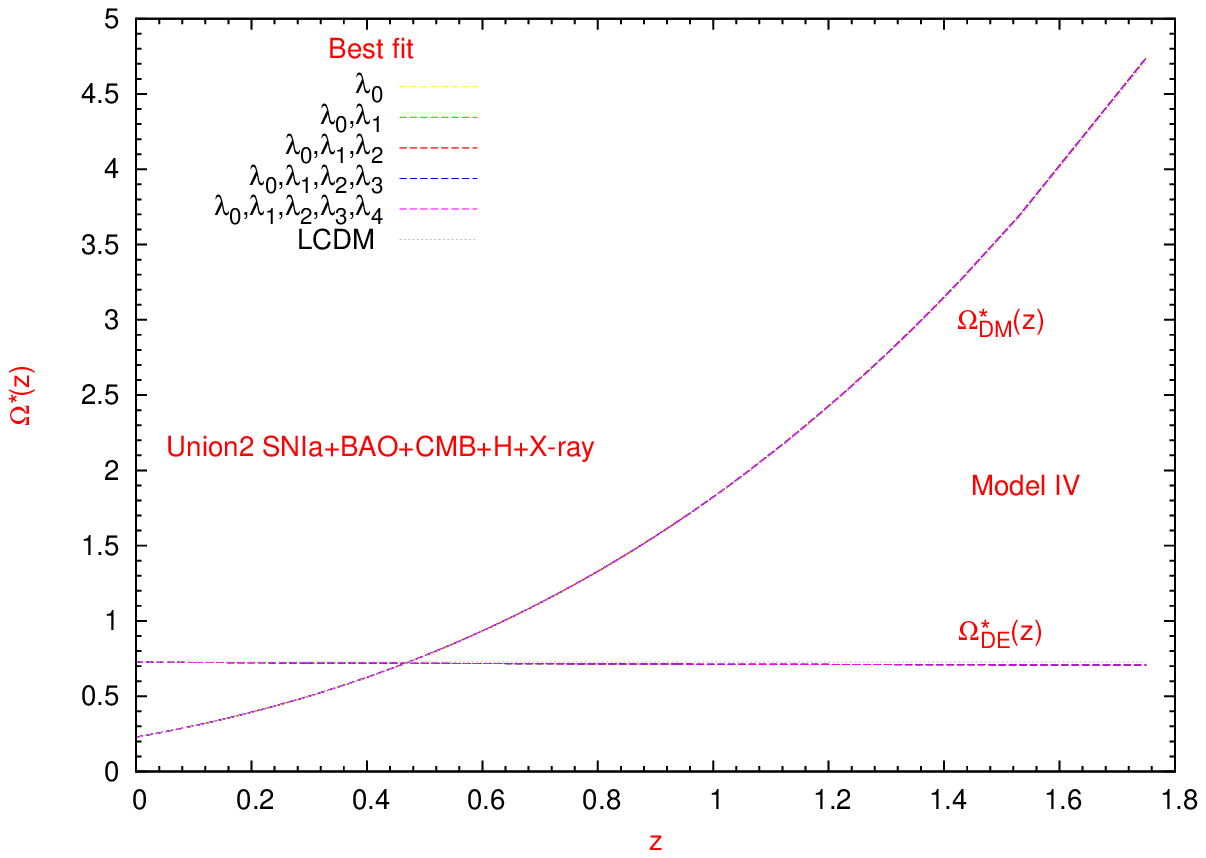}
  \caption{Same explanation as Figure \ref{ErrorsTotalExchangeModels} but now for the reconstruction of the dark matter and dark energy density parameters,
   $\Omega^{\star}_{DM}(z)$, $\Omega^{\star}_{DE}(z)$, as a function of the redshift for the model II (left above panel), III (right above panel)
   and IV (left below panel) respectively.
   Note that the density parameter of dark energy is definite positive for all the range of redshift considered in the reconstruction.}
  \label{TotalCoincidenceModels}
\end{figure}
\end{center}

The Figure \ref{TotalCoincidenceModels} shows the reconstruction of the dark matter and dark energy density parameters
$\Omega^{\star}_{DM}(z)$, $\Omega^{\star}_{DE}(z)$ as a function of the redshift for the models II, III and IV described above.

\begin{center}
\begin{figure}
   \includegraphics[width=8cm, height=60mm, scale=0.90]{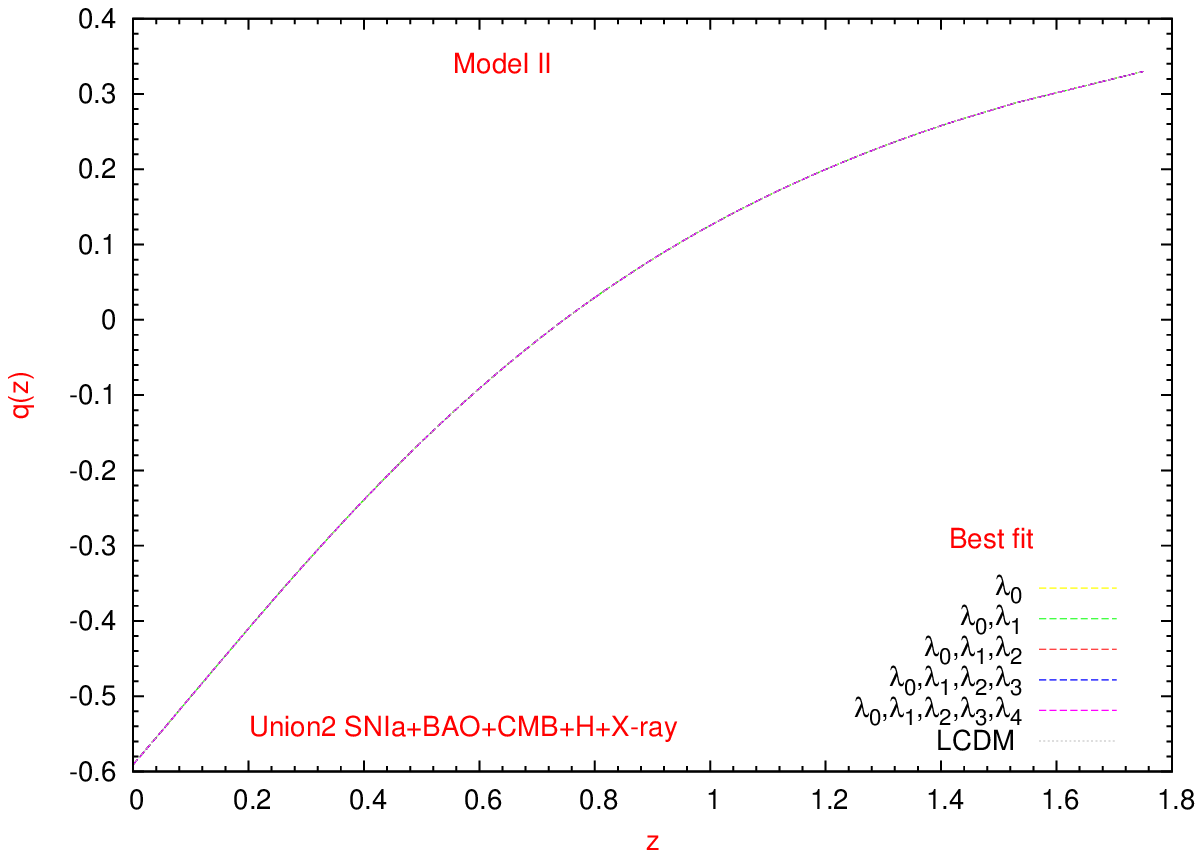}
   \includegraphics[width=8cm, height=60mm, scale=0.90]{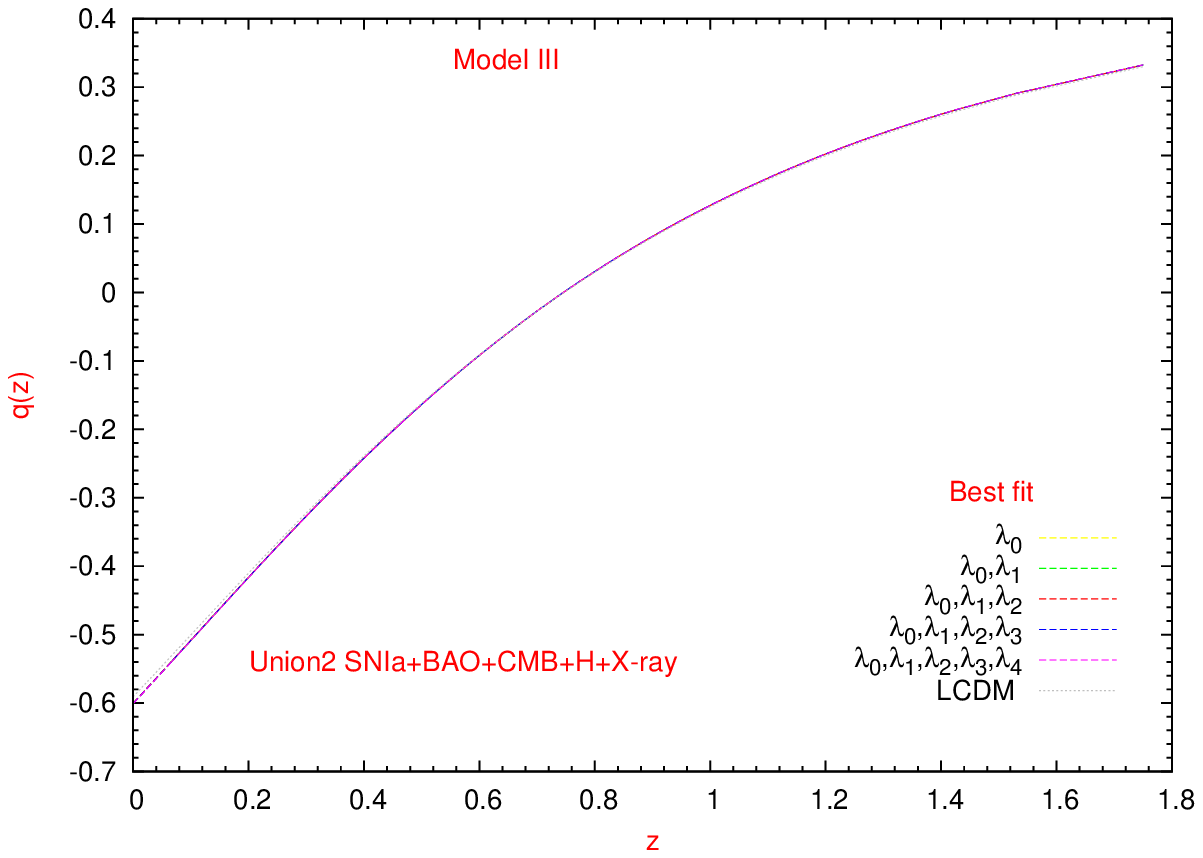}
   \includegraphics[width=8cm, height=60mm, scale=0.90]{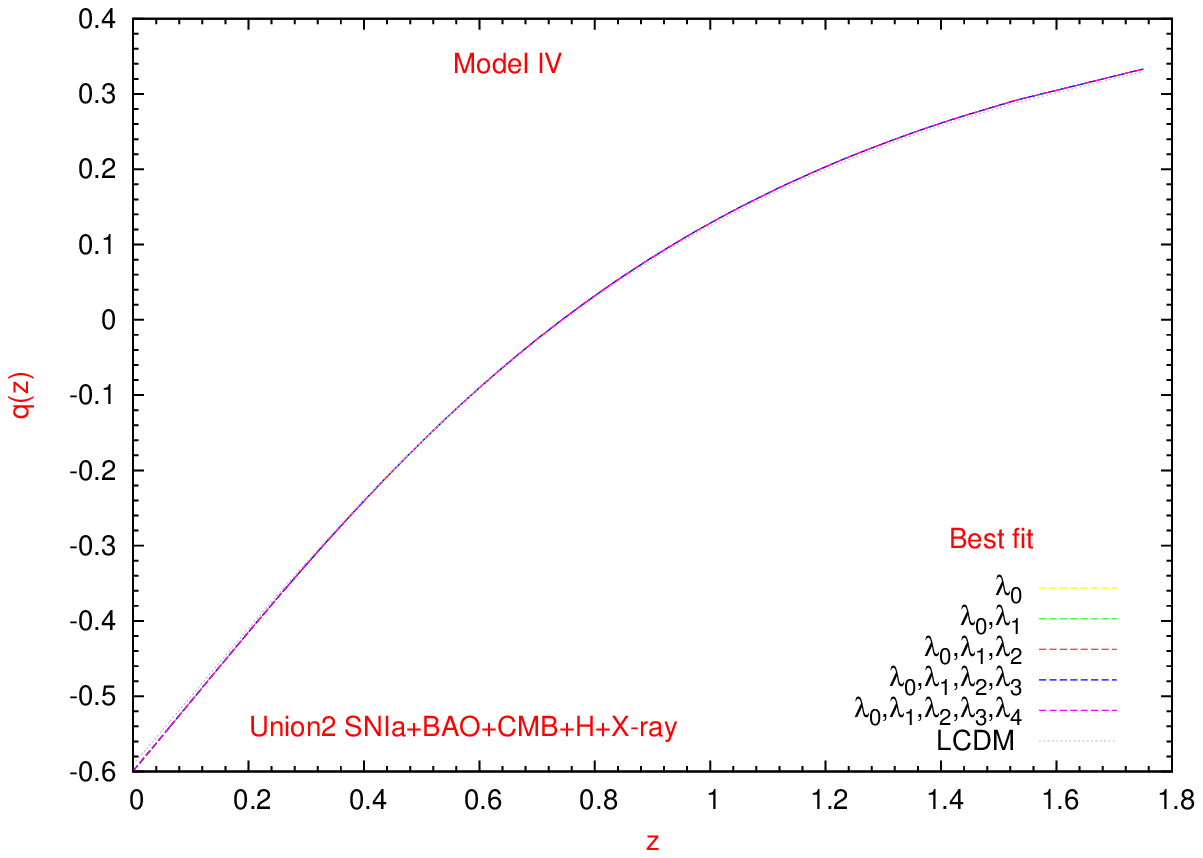}
  \caption{Similar explanation as Figure \ref{TotalExchangeModels} for the reconstruction of the deceleration parameter $q(z)$ as a function of the redshift
  for the model II (left above panel), III (right above panel) and IV (left below panel) respectively. At the same time, they are compared with the corresponding
  curve for the LCDM (Lambda Cold Dark Matter) model.}
  \label{TotalDecelerateModels}
\end{figure}
\end{center}

The Figure \ref{TotalDecelerateModels} shows the reconstruction of the deceleration parameter $q(z)$ as a function of
the redshift for the models II, III and IV respectively.

\begin{center}
\begin{figure}
   \includegraphics[width=8cm, height=60mm, scale=0.90]{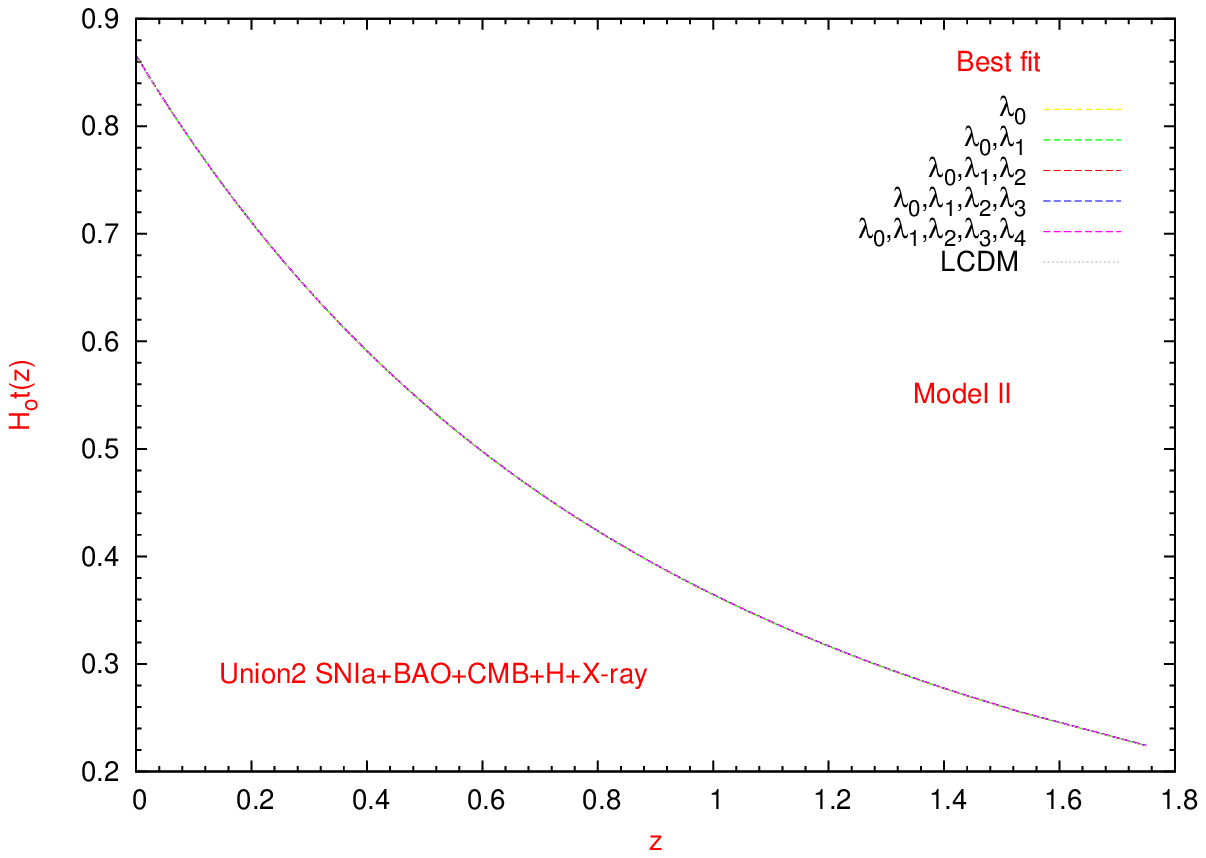}
   \includegraphics[width=8cm, height=60mm, scale=0.90]{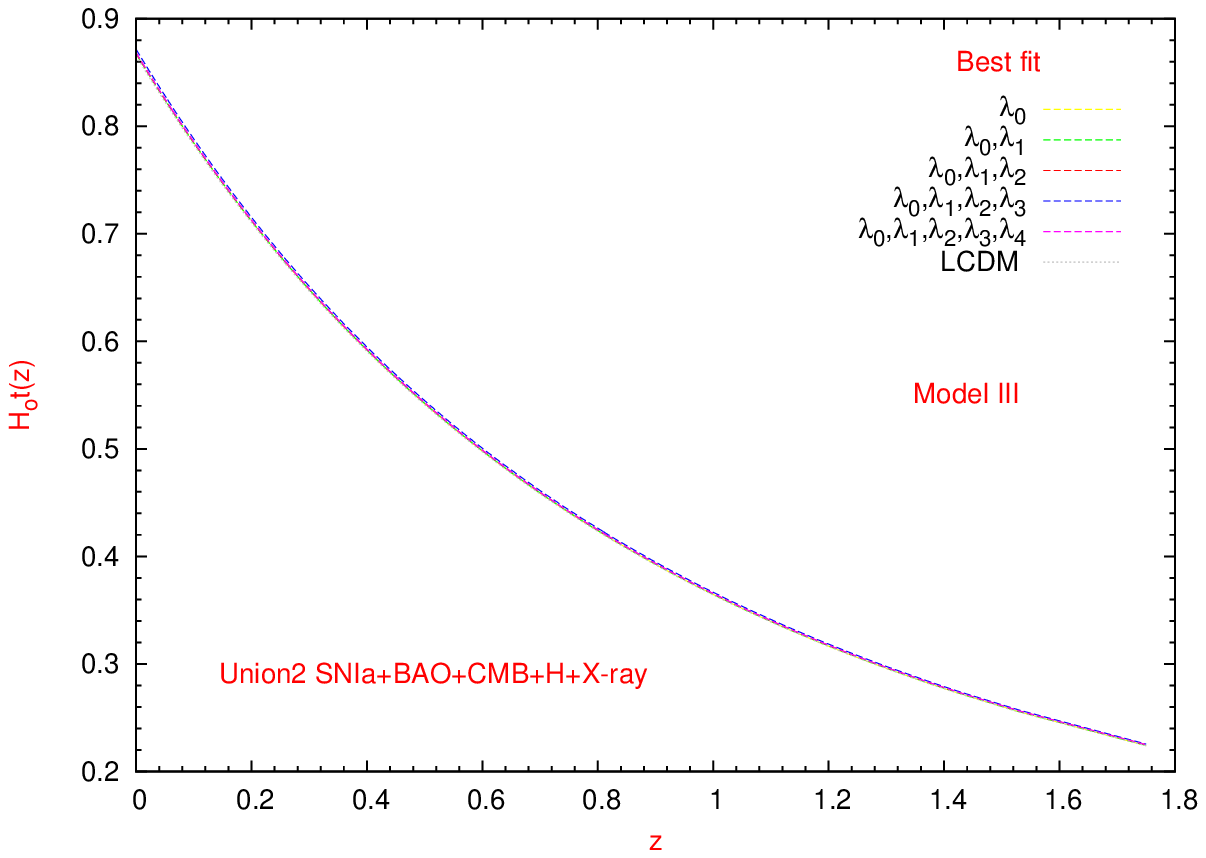}
   \includegraphics[width=8cm, height=60mm, scale=0.90]{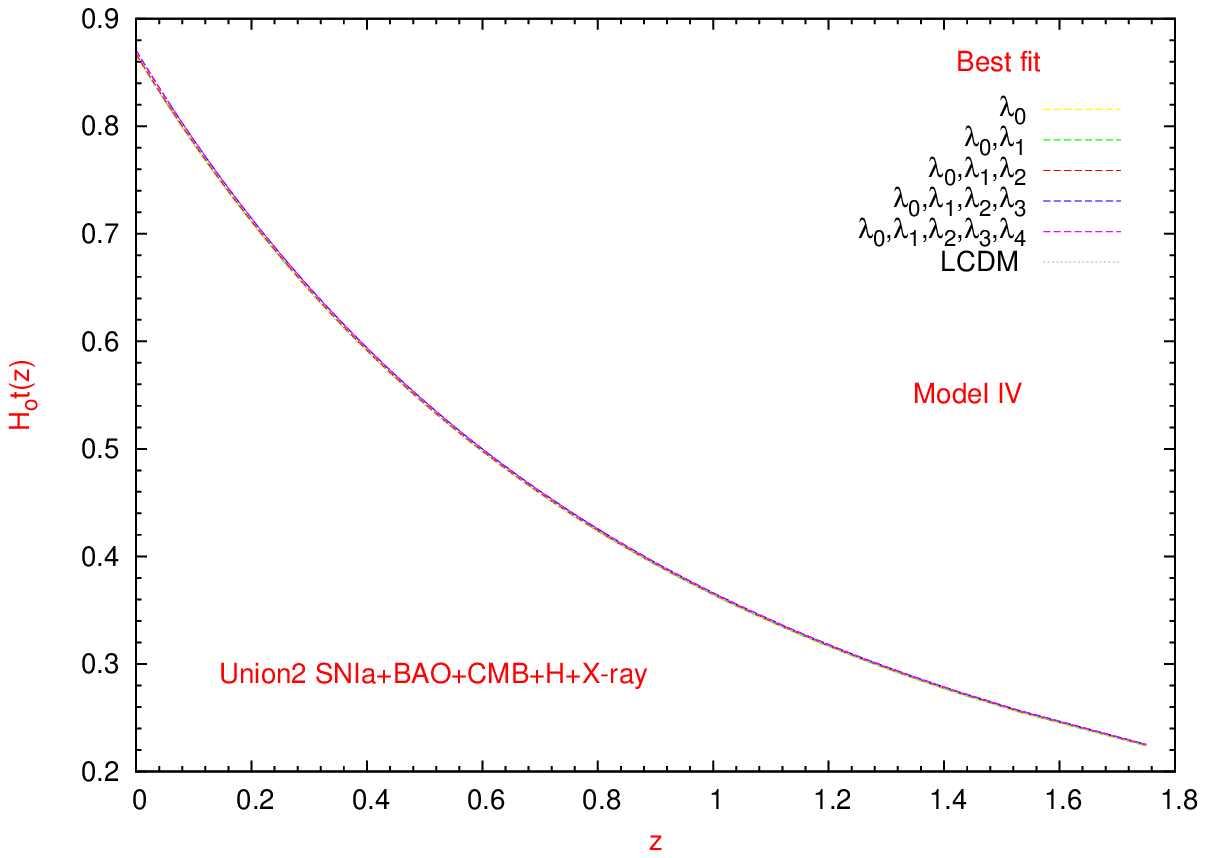}
 \caption{Same as Figure \ref{TotalExchangeModels}, shows the reconstruction of the age of the universe $H_0 t(z)$ as a function of the redshift
  for the model II (left above panel), III (right above panel) and IV (left below panel) respectively. Here, They also compared with the corresponding
  curve for the LCDM (Lambda Cold Dark Matter) model.}
  \label{TotalAgeModels}
\end{figure}
\end{center}

The Figure \ref{TotalAgeModels} shows the reconstruction of the age of the universe $H_0 t(z)$ as a function of
the redshift for the models II, III and IV described respectively.

\begin{center}
\begin{figure}
   \includegraphics[width=8cm, height=60mm, scale=0.90]{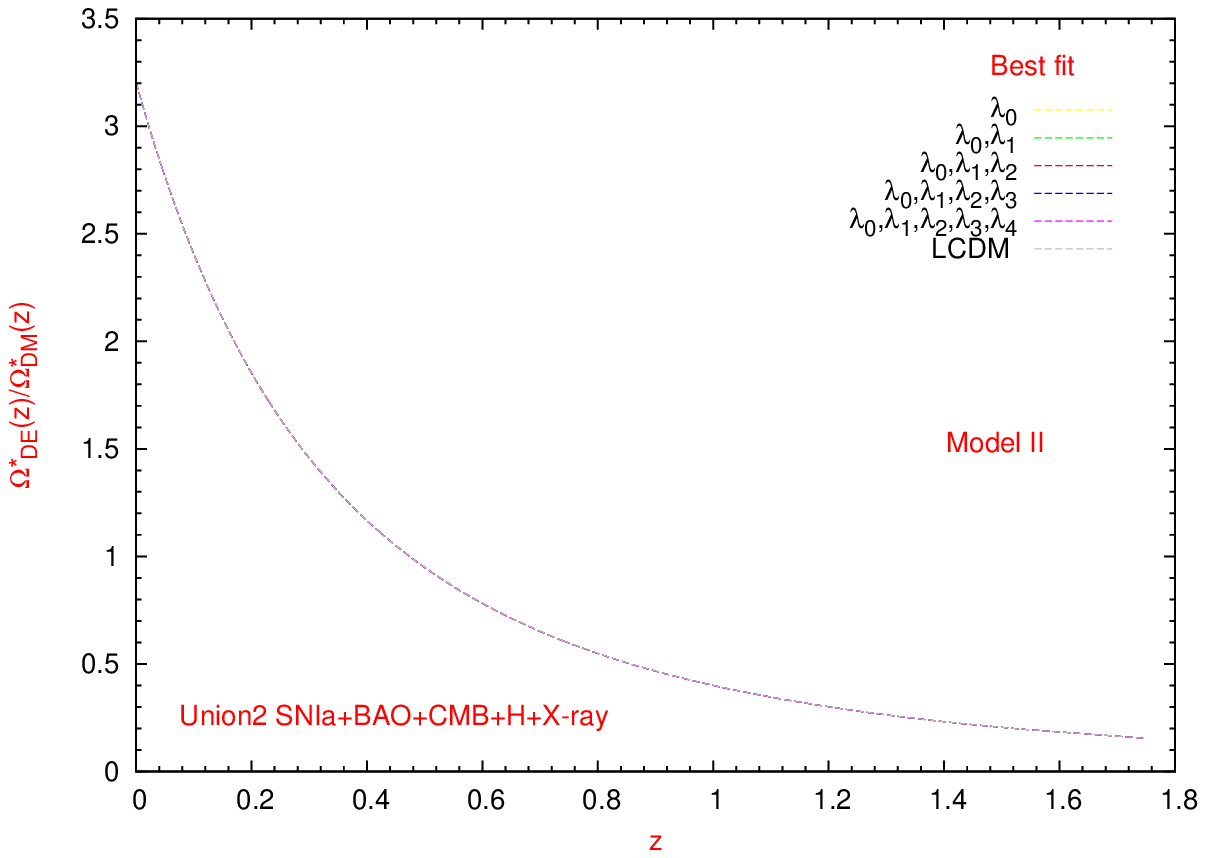}
   \includegraphics[width=8cm, height=60mm, scale=0.90]{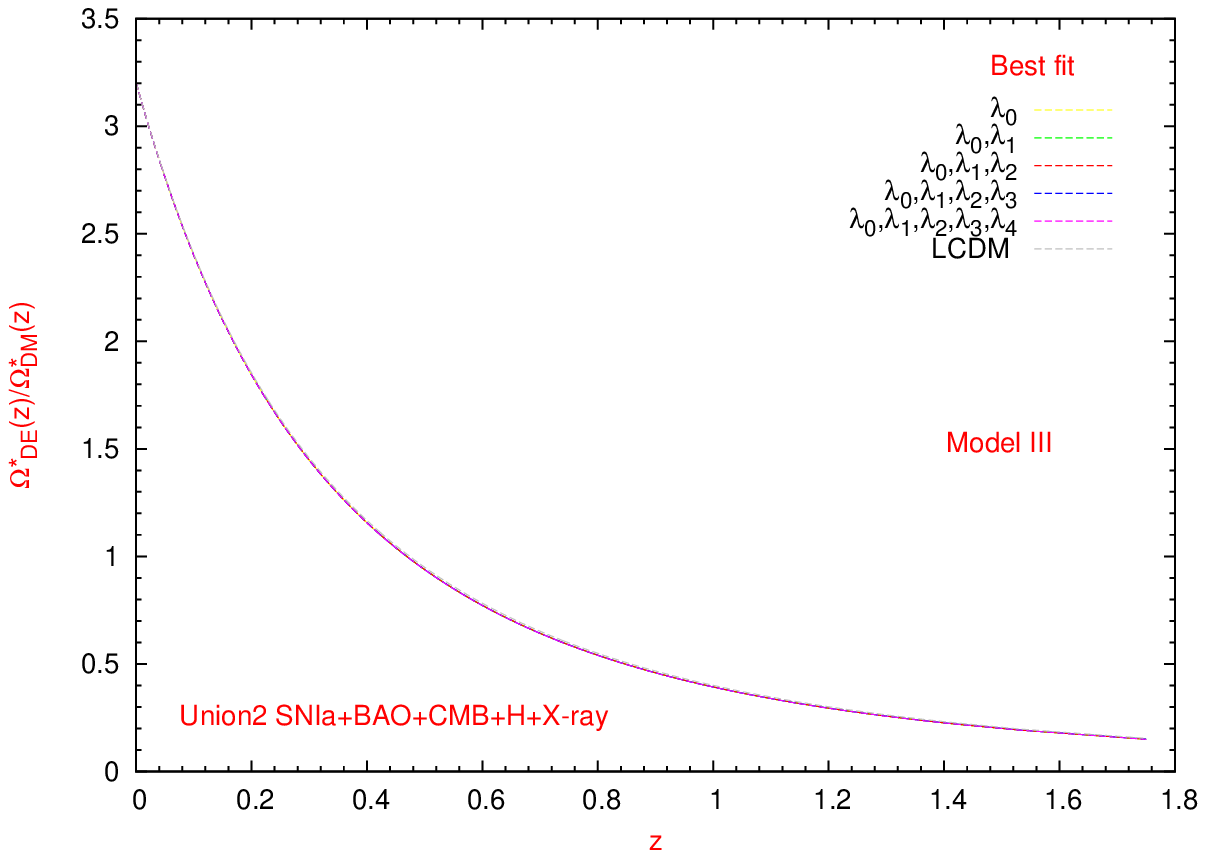}
   \includegraphics[width=8cm, height=60mm, scale=0.90]{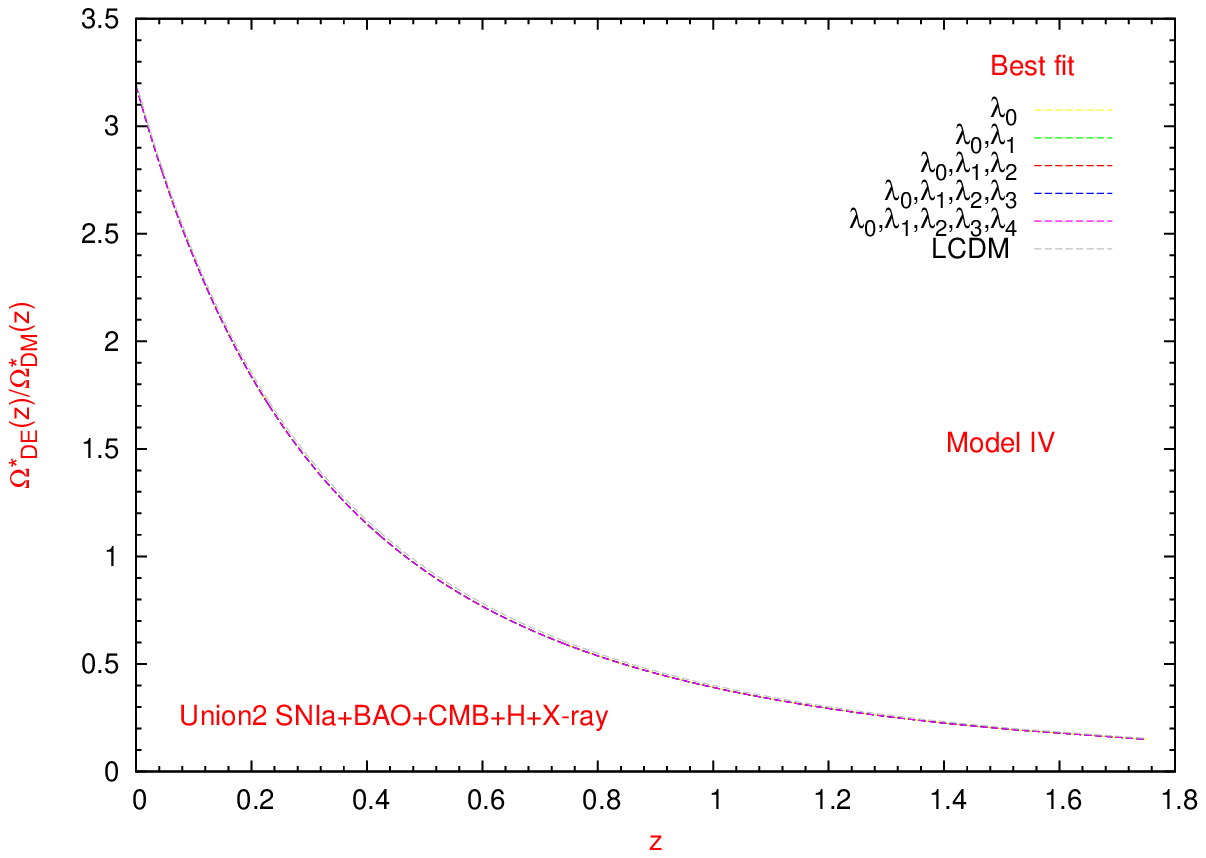}
   \caption{Superposition of the best estimated curves for the rate between dark density parameters $\Omega^{\star}_{DE}(z)/\Omega^{\star}_{DM}(z)$ for
  the model II (left above panel), III (right above panel), IV (left below panel).
  In the above figures, the different colored curves show the best estimates using the expansion in terms of the first
  $N=1,2,3,4$ Chebyshev polynomials respectively. They also compared with the corresponding curve for the LCDM (Lambda Cold Dark Matter) model.}
  \label{TotalRelationModels}
\end{figure}
\end{center}

The Figure \ref{TotalRelationModels} shows the reconstruction of the rate between dark density parameters
$\Omega^{\star}_{DE}(z)/\Omega^{\star}_{DM}(z)$ as a function of the redshift for the models II, III and IV described above.

All these Figures show the superposition of the best estimates for every cosmological variable in terms of the parameters $\lambda_n$ corresponding to
the coefficients of the polynomial expansion (\ref{eq:Coupling}) ranging from $N=1$ to $4$. The Tables \ref{BestestimationII}, \ref{BestestimationIII} and
\ref{BestestimationIV}, show the best fitted parameters and the minimum of the function $\tilde{\chi}_{min}^2$ for the models II, III and IV
respectively. From these tables, we can note the fast convergence of the best estimates when the numbers of parameters $N$
is increased in the expansion (\ref{eq:Coupling}).

\begin{center}
\begin{figure}
 \includegraphics[width=8cm, height=60mm, scale=0.90]{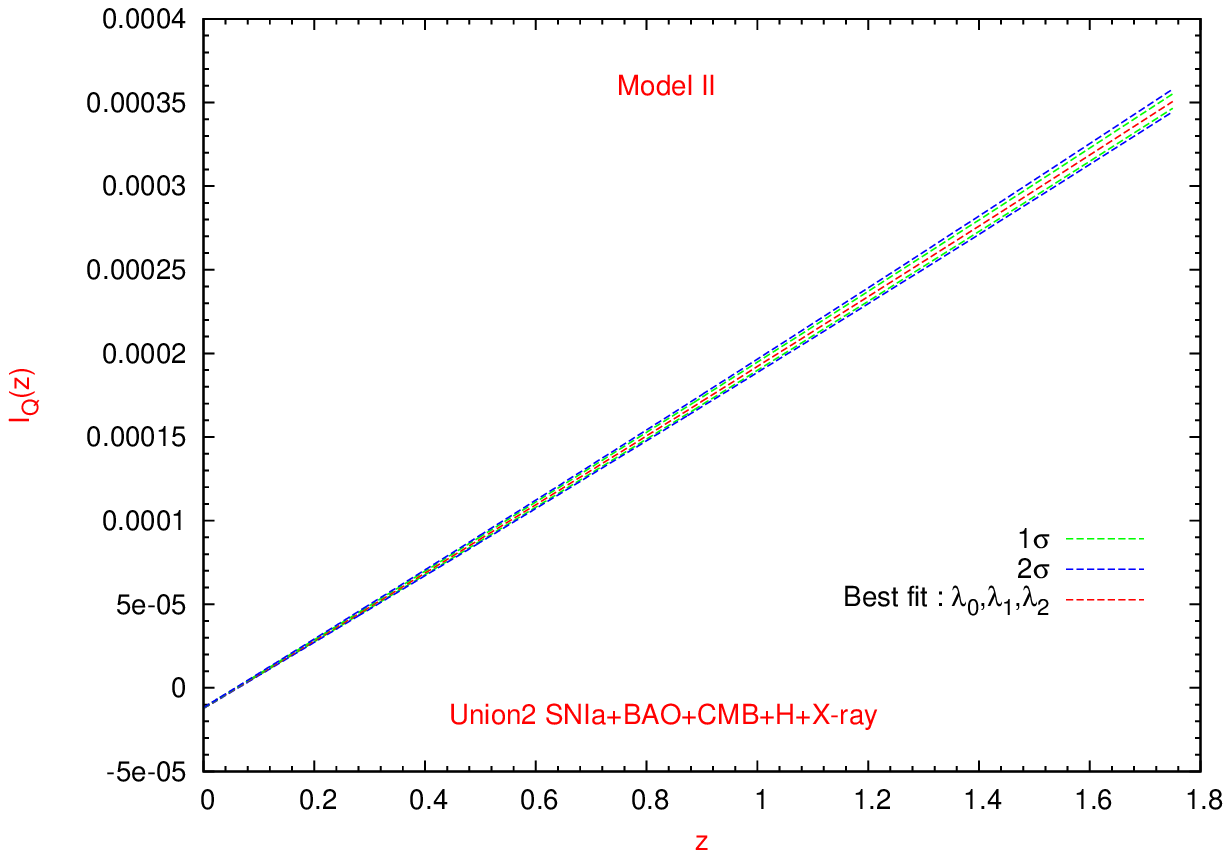}
 \includegraphics[width=8cm, height=60mm, scale=0.90]{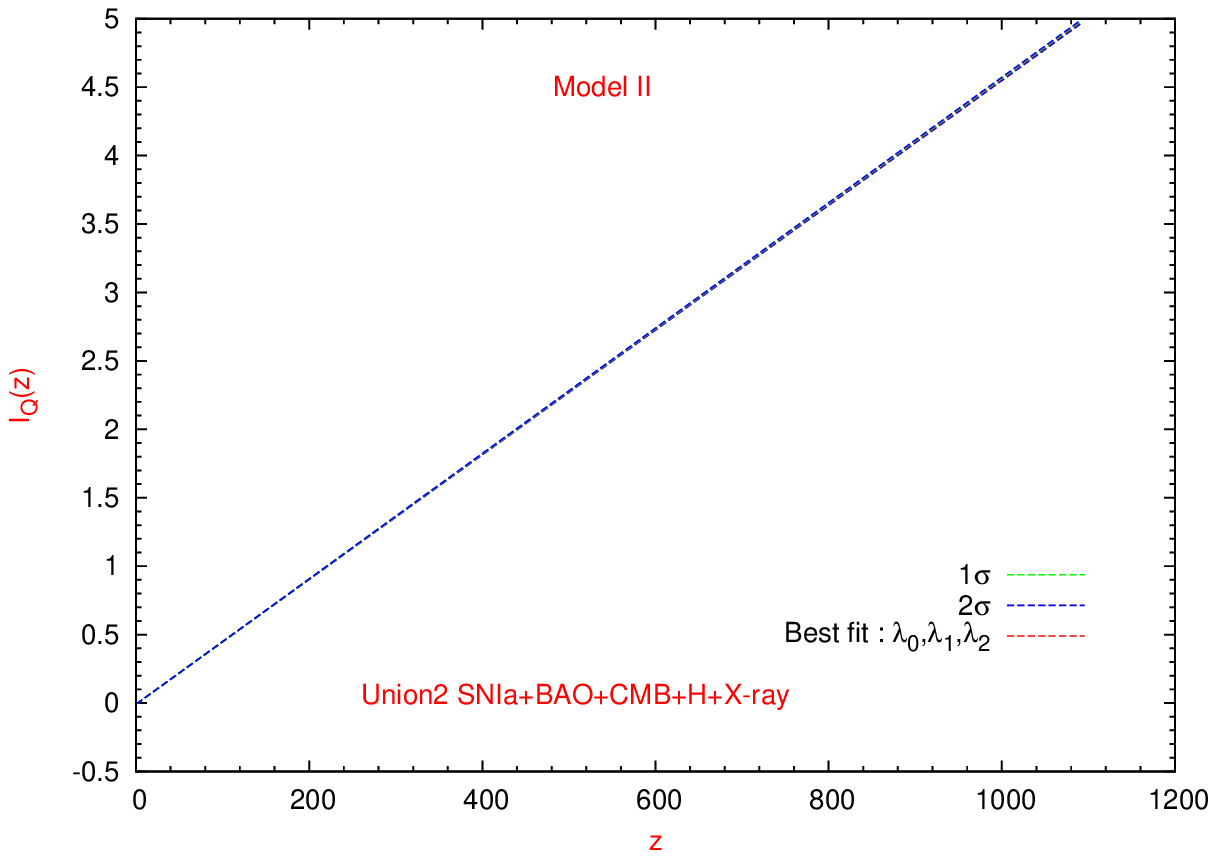}
 \includegraphics[width=8cm, height=60mm, scale=0.90]{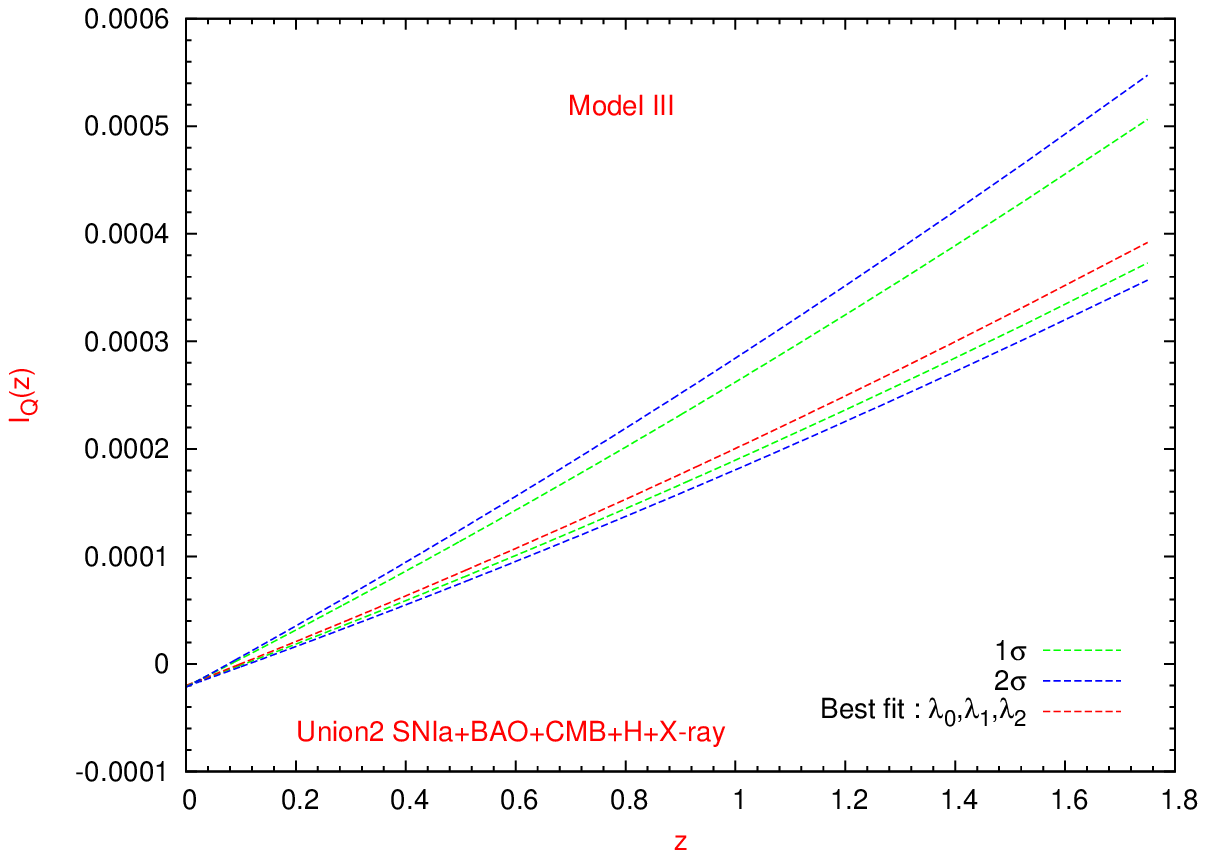}
 \includegraphics[width=8cm, height=60mm, scale=0.90]{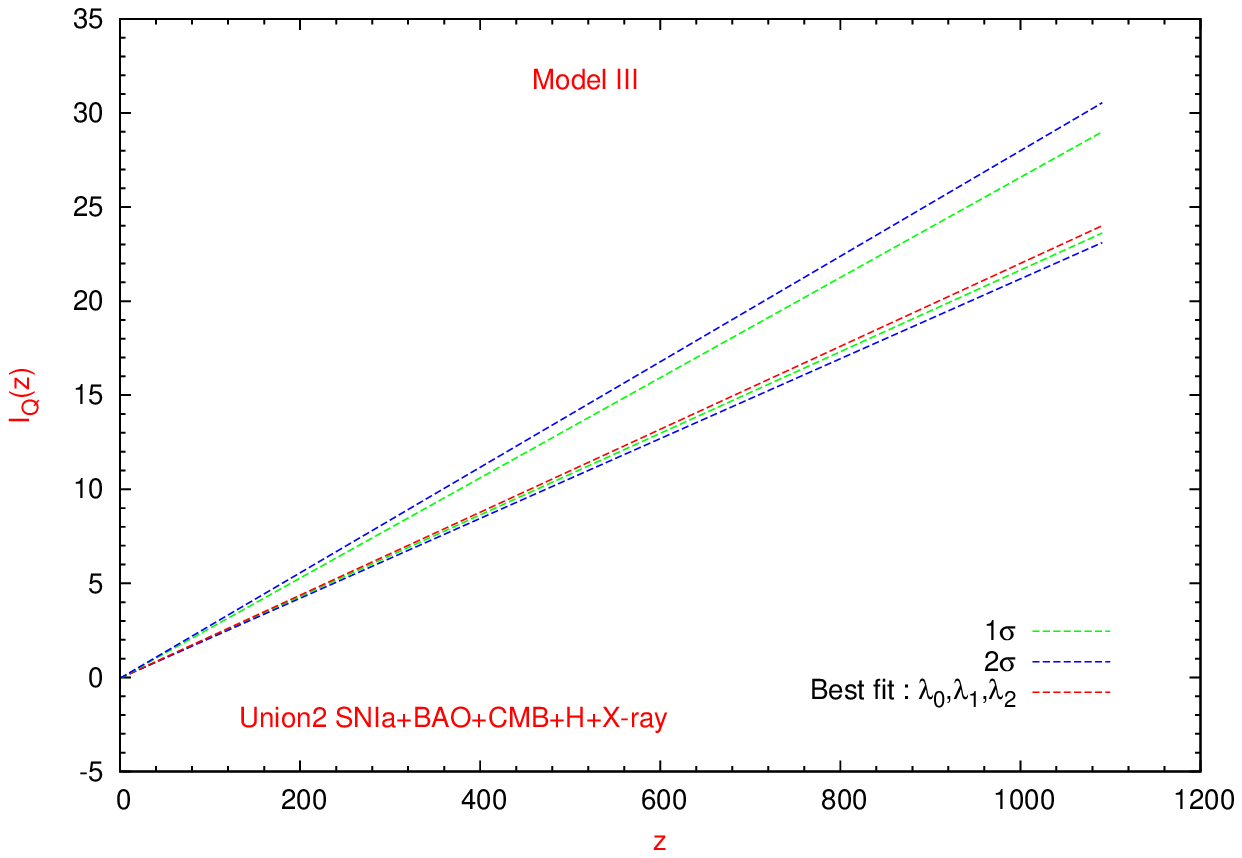}
 \includegraphics[width=8cm, height=60mm, scale=0.90]{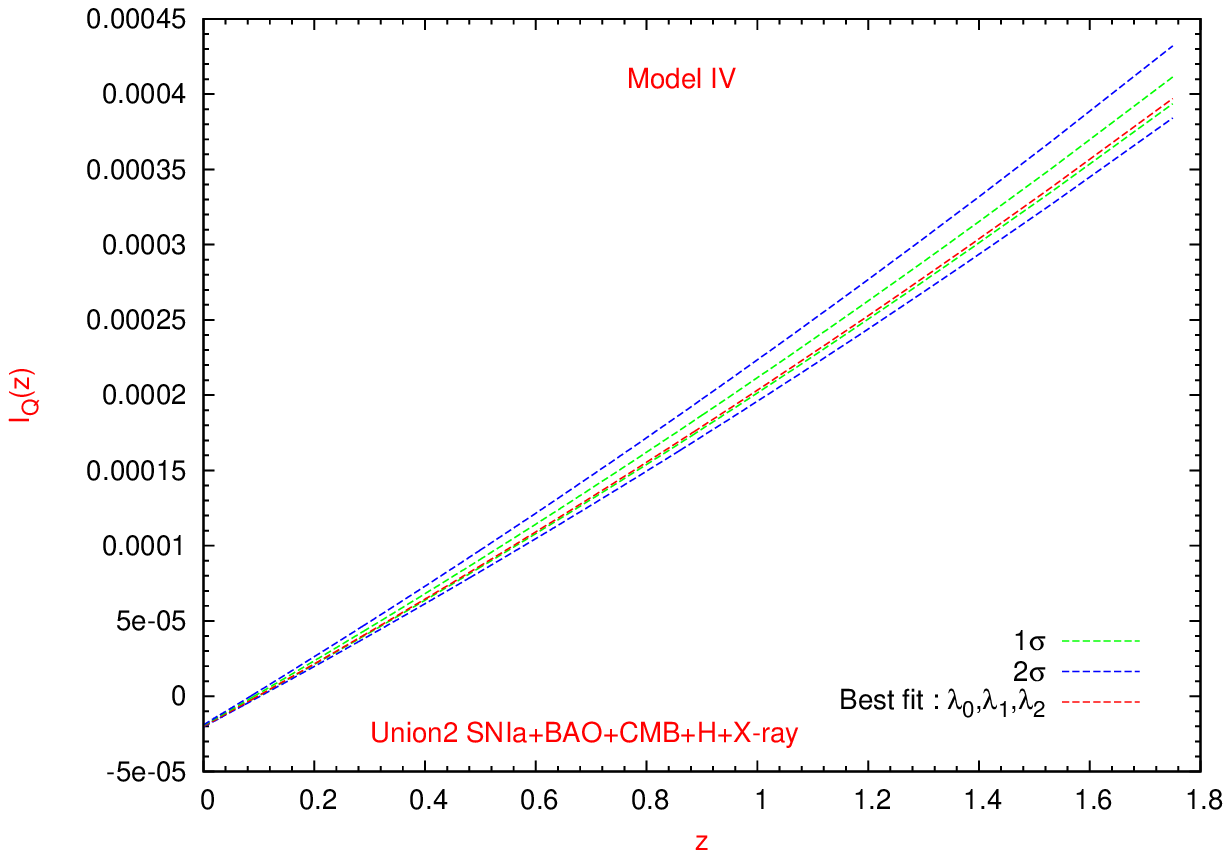}
 \includegraphics[width=8cm, height=60mm, scale=0.90]{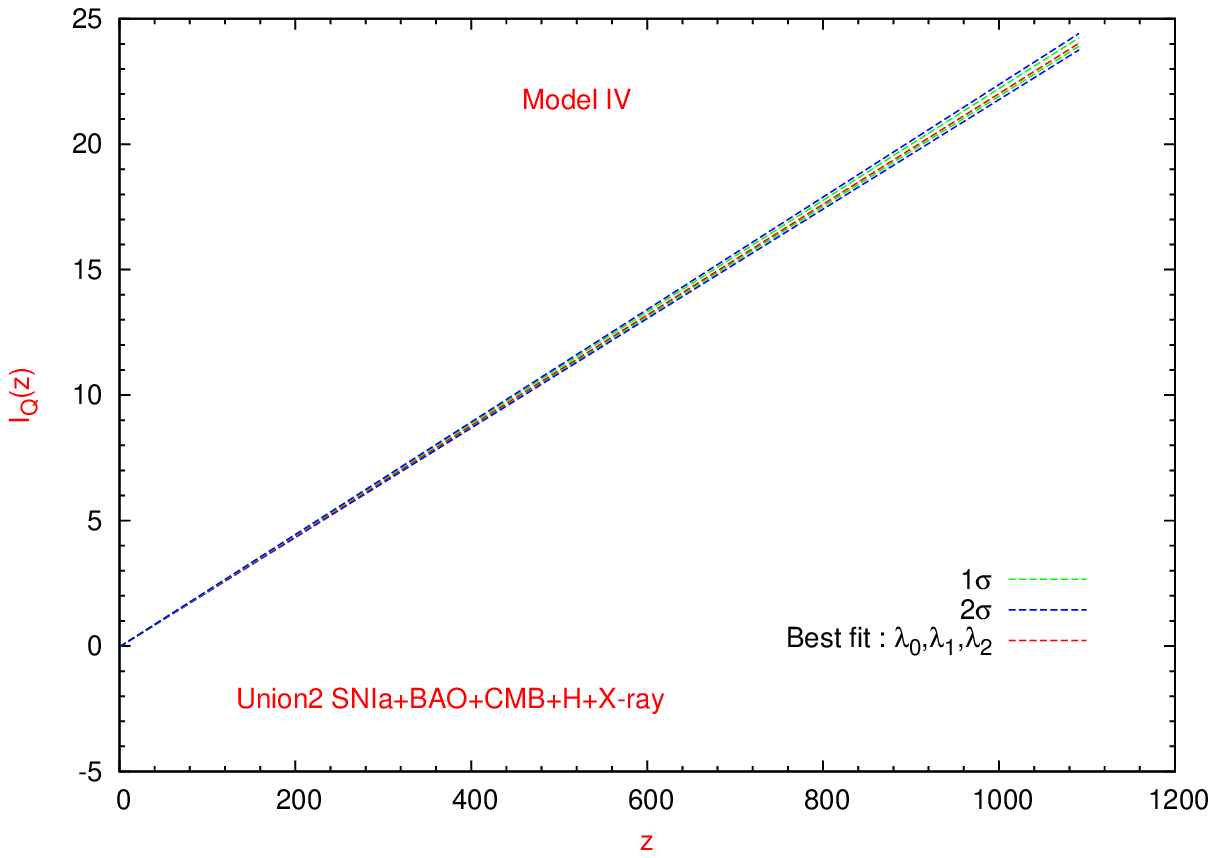}
 \caption{Comparison of the  best fitted reconstructed dimensionless interaction function
 ${\rm I}_{\rm Q}(z)$ and its errors, using the expansion of ${\rm I}_{\rm Q}(z)$ in terms of the parameters $\lambda_0$, $\lambda_1$, $\lambda_2$ and
 the corresponding Chebyshev polynomials for the  models II (left above panel), III
 (right above panel) and IV (left below panel) corresponding to  a dark energy equation of state parameter (II) $w=-1$,
 (III) $w=$ constant (with $\Omega_{DM}^0=0.227$) and (IV) $w=$ constant and $\Omega_{DM}^0=$ constant, respectively.
 Red lines show best fitted reconstructed results, while green lines and blue lines show reconstructed errors within the $1\sigma$ and $21\sigma$ confidence level errors,
 obtained from a combination of SNeIa + BAO + CMB + H + X-ray dataset.  Otherwise, note that within the $1\sigma$ and $2\sigma$ errors it could be the possibility that
 the crossing of the noninteracting ${\rm I}_{\rm Q}(z)=0$ line happens before the present.}
 \label{ErrorsTotalExchangeModels}
\end{figure}
\end{center}

The Figure \ref{ErrorsTotalExchangeModels} shows the reconstruction of the dimensionless interaction function ${\rm I}_{\rm Q}(z)$ as a
function of the redshift, using the best estimates for $N=2$ and their confidence intervals at $1\sigma$ and $2\sigma$ for the models II
(corresponding to a dark energy EOS parameter $w=-1$), III (corresponding to a dark energy EOS parameter $w=$ constant and $\Omega_{DM}^0$ fixed)
and IV (corresponding to a dark energy EOS parameter $w=$ constant and $\Omega_{DM}^0$ as a free parameter to be estimated) respectively.

\begin{center}
\begin{figure}
 \includegraphics[width=8cm, height=60mm, scale=0.90]{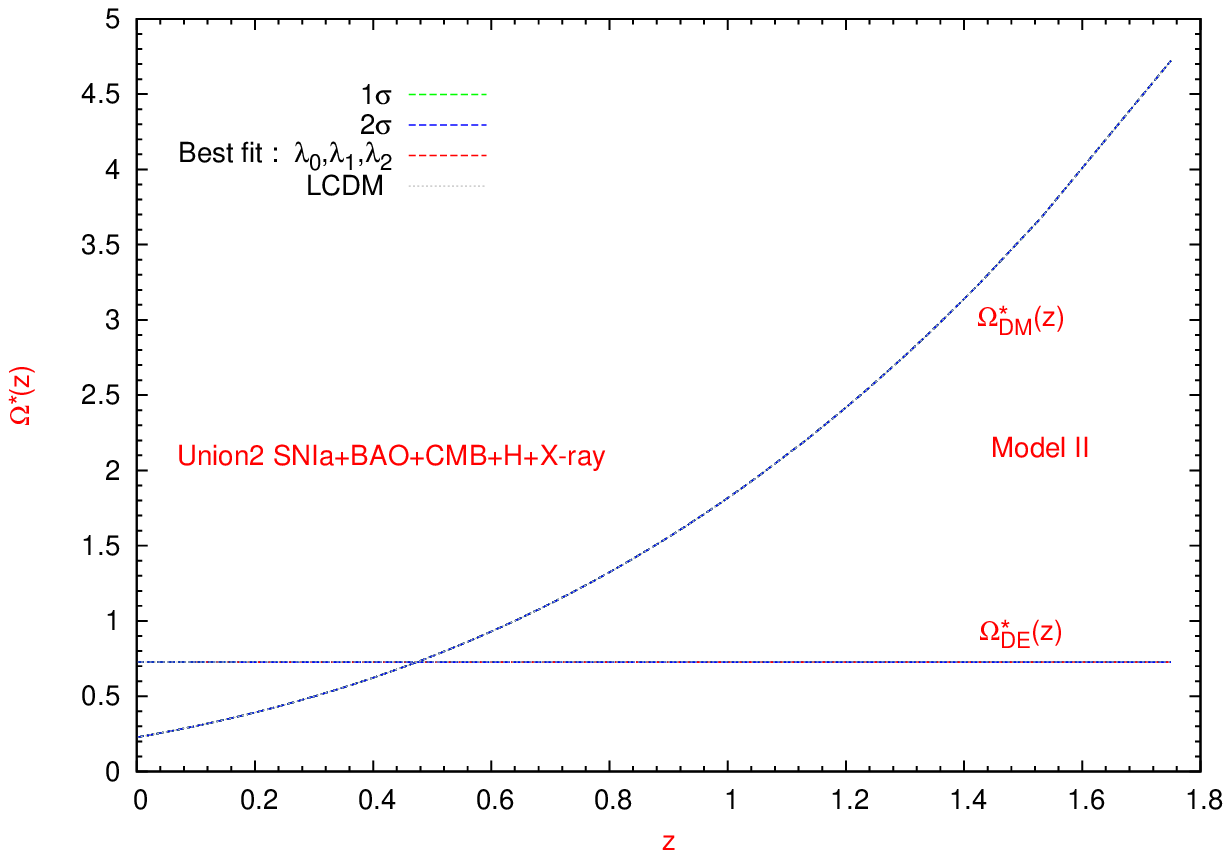}
 \includegraphics[width=8cm, height=60mm, scale=0.90]{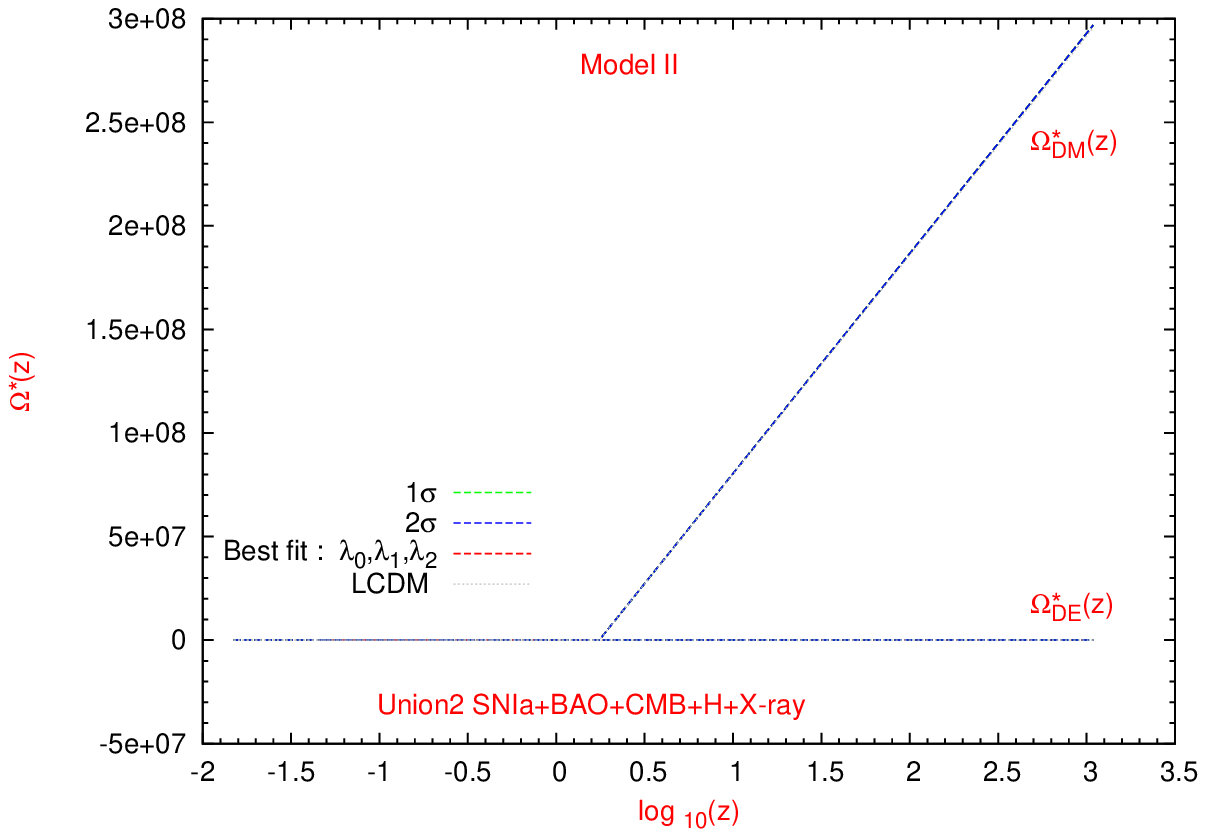}
 \includegraphics[width=8cm, height=60mm, scale=0.90]{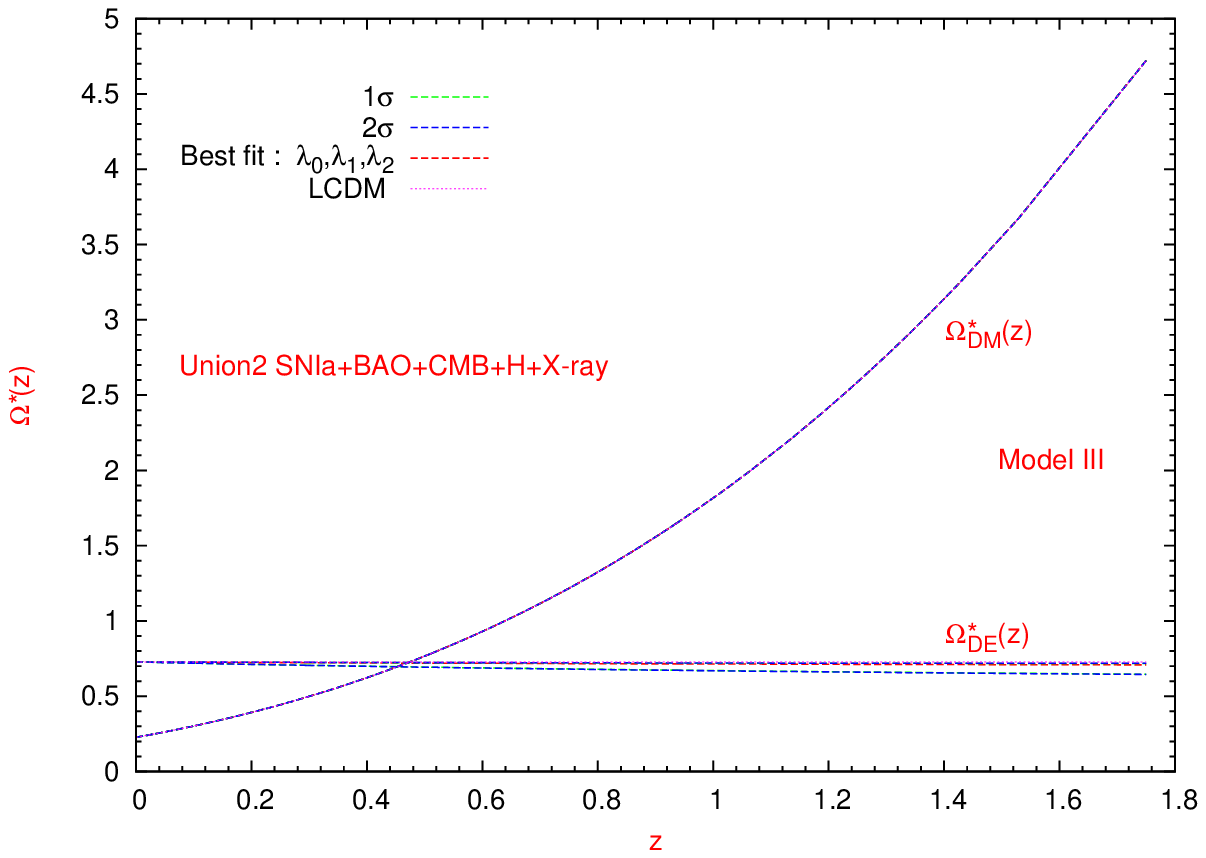}
 \includegraphics[width=8cm, height=60mm, scale=0.90]{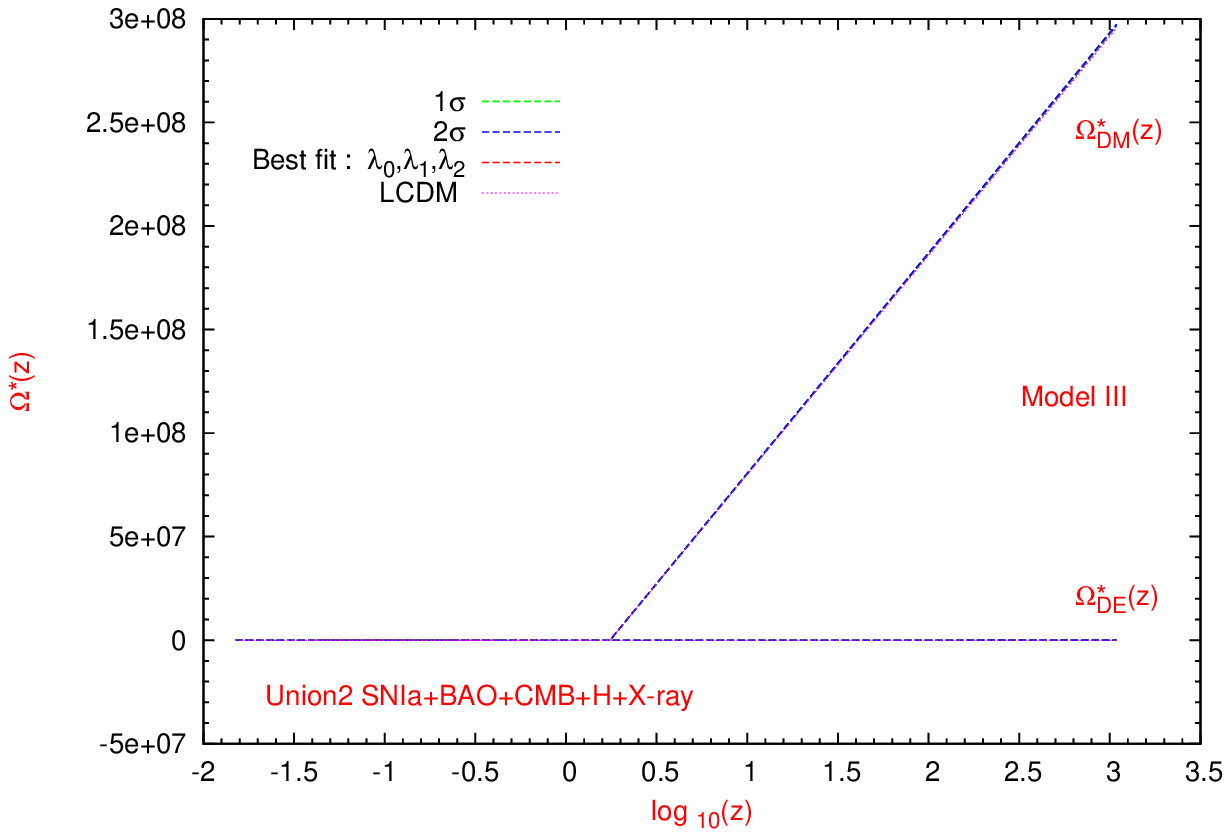}
 \includegraphics[width=8cm, height=60mm, scale=0.90]{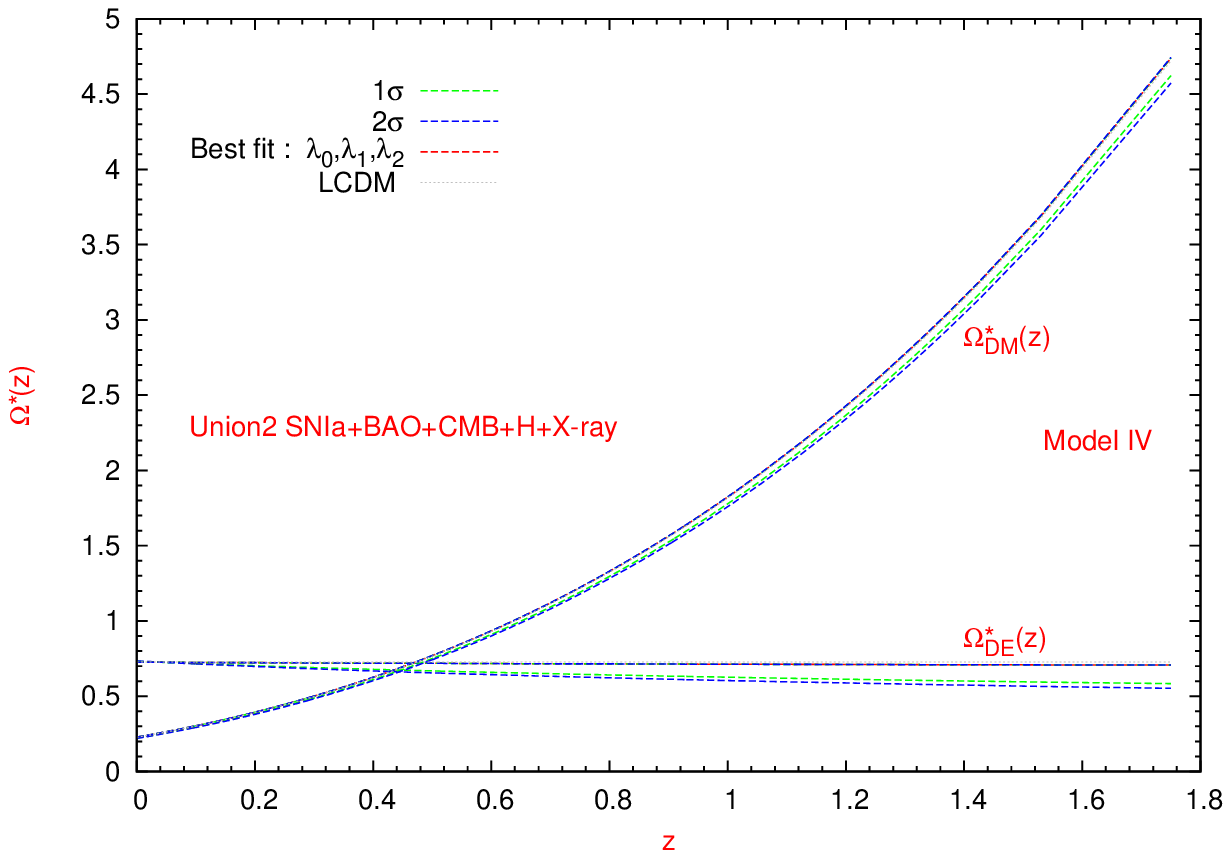}
 \includegraphics[width=8cm, height=60mm, scale=0.90]{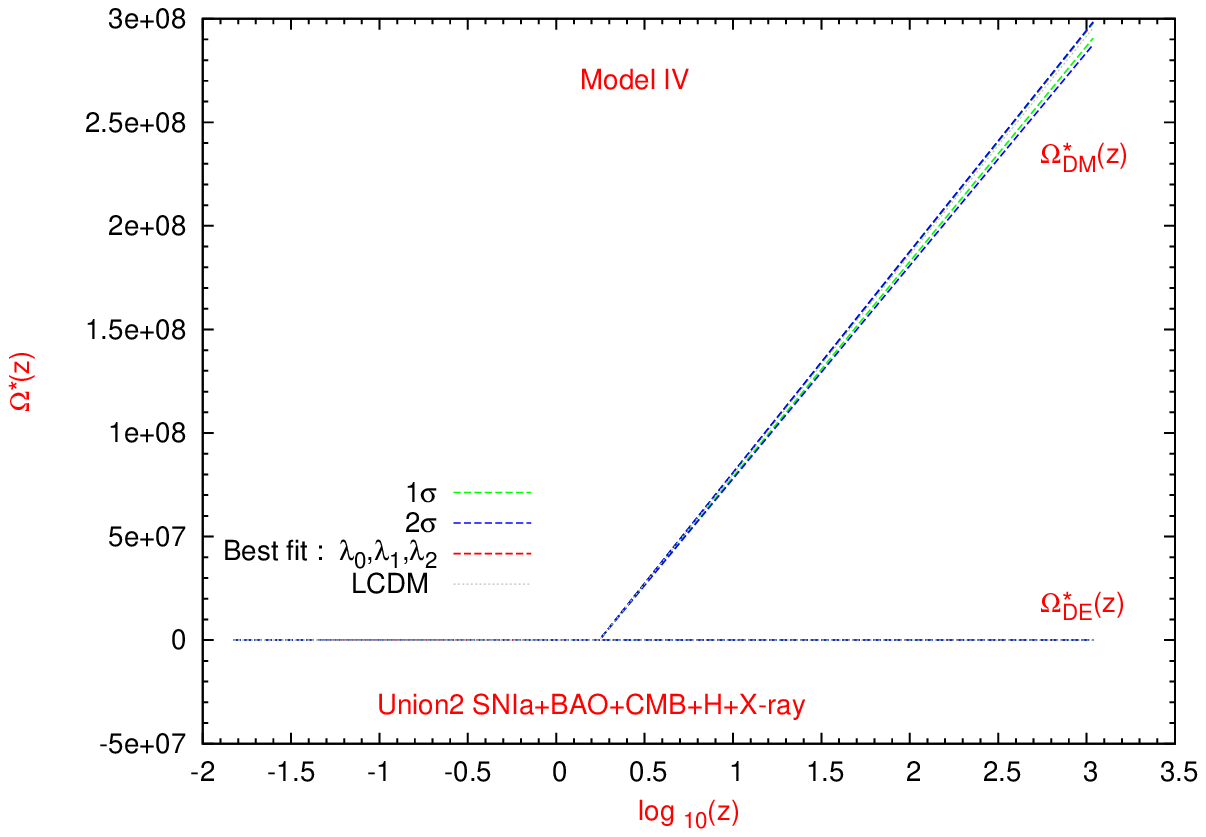}
 \caption{Same as Figure \ref{ErrorsTotalExchangeModels}, comparison of the  best fitted reconstruction and its errors of the dark matter $\Omega^{\star}_{DM}(z)$ and dark energy $\Omega^{\star}_{DE}(z)$ density parameters, in terms of the parameters $\lambda_0$, $\lambda_1$, $\lambda_2$ and the corresponding Chebyshev polynomials for the models II (above panel), III (right above panel) and IV (left below panel),respectively.
 Red lines show best fitted reconstructed results, while green lines and blue lines show reconstructed errors within the $1\sigma$ and $21\sigma$ confidence level errors,
 determined from a combination of SNeIa + BAO + CMB + H + X-ray dataset. In addition, note that within the $1\sigma$ and $2\sigma$ errors, our best fitted results and its errors are consistent with the constraints on $\Omega^{\star}_{DM}(z)$ and $\Omega^{\star}_{DE}(z)$ in the LCDM model. Furthermore, from the last Figure (right below panel) at early time, the $1\sigma$ and $2\sigma$ constraints for $\Omega^{\star}_{DM}(z)$ are above of the corresponding for the LCDM model, this considerably alleviates the coincidence problem albeit it does not solve it in full.}
 \label{ErrorsTotalCoincidenceModels}
\end{figure}
\end{center}

The Figure \ref{ErrorsTotalCoincidenceModels} shows the reconstruction of the dark matter and dark energy density parameters
$\Omega^{\star}_{DM}(z)$, $\Omega^{\star}_{DE}(z)$ as a function of the redshift, using the best estimates for $N=2$ and their confidence intervals
at $1\sigma$ and $2\sigma$ for the models II, III and IV described above.

\begin{center}
\begin{figure}
 \includegraphics[width=8cm, height=60mm, scale=0.90]{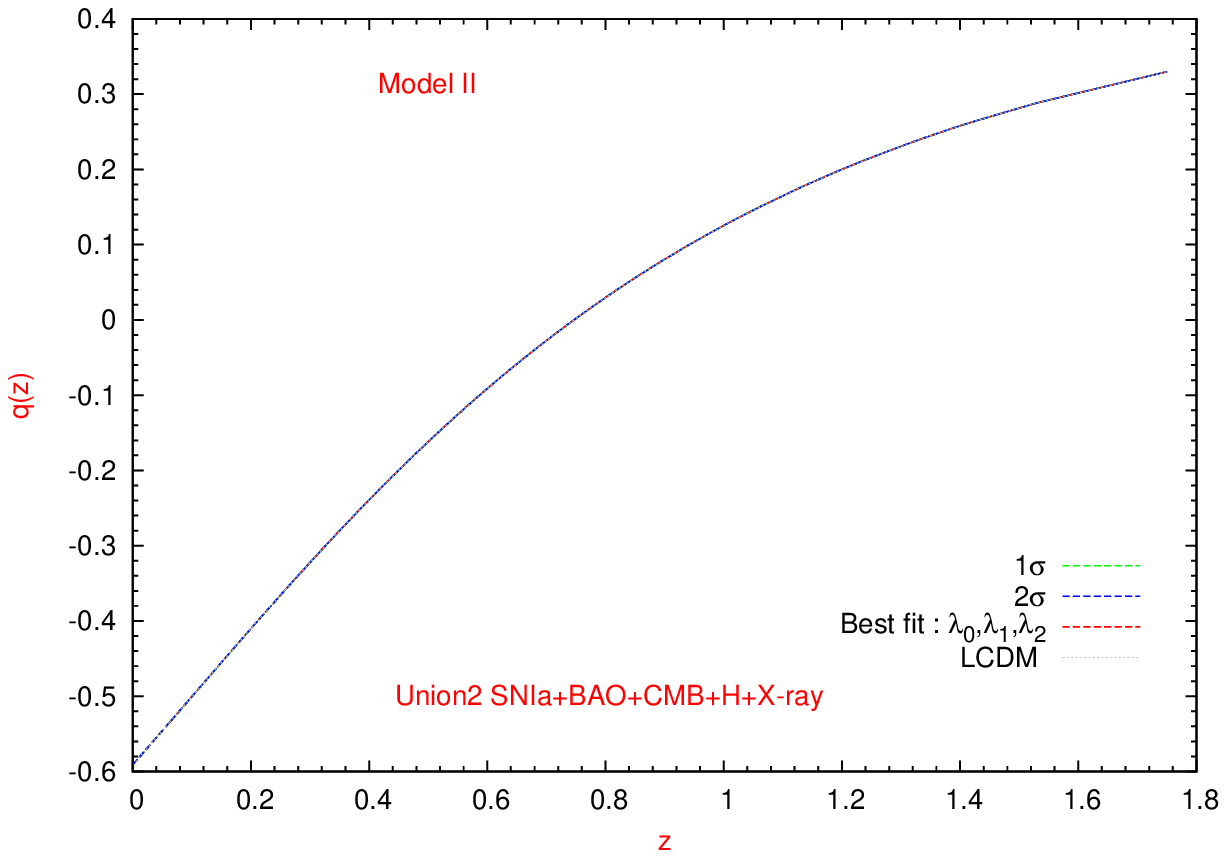}
 \includegraphics[width=8cm, height=60mm, scale=0.90]{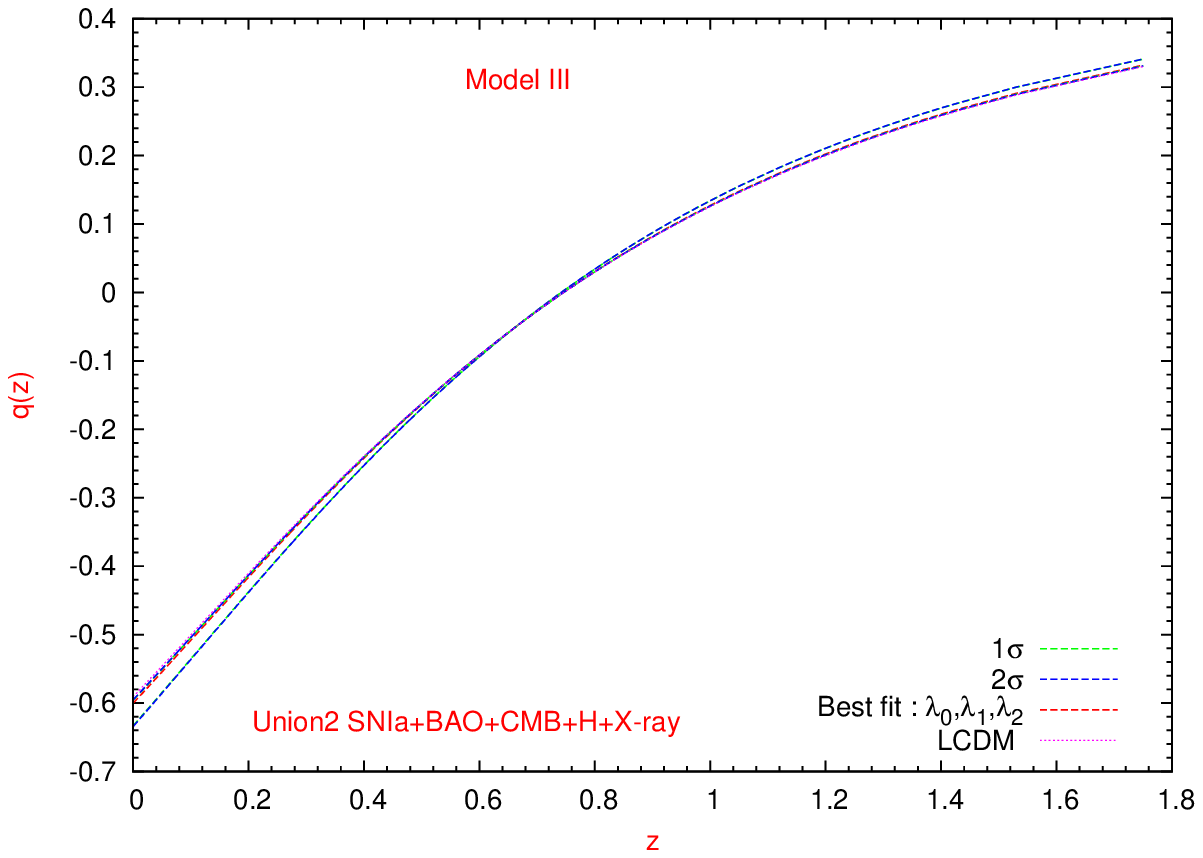}
 \includegraphics[width=8cm, height=60mm, scale=0.90]{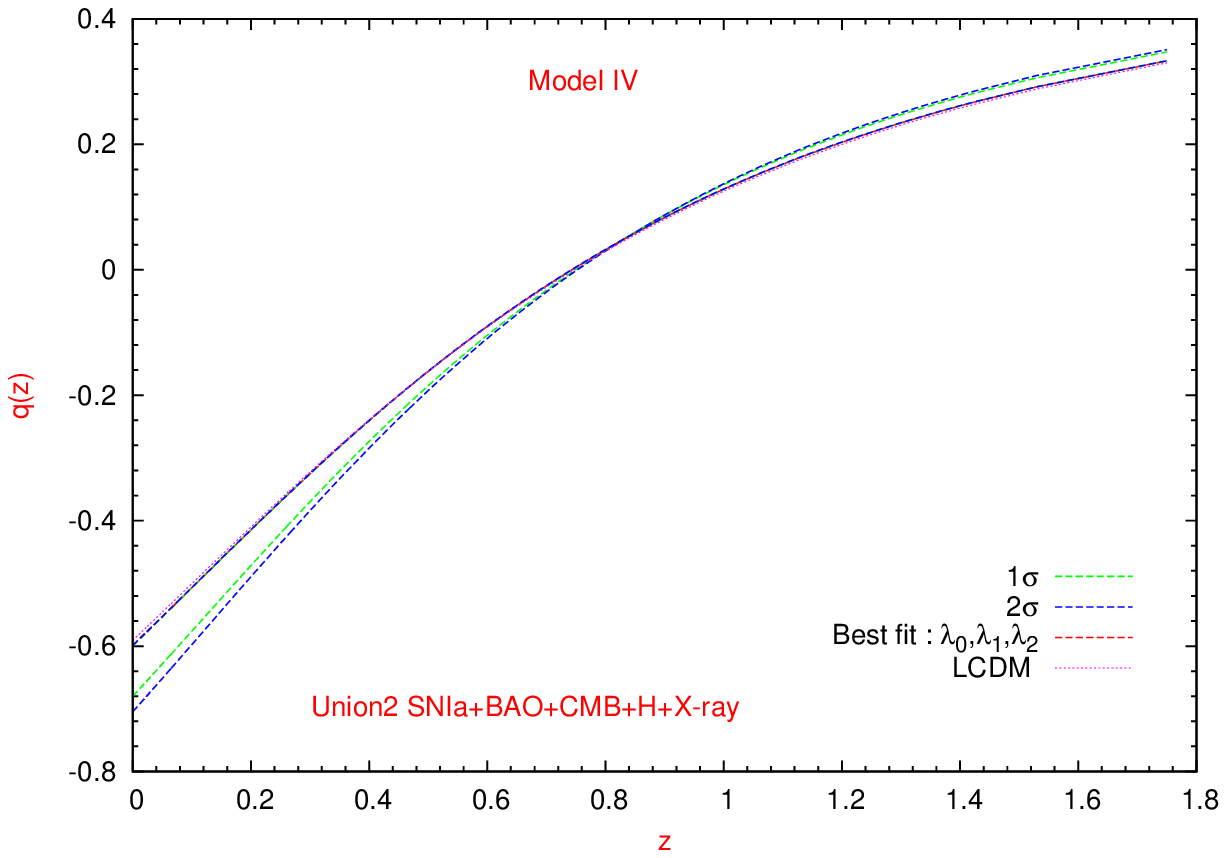}
 \caption{Same as Figure \ref{ErrorsTotalExchangeModels}. Comparison and evolution of the  best fitted reconstructed and its errors of the deceleration parameter $q(z)$,
 in terms of the parameters $\lambda_0$, $\lambda_1$, $\lambda_2$ and the corresponding Chebyshev polynomials for the models II (left above panel), III (right above panel) and IV (left below panel),respectively.
 Red lines show best fitted reconstructed results, while green lines and blue lines show reconstructed errors within the $1\sigma$ and $21\sigma$ confidence level errors,
 determined from a combination of SNeIa + BAO + CMB + H + X-ray dataset. In addition, note that within the $1\sigma$ and $2\sigma$ errors, our result and its errors are consistent with the constraints on $q(z)$ in the LCDM model. All Figures show a strong evidence of acceleration in the recent past which is consistent with a previous study in \cite{Gong2006}. Also the strongest evidence of acceleration again happens around the redshift $z~0.2$. The transition redshift when the universe underwent the transition from deceleration to acceleration is found to be $z~0.7$ at the $1\sigma$ level. Furthermore, our best fit results and its errors at $1\sigma$ and $2\sigma$ are in LCDM model.}
 \label{ErrorsTotalDecelerateModels}
\end{figure}
\end{center}

The Figure \ref{ErrorsTotalDecelerateModels} which shows the reconstruction and the behavior of the deceleration parameter $q(z)$ as a function of
the redshift and their confidence intervals at $1\sigma$ and $2\sigma$ for the models II, III and IV respectively.

\begin{center}
\begin{figure}
 \includegraphics[width=8cm, height=60mm, scale=0.90]{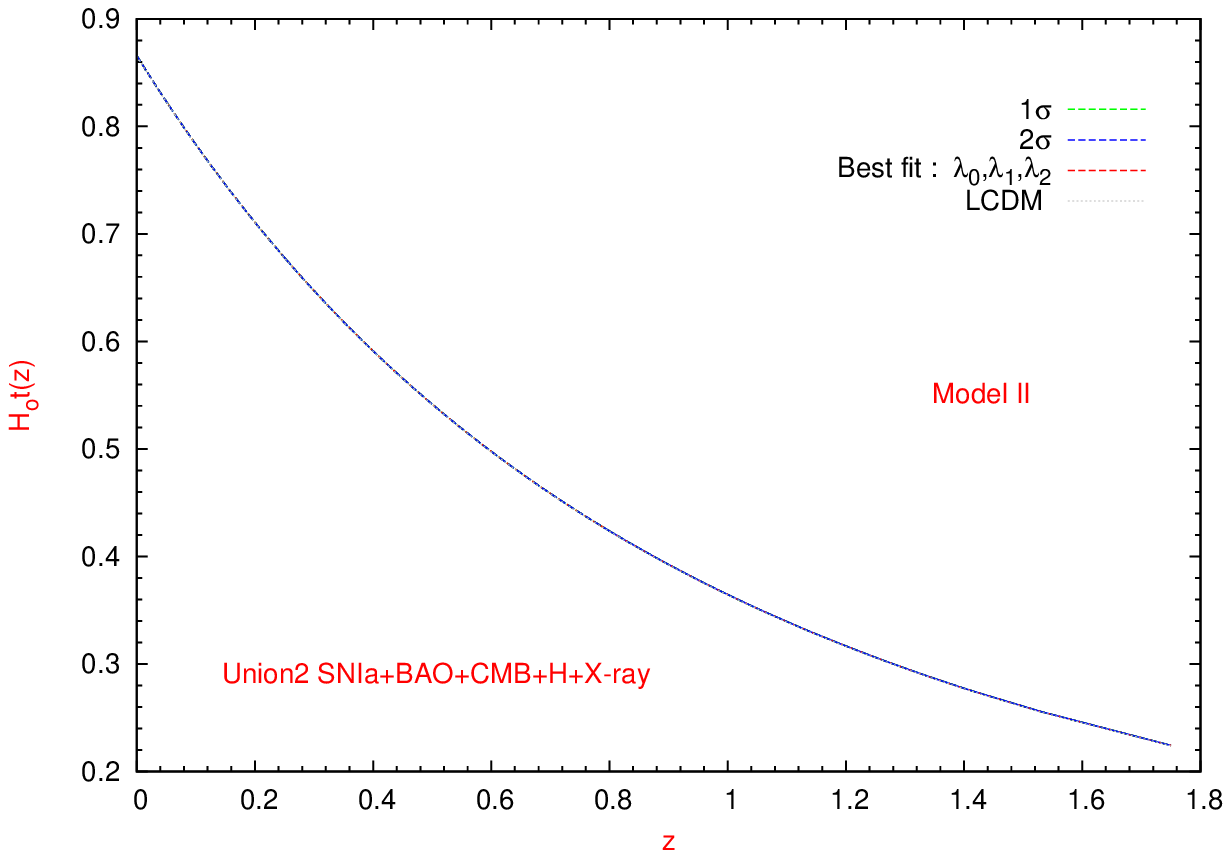}
 \includegraphics[width=8cm, height=60mm, scale=0.90]{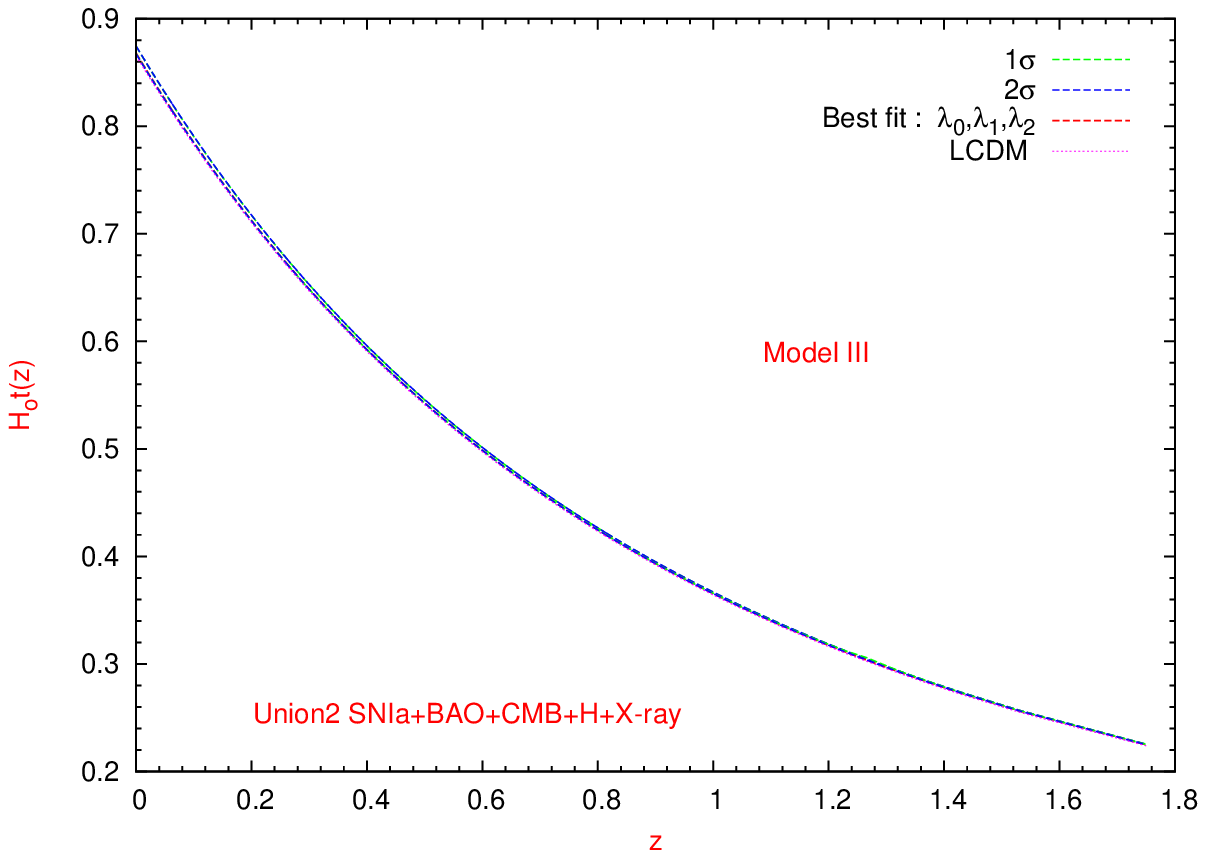}
 \includegraphics[width=8cm, height=60mm, scale=0.90]{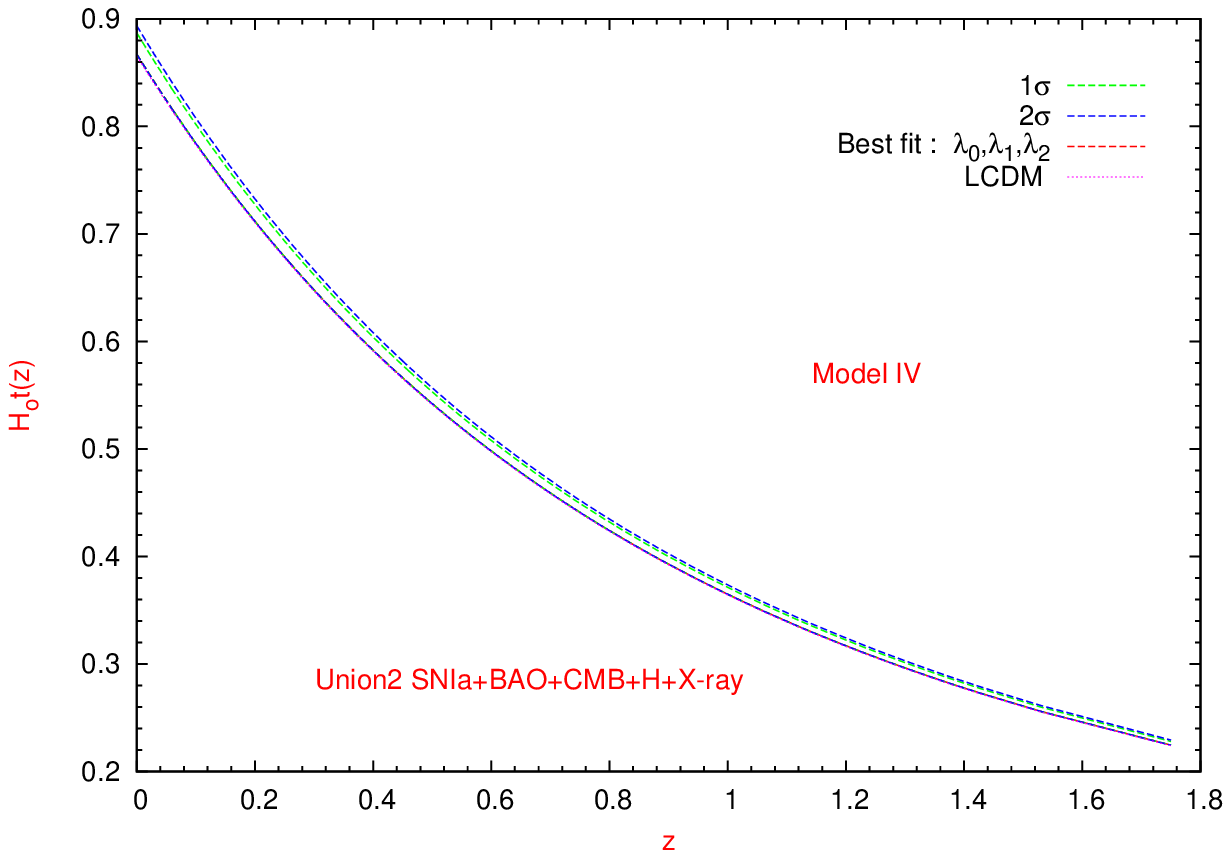}
 \caption{Same as Figure \ref{ErrorsTotalExchangeModels}. Evolution of the best fit reconstructed and its errors of the age of the universe $H_0 t(z)$,
 in terms of the parameters $\lambda_0$, $\lambda_1$, $\lambda_2$ and the corresponding Chebyshev polynomials for the models II (left above panel), III (right above panel) and IV (left below panel),respectively.
 Red lines show best fit reconstructed results, while green lines and blue lines show reconstructed errors within the $1\sigma$ and $21\sigma$ confidence level errors,
 determined from a combination of SNeIa + BAO + CMB + H + X-ray dataset. In addition, note that within the $1\sigma$ and $2\sigma$ errors, our result and its errors are  consistent with the constraints on $H_0 t(z)$ in the LCDM model. Furthermore, our best fit results and its errors at $1\sigma$ and $2\sigma$ are in LCDM model.}
 \label{ErrorsTotalAgeModels}
\end{figure}
\end{center}

The Figure \ref{ErrorsTotalAgeModels} shows the reconstruction of the age of the universe $H_{0}t(z)$ as a function of
the redshift and their confidence intervals at $1\sigma$ and $2\sigma$ for the models II, III and IV already described respectively.

\begin{center}
\begin{figure}
 \includegraphics[width=8cm, height=60mm, scale=0.90]{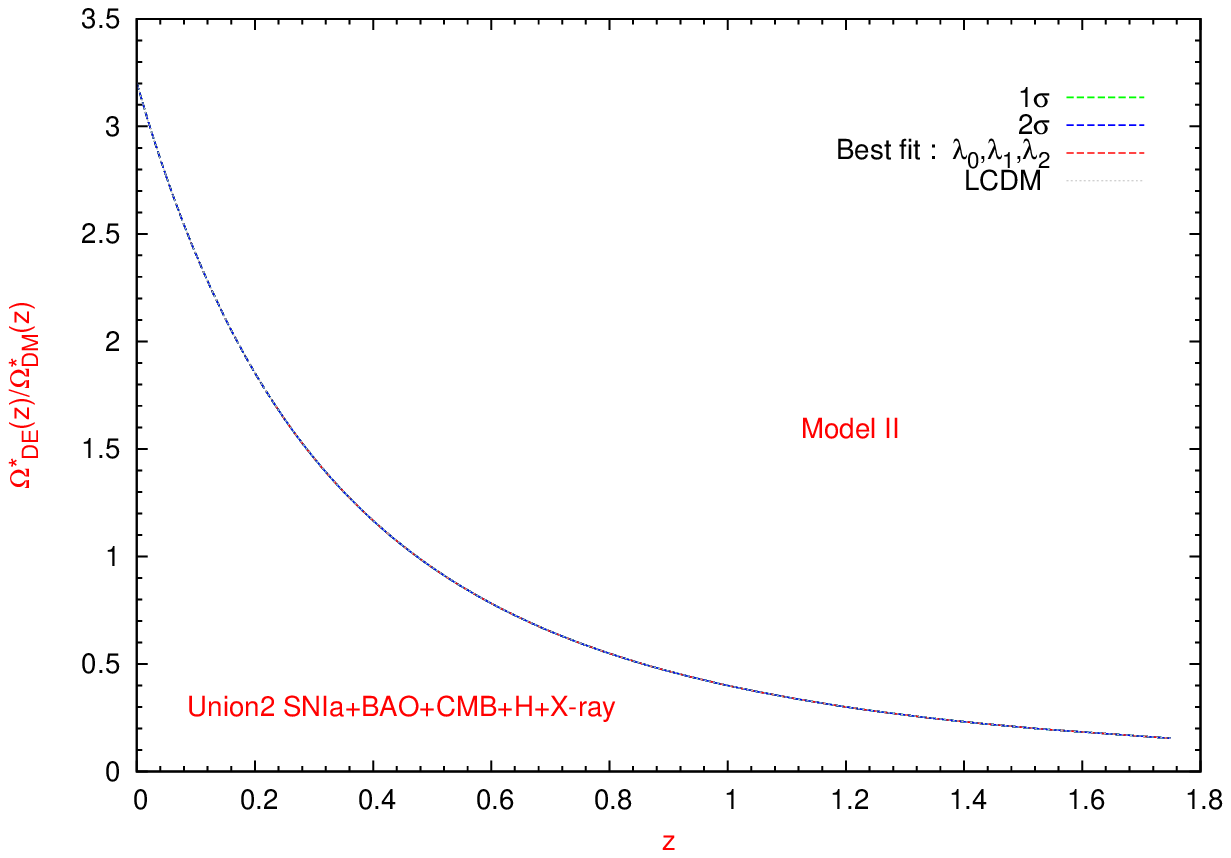}
 \includegraphics[width=8cm, height=60mm, scale=0.90]{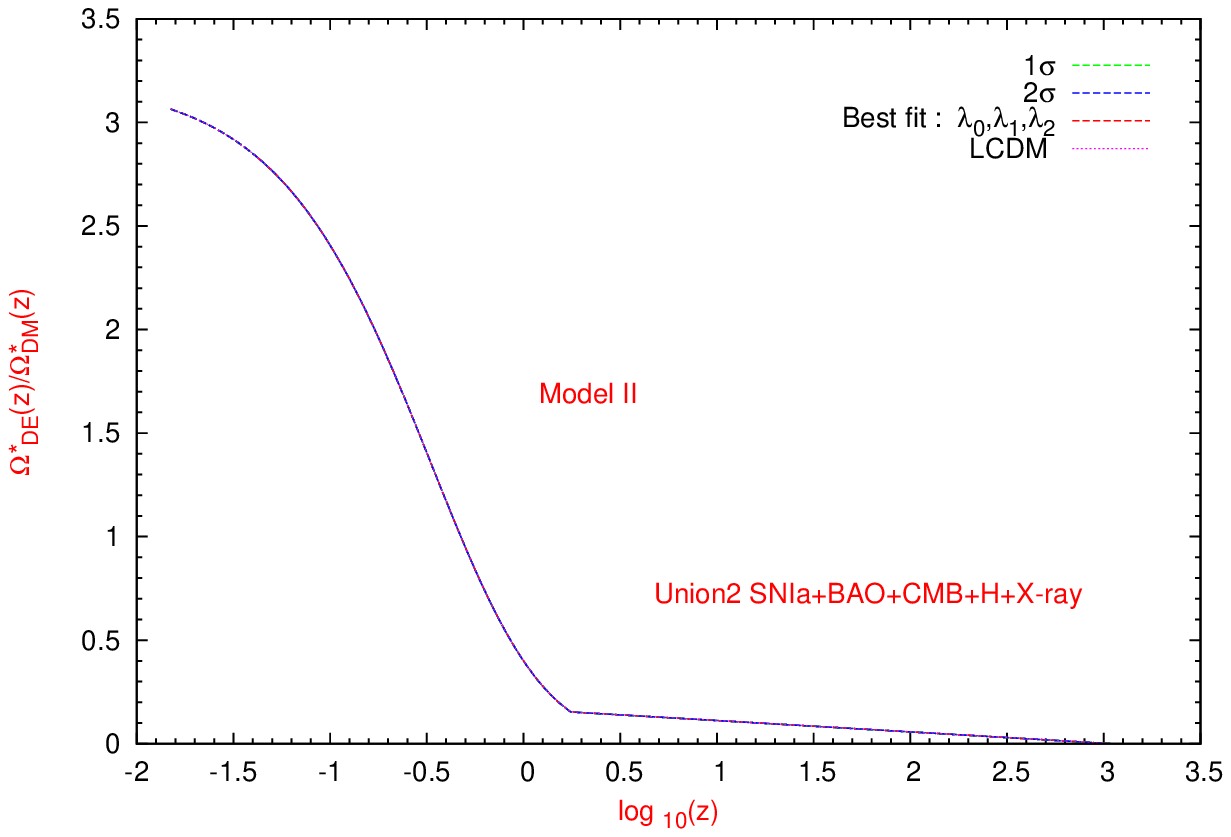}
 \includegraphics[width=8cm, height=60mm, scale=0.90]{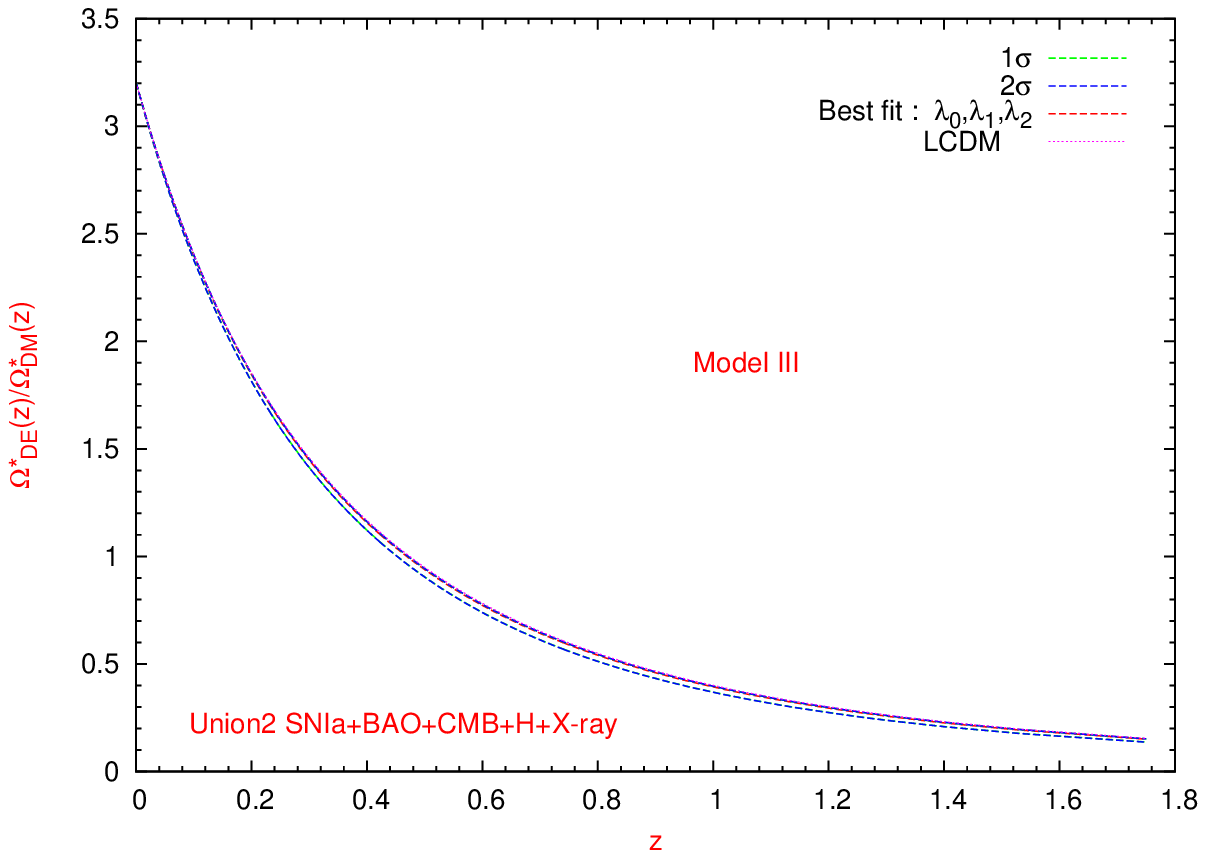}
 \includegraphics[width=8cm, height=60mm, scale=0.90]{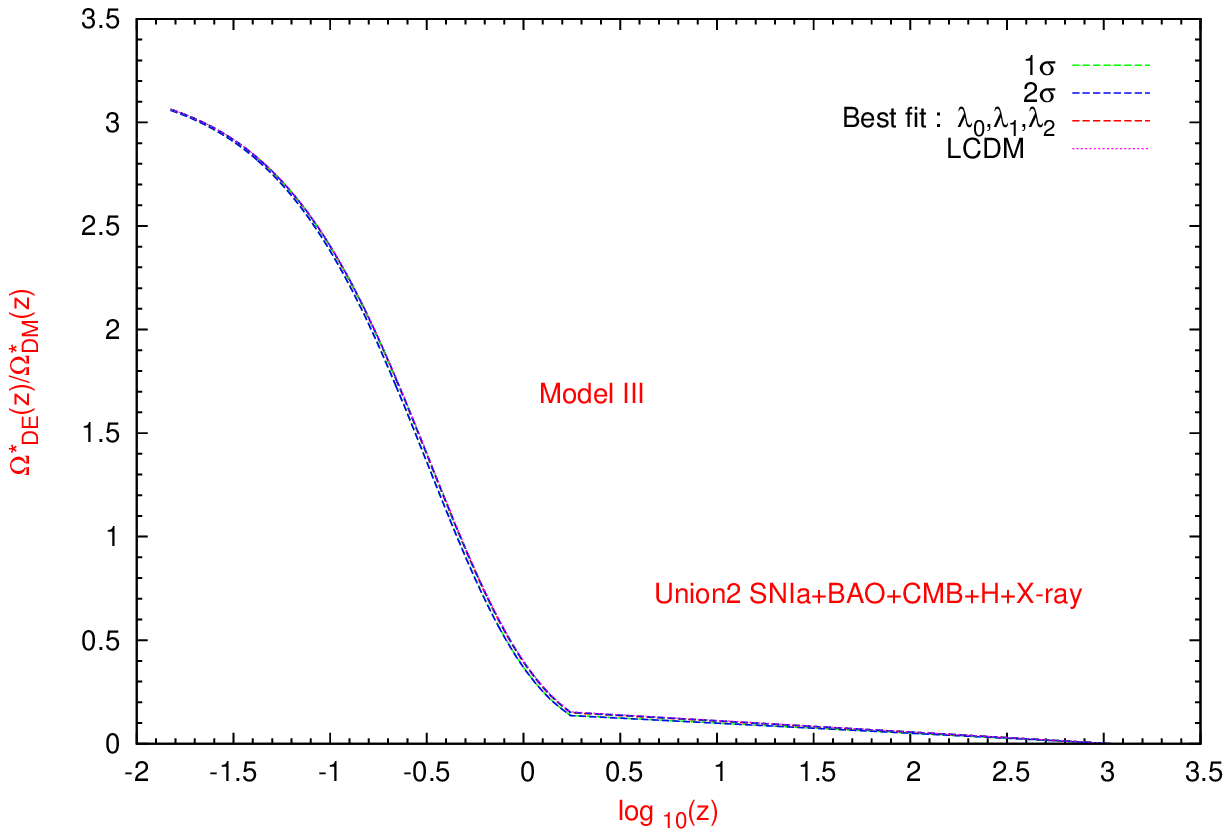}
 \includegraphics[width=8cm, height=60mm, scale=0.90]{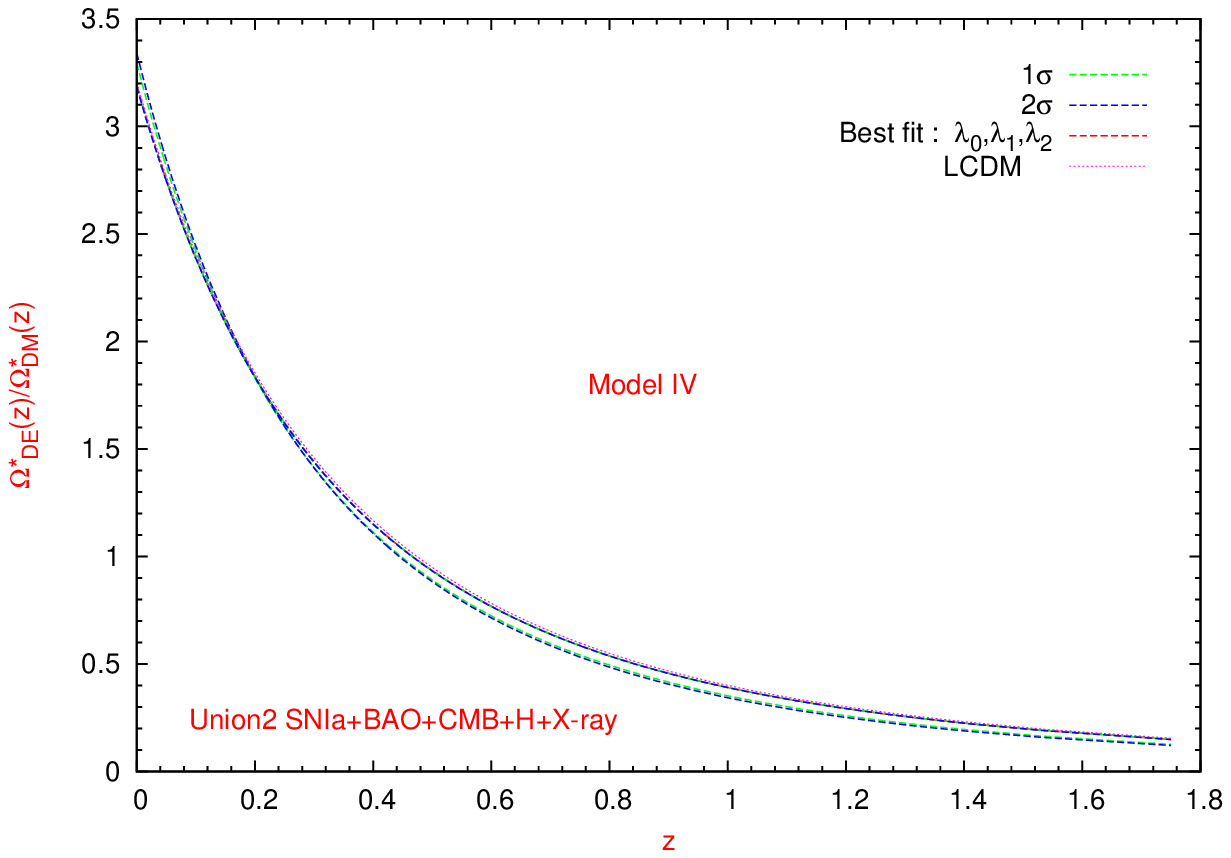}
 \includegraphics[width=8cm, height=60mm, scale=0.90]{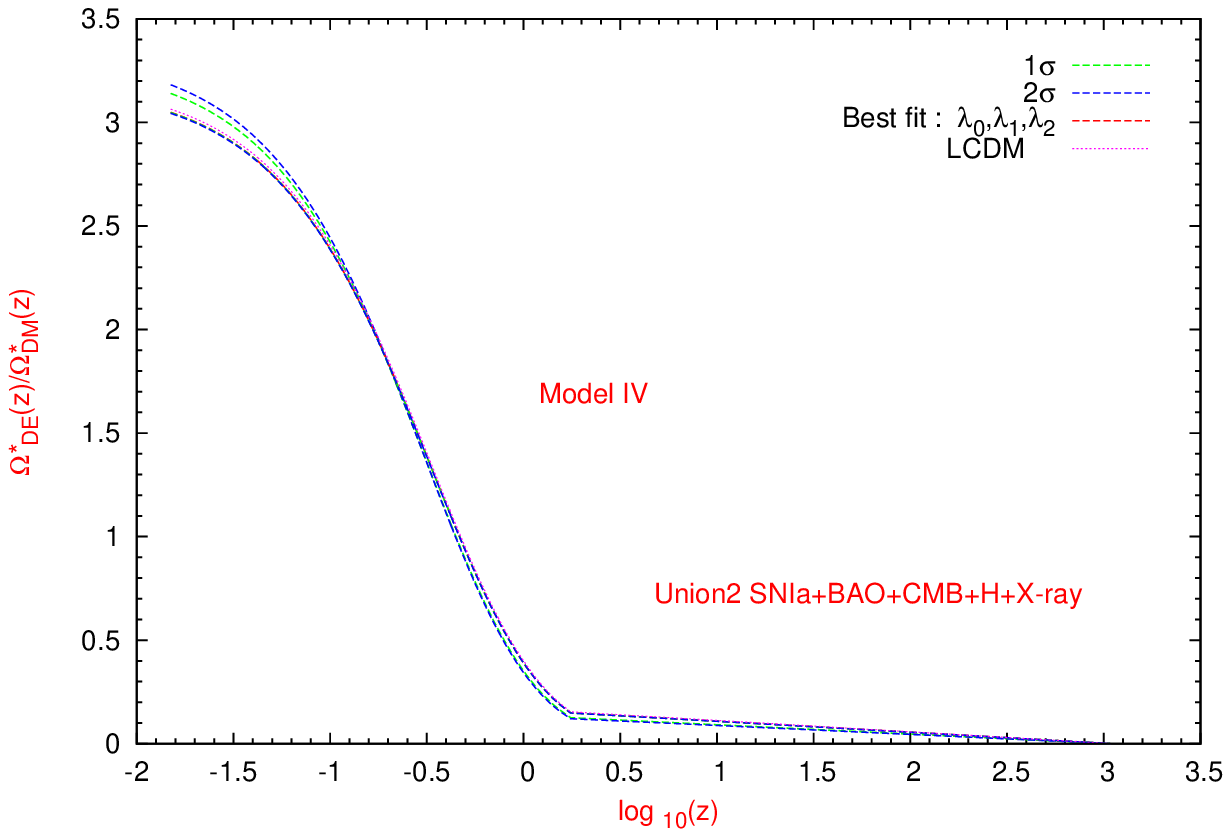}
 \caption{Comparison of the best fit reconstructed and its errors for the rate between dark density parameters $\Omega^{\star}_{DE}(z)/\Omega^{\star}_{DM}(z)$ in terms of the parameters $\lambda_0$, $\lambda_1$, $\lambda_2$ and the corresponding Chebyshev polynomials for the models II (left above panel), III (right above panel) and IV (left below panel),respectively.
 Red lines show best fit reconstructed results, while green lines and blue lines show reconstructed errors within the $1\sigma$ and $21\sigma$ confidence level errors,
 determined from a combination of SNeIa + BAO + CMB + H + X-ray dataset. In addition, note that within the $1\sigma$ and $2\sigma$ errors, our result and its errors are consistent with the constraints in the LCDM model. Furthermore, our best fit results and its errors at $1\sigma$ and $2\sigma$ are in LCDM model.}
 \label{ErrorsTotalRelationModels}
\end{figure}
\end{center}

The Figure \ref{ErrorsTotalRelationModels} shows the reconstruction of the rate between dark density parameters
$\Omega^{\star}_{DE}(z)/\Omega^{\star}_{DM}(z)$ as a function of the redshift and their confidence intervals at $1\sigma$ and $2\sigma$ for the
models II, III and IV already mentioned.

These last Figures show the superposition of the best estimates for every cosmological variable and their confidence intervals at $1\sigma$ and $2\sigma$.
The Tables \ref{BestestimationIIerrors}, \ref{BestestimationIIIerrors} and \ref{BestestimationIVerrors}, show the best fitted parameters with their respective $1\sigma$ and $2\sigma$.

\begin{center}
\begin{figure}
   \includegraphics[width=8cm, height=60mm, scale=0.90]{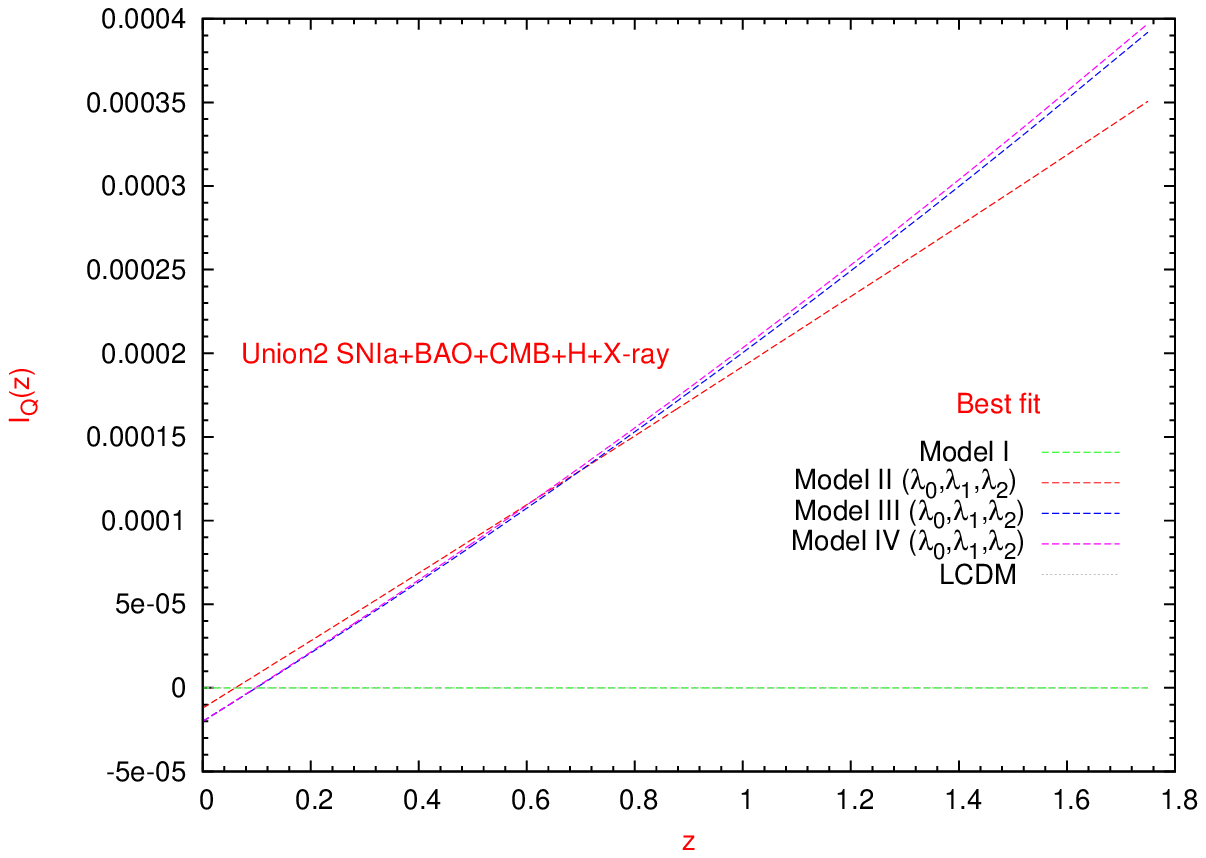}
   \includegraphics[width=8cm, height=60mm, scale=0.90]{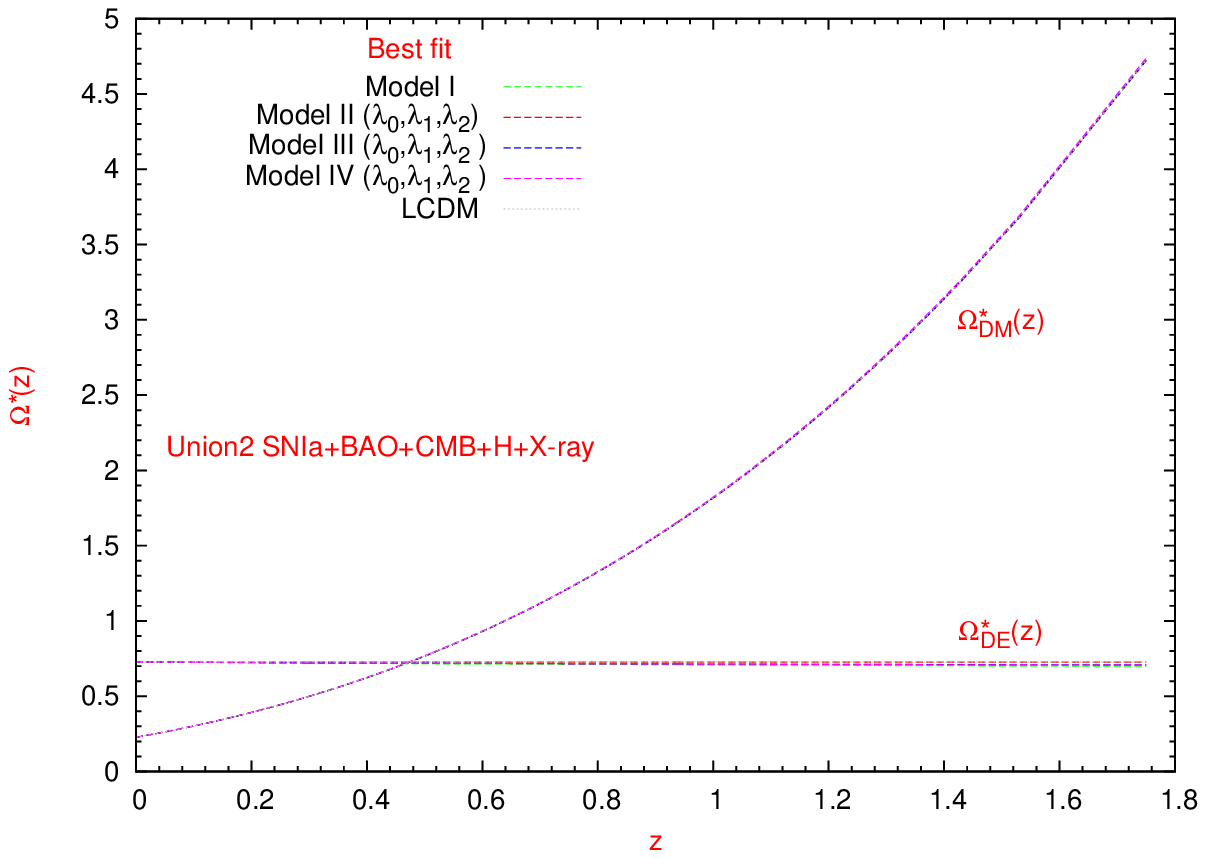}
   \includegraphics[width=8cm, height=60mm, scale=0.90]{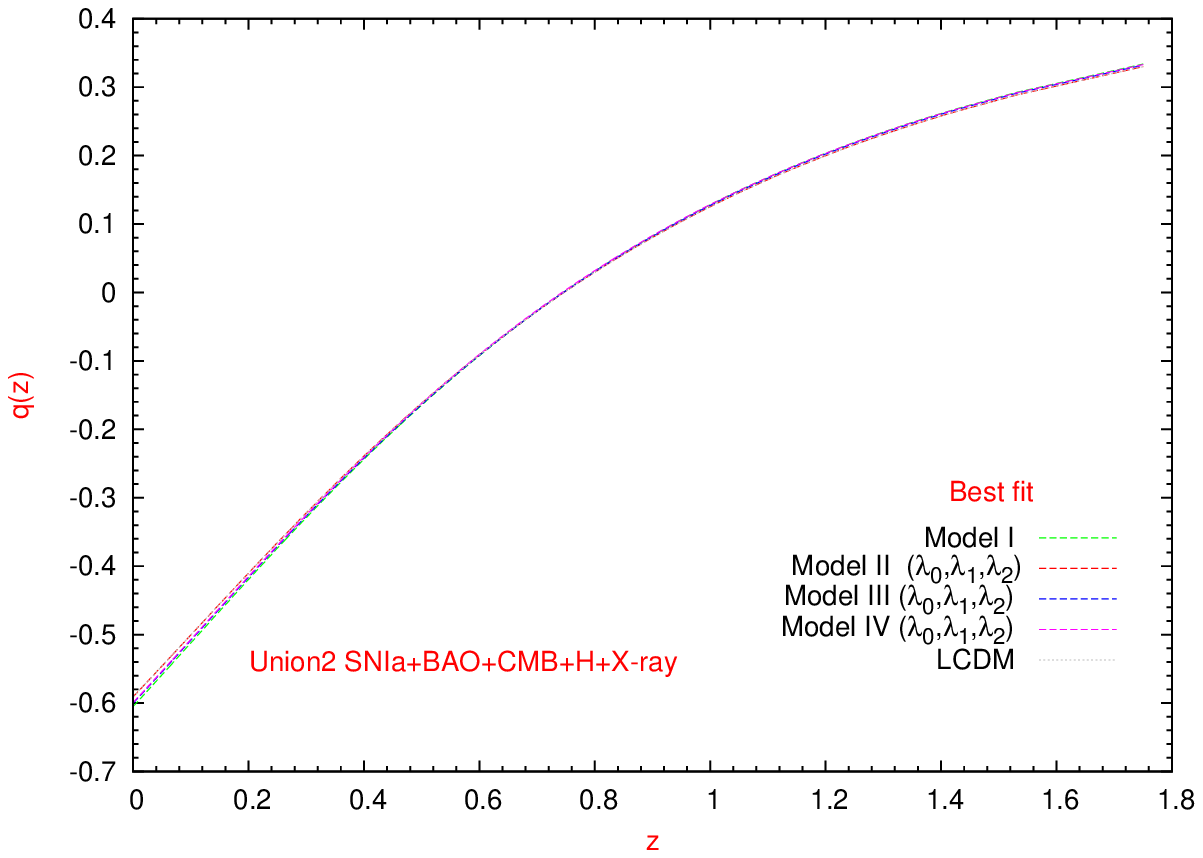}
   \includegraphics[width=8cm, height=60mm, scale=0.90]{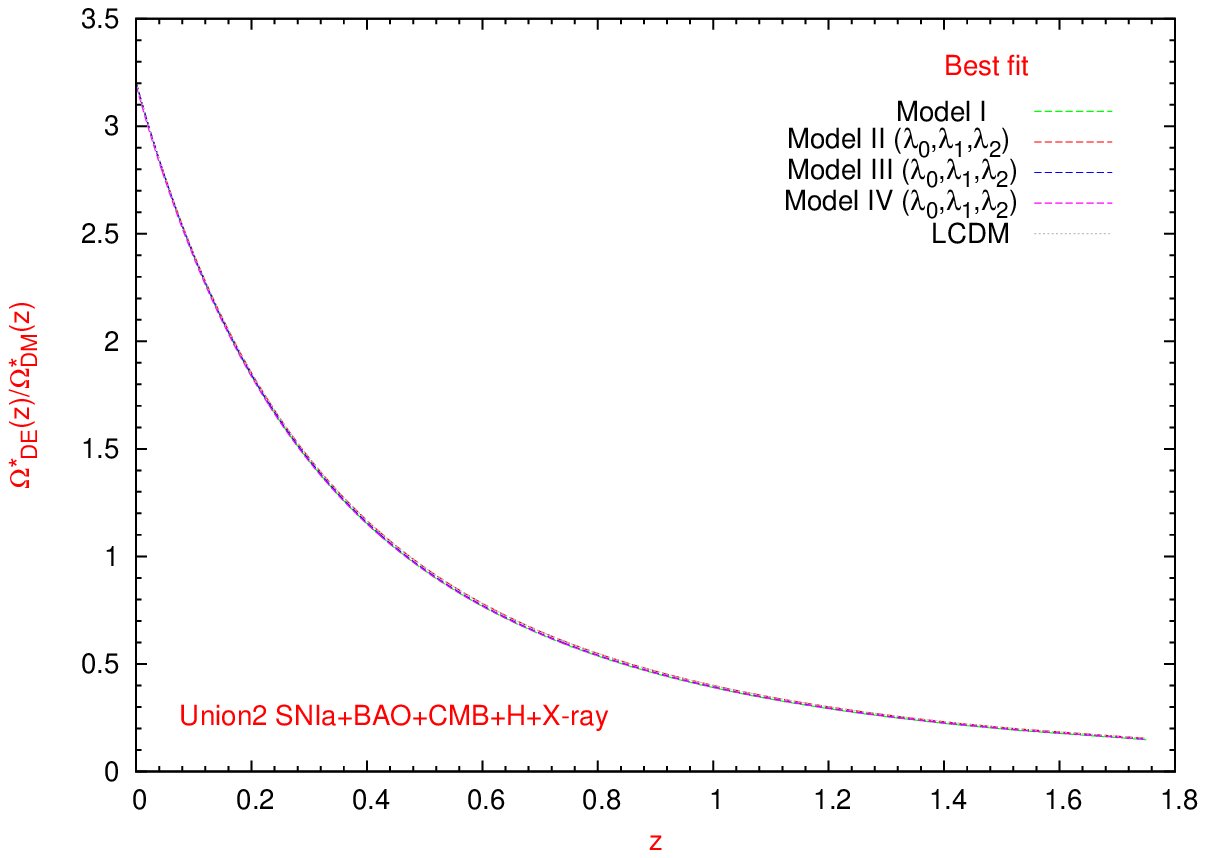}
 \caption{Superposition of the best estimates for the dimensionless interaction function
  ${\rm I}_{\rm Q}(z)$ (left above panel), the density parameters
  $\Omega^{\star}_{DM}(z), \, \Omega^{\star}_{DE}(z)$ (right above panel) and
  the deceleration parameter $q(z)$ (left below panel) as a function of the redshift
  for the models I (green line), II (red line), III (blue line) and IV (black line).
  By comparison, the LCDM model (pink line) is shown.
  The curves show the best estimates using the expansion of all the functions
  in terms of the first $n=2$ Chebyshev polynomials.
  Note that the reconstruction of the best estimate of the dimensionless
  interaction function ${\rm I}_{\rm Q}(z)$ for the models II, III and
  IV produces roughly the same curve and that the density parameter of dark
  energy is definite positive for all the range of redshift
  considered in the reconstruction.}
  \label{Total4Models}
\end{figure}
\end{center}

From Figures \ref{TotalExchangeModels} to \ref{ErrorsTotalExchangeModels} we notice that, for all
interacting models, the best estimates for the interaction function ${\rm I}_{\rm Q}(z)$ cross marginally the noninteracting line ${\rm
I}_{\rm Q}(z) = 0$ during the present cosmological evolution (at around $z \approx 0.09$) changing sign from positive values at the
past (energy transfers from dark energy to dark matter) to negative values at the present (energy transfers from dark matter to dark
energy). However, taking in account the errors corresponding to the fit using three parameters ($N=2$), we see that within the $1\sigma$ and $2\sigma$ errors,
it exists the possibility of crossing of the noninteracting line in the recent past at around the range $z \in (0.08, 0.12)$.
Crossings of the noninteracting line $Q(z) = 0$ have been recently reported at the references \cite{Cai-Su} (with an interacting term
$Q(z)$ proportional to the Hubble parameter) and \cite{Cueva-Nucamendi2012}. The direction of the change found in our work is in the same direction to the results published by these references where a crossing (from positive values at the past to negative values at the present was found at $z \simeq 0-0.12$.
On the other hand, we did not find the oscillatory behavior of the interaction function found by Cai and Su \cite{Cai-Su} who, using
observational data samples in the range $z \in [0, 1.8]$, fitted a scheme in which the whole redshift range is divided into a
determined numbers of bins and the interaction function set to be a constant in each bin.

From Figures \ref{TotalCoincidenceModels} to Figures \ref{ErrorsTotalCoincidenceModels} show that, for all the interacting models studied in this work,
the best estimates for the dark energy density parameter $\Omega^{\star}_{DE}(z)$ become to be definite positive at all the range of redshifts considered in the
data samples. However, this statement is conclusive because within the $1\sigma$ and $2\sigma$ errors for the fit with three parameters ($N=2$),
the $\Omega^{\star}_{DE}(z)$ becomes to be positive in all the range of redshifts considered (remember that we are using five data sets and then the space parameters permitted is remarkably reduced). According to the Model 4, note also that the behavior of dark matter $\Omega^{\star}_{DM}(z)$ density parameter, for $1\sigma$ and $2\sigma$ errors exist the possibility of being smaller compared to the noninteracting model (LCDM model).

From Figures \ref{TotalDecelerateModels} to Figure \ref{ErrorsTotalDecelerateModels}, show that, for all models, a transition from a deceleration
era at early times dominated by the dark and baryonic matter density to an acceleration era at late times corresponding to the present domain of the dark energy density.
This transition deceleration-acceleration to take place at redshift at around $z \approx 0.7$ (it was also obtained by \cite{campo-olivares2006}), while that
the rhythm of current universe at around $q_{0} \approx 0.6$. It is clear for our constraint, that the universe tends to an early time to accelerate and a milder
expansion rhythm at present.

We can see from the figure \ref{TotalCoincidenceModels} to \ref{ErrorsTotalCoincidenceModels} that, at the present, the dark energy density parameter becomes
$\Omega^{0}_{DE}(z) \approx 0.7$, which is sufficiently large to generate a non negligible dimensionless interacting term of the order of ${\rm I}^{0}_{\rm Q} \approx
-10^{-5}$, as is shown in the Figures \ref{TotalExchangeModels} and \ref{ErrorsTotalExchangeModels}. In fact, in these same figures we can appreciate that in the
interval of redshifts $z \in [0, 1090.89]$, the dimensionless interaction is in the range ${\rm I}_{\rm Q} \in [-10^{-5}, 50]$ for the Model II,
${\rm I}_{\rm Q} \in [-10^{-5}, 35]$ for the Model III and ${\rm I}_{\rm Q} \in [-10^{-5}, 25]$ for the Model IV respectively, corresponding at $2\sigma$ error.
The order of magnitude of this interaction is in agreement with the local constraints put on the strength of a constant dimensionless interaction derived from
the fit to a data sample of virial masses of relaxed galaxies clusters obtained using weak lensing, x-ray and optical data \cite{abramo}.

On the contrary, a recent study fitting CMB anisotropy data from the seven-year Wilkinson Microwave Anisotropy Probe (WMAP)
\cite{Komatsu2011}, the BAO distance measurements \cite{Percival}, the Constitution sample of SnIa \cite{Riess1998} and constraints on
the present-day Hubble constant, put stronger constraints on the magnitude of such dimensionless strength of the order of $\xi
\approx 10^{-2}-10^{-4}$ \cite{abramo2}.

Another important aspect concerns to study the coincidence problem, then from the Figures \ref{TotalCoincidenceModels} and \ref{ErrorsTotalCoincidenceModels}, we note that the
Models II and III, do not have the possibility of alleviating this problem. Their behaviors are in a degree similar to the LCDM model.
On the contrary, our Model IV, has the possibility of alleviating it at $2\sigma$ error. In the behavior of dark matter $\Omega^{\star}_{DM}(z)$ density parameter, within $1\sigma$ and $2\sigma$ errors, exist the possibility of being smaller compared to the corresponding noninteracting model (LCDM model).

It is interesting to note that from the Figure \ref{TotalAgeModels} to Figure \ref{ErrorsTotalAgeModels} and according to the
results of the tables \ref{ageuniverse2}, \ref{ageuniverse3} and \ref{ageuniverse4}, the best fitted values predicted for our models are consistent with the requirement that the universe be older that any of its constituents at a given redshift \cite{Dunlop1996}, \cite{Saini2000}.

From all our analysis, and from the Figure \ref{TotalModel3} and \ref{TotalModel4}, \ref{TotalCoincidenceModels} and \ref{ErrorsTotalCoincidenceModels},
show that the $1\sigma$ and $2\sigma$ constraints on the EOS parameter $w$ contain high probability of being in the phantom region, but our results are even consistent and compatible with the LCDM model at $1\sigma$ error.

In general when more cosmic observational data sets are added to constrain our model parameters space, the degeneracies between model parameters will be broken.
The reason comes from the fact that the constraints on our models are more stringent that the results obtained in reference \cite{Cueva-Nucamendi2012}.

\section{Conclusions.} \label{SectionConclusions}
In this paper, we developed theoretically a novel method for the reconstruction of the interaction function between dark matter and dark energy assuming an expansion of the general interaction term proportional to the Hubble parameter in terms of Chebyshev polynomials which form a complete set of orthonormal functions. To show how the method works, we applied it to the reconstruction of the interaction function expanding it in terms of only the first $N$  Chebyshev polynomials (with $N=1,2,3,4$) and fitted for the coefficients of the expansion assuming three models: (a) a DE equation of the state parameter $\omega =-1$ (an interacting cosmological $\Lambda$), (b) a DE equation of the state parameter $\omega=$ constant and (c) a DE equation of the state parameter $\omega=$ constant and $\Omega_{DM}=$ constant, respectively.
The fit of the free parameters of every model is done using the Union2 SNe Ia data set from ``The Supernova Cosmology Project'' (SCP) composed by 557 type Ia supernovae \cite{AmanullahUnion22010}, the Baryon Acoustic Oscillation (BAO), the Cosmic Microwave Background (CMB) data from 7-year WMAP, the Observational Hubble data and the Cluster X-ray Gas Mass Fraction.

Our principal results can be summarized as follows:
\begin{enumerate}
    \item Compared with the results of reference \cite{Cueva-Nucamendi2012}, our fitting results show a faster convergence
    of the best fitted values for the several cosmological variables considered in this paper when the numbers of parameters
    $N$ is increased in the expansion (\ref{eq:Coupling}).
    \item The best estimates for the interaction function ${\rm I}_{\rm Q}(z)$ prefer to cross the noninteracting line
    ${\rm I}_{\rm Q}(z)=0$ during the present cosmological evolution. This conclusion is independent of the numbers of coefficients (up to $N=4$ in this work) used in the expansion of ${\rm I}_{\rm Q}(z)$.
    The crossing implies a change of sign of ${\rm I}_{\rm Q}(z)$ from positive values at the past (energy transfers from dark energy to dark matter) to negative values at the present (energy transfers from dark matter to dark energy). This direction of decay is similar to the results found in the recent literature \cite{Cueva-Nucamendi2012} and is in disagreement with the oscillatory behavior reported in \cite{Cai-Su}.
    \item The statement above is conclusive because the existence of crossing of the noninteracting line ${\rm I}_{\rm Q}(z) = 0$ in some moment of the recent past is
    totally contained inside the $1\sigma$ and $2\sigma$ constraints given by current observations.
    \item We can state that is reasonable to expect that DE and DM interact via a small but calculable coupling and to be consistent with the observations at $2\sigma$ error.
    In this aspect, The the Cosmic Microwave Background (CMB) data provides a stringent constraints on the coupling.
    \item We confirm that adding the Baryon Acoustic Oscillation (BAO), the Cosmic Microwave Background (CMB), the Observational Hubble (H) and the Cluster X-ray
    Gas Mass Fraction (X-ray) data sets, generally reduces the allowed range of the model parameters. It is to say are strongly reduced compared with the results of the
    ref \cite{Cueva-Nucamendi2012}, which were obtained using only Union2 SNe Ia data.
    \item For all the interacting models studied in this work, the best estimates for the dark energy density parameter $\Omega^{\star}_{DE}(z)$ becomes positive
    definite in the range of redshifts considered in this work. This statement is conclusive because, within the $1\sigma$ and $2\sigma$ errors for the fit with three
    parameters ($N=2$), $\Omega^{\star}_{DE}(z)$ becomes positive in all the range of redshifts considered in the observational data sets.
    \item The $1\sigma$ and $2\sigma$ confidence intervals, for the EOS parameter $\omega$ considered in the marginalized models III and IV, preferred to be in the
    phantom region ($w<-1$). But exists a small probability for the EOS parameter $\omega$ of being in the quintessence region. This conclusion is consistent and is within
    of LCDM model.
    \item According to the Figures \ref{TotalDecelerateModels} and \ref{ErrorsTotalDecelerateModels} and to our results, we confirm that there is strong evidence that the
     universe is accelerated in the recent past, which is consistent with studies in \cite{Shapiro2006}, \cite{Gong2006}.
    \item we therefore conclude that our models are consistent with the properties of the cosmological variables and its results are also consistent with the LCDM model
    at $2\sigma$ error.
\end{enumerate}

Finally, we emphasize the importance of more accurate and extensive measurements. These observations will certainly provide a complementary tool to test the reality
of the current cosmological parameters, of the term of interaction between dark sectors and the existence of the crossing of the noninteracting ${\rm I}_{\rm Q}(z)=0$ line
and place stringent constraints on them, and as a consequence, to distinguish among the many alternatives models. We believe that the analysis, combined with the results
give us the possibility of understanding all that.
With the successful experience achieved to reconstruct the general interaction term between dark matter and dark energy, our next step is to extend our study to do a double reconstruction as much
the interaction term ${\rm I}_{\rm Q}(z)$ as dark energy equation of state parameter $\omega$ simultaneously, within the dark sectors. More efforts will be required on this way to carry out it.

\section{Acknowledgements.}
This work was in part supported by grants SNI-20733, CIC-UMSNH
No.4.8, UMSNH-CA-22. F. Cueva thanks support by CONACYT-SEP.

\end{document}